\documentclass[twocolumn,english,showpacs,preprintnumbers,amsmath,amssymb,floatfix,nofootinbib]{revtex4-1}

\usepackage[T1]{fontenc}
\usepackage[latin9]{inputenc}
\usepackage{color}
\usepackage{array}
\usepackage{amstext}
\usepackage{graphicx}
\usepackage{esint}
\usepackage{rotating}
\usepackage{slashed}
\usepackage{siunitx}
\usepackage[font=small,labelfont=bf]{caption}
\usepackage{soul,xcolor}

\usepackage{xcolor}

\usepackage[colorlinks = true,
linkcolor = magenta,
urlcolor  = blue,
citecolor = red,
anchorcolor = blue]{hyperref}

\usepackage{ulem}

\makeatletter

\begin{document}

\setstcolor{blue}


%
%
\title{ 
Neural Network QCD analysis of charged hadron Fragmentation Functions in the presence of SIDIS data
} 
%
%

\author{Maryam Soleymaninia$^{1}$}
\email{Maryam\_Soleymaninia@ipm.ir}

\author{Hadi Hashamipour$^{1}$}
\email{H\_Hashamipour@ipm.ir}

\author{Hamzeh Khanpour$^{2,1,3}$}
\email{Hamzeh.Khanpour@cern.ch}

\affiliation {
$^{1}$School of Particles and Accelerators, 
Institute for Research in Fundamental Sciences (IPM), 
P.O.Box 19395-5531, Tehran, Iran 
\\
$^{2}$Department of Physics, University of Science and 
Technology of Mazandaran, P.O.Box 48518-78195, 
Behshahr, Iran 
\\
$^{3}$Department of Theoretical Physics, Maynooth University, Maynooth, Co. Kildare, Ireland
}

\date{\today}

%
\begin{abstract}

In this paper, we present a QCD analysis to extract the 
Fragmentation Functions (FFs) of unidentified light charged 
hadron entitled as {\tt SHK22.h} from high-energy lepton-lepton 
annihilation and lepton-hadron scattering data sets. 
This analysis includes the data from all available single 
inclusive electron-positron annihilation (SIA) processes 
and semi-inclusive deep-inelastic scattering (SIDIS) 
measurements for the unidentified light charged hadron productions.
The SIDIS data which has been measured by the COMPASS 
experiment could allow the flavor dependence of 
the FFs to be well constrained.  
We exploit the analytic derivative of the Neural 
Network (NN) for fitting of FFs at 
next-to-leading-order (NLO) accuracy in the perturbative QCD (pQCD). 
The Monte Carlo method is implied for all 
sources of experimental uncertainties and the 
Parton distribution functions (PDFs) as well. 
Very good agreements are achieved between the {\tt SHK22.h} FFs set 
and the most recent QCD fits available in literature, namely {\tt JAM20} and {\tt NNFF1.1h}.  
In addition, we discuss the impact arising from the inclusion of  
SIDIS data on the extracted light-charged hadron FFs.
The global QCD resulting at NLO for charged hadron FFs provides 
valuable insights for applications in present and future 
high-energy measurement of charged hadron final state processes. 

\end{abstract}
%


\maketitle
\tableofcontents{}

%
\section{Introduction}\label{sec:introduction}
%

In perturbative quantum chromodynamic (pQCD), the 
hard-scattering processes in which a hadron is 
observed in the final state, include an integral 
part in which to be called as Fragmentation 
Functions (FFs) in the theoretical framework.
They are process-independent and universal quantity, 
and they show a non-perturbative transition of a parton into a hadron. 
FFs depend on the fraction of the longitudinal 
momentum of the parton taken by the hadron and the scale of energy~\cite{Metz:2016swz}.
FFs have a critical role in the current experimental
programs at Jefferson Lab, Future Electron-Ion Collider 
(EIC)~\cite{Accardi:2012qut,Aguilar:2019teb}, Future Circular Collider 
(FCC)~\cite{FCC:2018byv,FCC:2018evy} and LHC, 
and this aspect of their role leads to the main 
motivation for studying the collinear FFs in several phenomenological studies. 

Since FFs are non-perturbative quantities, they need to be determined from a QCD analysis of the corresponding experimental data sets.
The core experimental data sets are the single-inclusive 
electron-positron annihilation (SIA) from several 
collaborations and at different range of center 
of mass energy from $10.5$ GeV up to the $M_Z$~\cite{Belle:2020pvy, BaBar:2013yrg, Braunschweig:1990yd,Aihara:1988su,Buskulic:1995aw,Abreu:1998vq,Ackerstaff:1998hz,Abe:2003iy}. 
In order to disentangle all the different 
flavors of FFs for quark and anti-quark, 
in addition to the SIA data sample, one 
needs to taken into account some other observable.
Hence, the determination of FFs in the global QCD analyses also 
include the data on semi-inclusive deep-inelastic scattering (SIDIS) processes~\cite{COMPASS:2016xvm,HERMES:2012uyd,COMPASS:2018lzp,COMPASS:2020oyf} 
and the single-inclusive hadron production in proton-proton collisions~\cite{ALICE:2020jsh,STAR:2013zyt,PHENIX:2015fxo,CMS:2012aa,CMS:2011mry}.

Several theoretical analyses have been exploited 
on SIA, SIDIS and pp collisions data sets in QCD analysis 
to constrain the FFs of identified light charged hadrons~\cite{Khalek:2021gxf,Abdolmaleki:2021yjf,Moffat:2021dji,Bertone:2017tyb,Soleymaninia:2020ahn,Soleymaninia:2020bsq}, unidentified light charged hadron~\cite{Moffat:2021dji,Soleymaninia:2018uiv,Bertone:2018ecm} and heavy hadrons~\cite{Delpasand:2020vlb,Benzke:2019usl,Salajegheh:2019ach}.

There are two methods to calculate the FFs of 
unidentified light charged hadron. 
In the first method, for every flavor, they can be calculated as a sum of 
the FF sets of all identified light charged hadrons 
produced in the fragmentation of the given parton. 
Alternatively, the FFs of unidentified charged hadrons are implemented 
independently from a QCD analysis 
included the unidentified charged hadron experimental data directly. 

Before discussing our analysis, we will first review the FF sets of 
unidentified charged hadrons which have been recently calculated. 
In our recent analysis entitled 
{\tt SGKS20}~\cite{Soleymaninia:2020bsq}, we implemented the FFs 
of unidentified charged hadron up to 
next-to-next-to-leading order (NNLO) by 
taking advantage of the first method and determined 
in a simultaneous fit the FF sets for pion, kaon, proton, 
and the residual light charged hadrons.
All the available SIA data for pion, kaon, 
and proton and unidentified charged hadrons production 
have been considered in this analysis.

Another recent analyses for light charged hadron 
has been done by {\tt JAM} Collaboration~\cite{Moffat:2021dji} up to the 
next-to-leading-order (NLO). 
Their analysis includes all available SIDIS 
and SIA data for pion, kaon, and unidentified charged hadrons 
to calculate the FFs of pion, kaon, and charged hadrons. 
In addition, they have used data from inclusive
DIS and Drell-Yan lepton-pair production to 
calculate the PDFs simultaneously with the FFs. 
Accounting for unidentified charged hadrons, they 
have used the first method and add a fitted 
residual correction to the sum.

The other analysis of unidentified charged-hadron FFs has been 
presented by {\tt NNPDF} Collaboration~\cite{Bertone:2018ecm}.
They have utilized the second method to calculate the FFs up to the
NLO accuracy. The proton-proton data for 
unidentified charge hadron production has been added by 
means of Bayesian reweighting to the 
analysis based only on SIA data sets.
They have tried to complement their analysis of 
Ref.~\cite{Nocera:2017gbk} with the measurements 
of the charged hadron spectra in $pp$ collisions. 
Their study demonstrated that the inclusion of $pp$ data 
in a FF fit could provide a stringent constraint
on the gluon distribution FF.

Another analysis of unidentified charged hadron, 
based on the second method, have been done by {\tt SGK18} in which  
the FFs of 
charged hadrons have been done up to NNLO  by 
including all the unidentified charged hadrons 
from SIA experimental data sets~\cite{Soleymaninia:2018uiv}.

The main aim of this paper is to revisit our previous 
QCD analysis in Ref.~\cite{Soleymaninia:2018uiv} to implement a global 
QCD analysis for FFs of charged hadrons by adding the 
SIDIS data sets to the data sample, and applying the Neural Network (NN) technique. 
In this analysis, the hadron FFs are fitted
directly from all the experimental data for 
unidentified light charged hadrons production from SIA and SIDIS processes. 
Our main goal in this study is the inclusion of COMPASS 
SIDIS experimental data~\cite{COMPASS:2016xvm} as the only data set 
for the charged hadron production from SIDIS process. 

In recent years, Machine Learning (ML) has spread 
through all subjects of particle physics, specially collider physics.
One of the encouraging areas of application of 
such methods is improving our knowledge of 
non-perturbative quantities of nucleon such as PDFs 
and FFs~\cite{Khalek:2021gxf,Bertone:2017tyb,NNPDF:2017mvq}. 
In the light of this fact, we decided to use such method based on the 
artificial Neural Networks (NNs) to extract the 
light charge hadron FFs form QCD analysis of
the corresponding data sets. 
The modern optimization techniques are utilized in 
this project to minimize the bias of FFs parametrization 
by taking advantage of the Neural Network 
and also the Monte Carlo sampling method as a proper 
statistical treatment of experimental data 
uncertainties to obtain the probability density distribution from the data. 
For this purpose, we use the publicly available  
code called {\tt MontBlanc} in this analysis
which can be obtained from~\cite{MontBlanc}. 

This code is devoted to the extraction of collinear 
distributions of fragmentation functions. 
The code is an open-source package that provides a 
framework for the determination of the FFs, 
for many different kinds of analyses in QCD. 
So far, it has been developed to determine the FFs of the pion 
from experimental data for SIA and SIDIS data sets~\cite{Khalek:2021gxf}, and in our most
recent study to determine the fragmentation functions of $\Xi ^-/\bar{\Xi}^+$~\cite{Soleymaninia:2022qjf}.
{\tt MontBlanc} can analyze
the SIA data up to NNLO and the SIDIS data up to NLO in perturbation theory.
The framework in this code is combination of the Monte Carlo 
method to map the uncertainty distributions of FFs 
and Neural Networks to parameterize the FFs.

The structure of the paper is as follows:
In Sec.~\ref{sec:Theoretical}, we review the theoretical formalism for the 
inclusive hadron production in electron-positron annihilation and 
semi-inclusive deep-inelastic scattering (SIDIS) process, and the 
time-like evolution equation.
The parametrization of FFs in terms of Neural Network are also 
discussed in detail in this section.
Sec.~\ref{sec:Fit} includes our fitting methodology.
We also illustrate the Monte Carlo methodology adopted 
in our analysis to calculate the 
uncertainties of FFs and the optimal fit. 
This section also summarizes the SIA and SIDIS experimental 
data sets analyzed in this study, 
and the possible tensions between the data sets also examined.
The main results of {\tt SHK22.h} 
 are presented in Sec.~\ref{sec:results}. 
This section includes the {\tt SHK22.h} fit quality and 
the numerical results for the differential 
cross sections, and detailed comparison of theory predictions with 
the analyzed experimental data sets. 
We present the {\tt SHK22.h} light charged hadron FFs and detailed comparisons with
other results available in the literature, namely {\tt JAM20} and {\tt NNFF1.1h} FFs. 
We also discuss in this section the impact arising from the inclusion of 
SIDIS data on the extracted
light-charged hadron FFs.
Finally, we summarize our conclusions in Sec.~\ref{sec:Summary} 
and outline possible future developments.

%
\section{Theoretical Setup}\label{sec:Theoretical}
%

In this section, we present the theoretical backgrounds of the  
standard collinear factorization 
and discuss the perturbative and non-perturbative parts of the
cross-sections measurements in the SIA and multiplicities in SIDIS processes. 
Then, we present the time-like evolution equations and the 
splitting functions used for the FFs.
Finally, we discuss the parametrization of the FFs in terms of the Neural Networks 
in the presence of the SIDIS data. 
The Neural Network architecture for all the fitted FFs and the Monte 
Carlo uncertainty propagation also will be 
discussed in this section.

%
\subsection{SIA and SIDIS factorization }\label{sec:Factor}
%

In the standard collinear factorization, we separate the QCD cross sections into the 
perturbative partonic hard factors which are convoluted with the 
non-perturbative partonic or hadronic distribution functions. 

In the present analysis, we consider the SIA process which is given by,

\begin{eqnarray}
\label{eq:SIA}
e^{+} + e^{-} 
\rightarrow 
(\gamma, Z^0) 
\rightarrow 
{\it h^\pm} + X \,,
\end{eqnarray}

and the semi-inclusive charged hadrons production in the lepton-nucleon 
deep inelastic scattering which can be written as,

\begin{eqnarray}
\label{eq:SIDIS}
{\ell} + N 
\rightarrow {\ell} + 
{\it h^{+}/h^{-}} +
X \,.
\end{eqnarray}

According to the collinear factorization theorem, the cross-sections for the  
two processes mentioned above can be written as, 

\begin{eqnarray}
\label{eq:SIA-cross}
\sigma ^{\rm {SIA}}&=&
\hat{\sigma} 
\otimes \rm {FF},\nonumber\\
\sigma ^{\rm {SIDIS}}&=&
\hat{\sigma} 
\otimes \rm {PDF} 
\otimes \rm {FF}\,,
\end{eqnarray}

where the $\hat{\sigma}$ indicates to the process dependent 
perturbative partonic cross-section. 
The parton distribution function (PDF) and FF are non-perturbation functions.

The details of the computation of the SIA cross-sections 
are provided in some studies available in 
the literature, and we refer the reader to 
the Refs.~\cite{Soleymaninia:2018uiv,Bertone:2017tyb} 
for a clear review.

The basic cross-sections for the charged hadron 
production with four-momentum $p_h$ in deep 
inelastic scattering of a lepton with momentum $l$ from a 
nucleon with momentum $p$ can be written as,

\begin{eqnarray}
\label{eq:sidis-cross-sec}
\nonumber
\frac{d\sigma^{h}} {dx\, 
dy\, dz_{h}} &=&
\frac{2\, \pi\alpha^2}
{Q^{2}}
\left[ \frac{(1 + 
(1-y)^{2})}{y} 2\, 
F_1^{h}(x, z_{h}, Q^{2}) 
\right. \\
&& \left. +
\frac{2 (1-y)}{y} 
F_{L}^{h}
(x, z_{h}, Q^{2}) 
\right]\,,
\end{eqnarray}

which are functions of Bjorken scaling variable $x=\frac{Q^2}{2p.q}$, 
the charged hadrons fragmentation scaling variable $z_h= \frac{p_h.p}{q.p}$, 
energy transfer or inelasticity $y=\frac{Q^2}{xs}$, 
and the four momentum transfer squared of the virtual photon $Q^2=-q^2$.
In this equation, the $\alpha$ indicates the fine-structure constant.

The structure functions $F_1^h$ and $F_L^h$ in Eq.~\ref{eq:sidis-cross-sec} 
are the relevant inclusive DIS 
structure functions in which at NLO accuracy are given by,

\begin{eqnarray}
\label{eq:F1}
\nonumber
F_{1}^{h}(x, z_{h}, Q^{2}) &
\!\!=\!\!& 
\frac{1}{2} \sum_{q,
\overline{q}} e_q^{2} 
\Bigg\{  q (x,Q^2)  
D^h_{q} (z_h,Q^2) \\
\nonumber
&& +\frac{\alpha_s(Q^2)} 
{2\pi} \bigg[ q
\otimes  C^1_{qq} 
\otimes D^h_{q} \\
\nonumber
&& + q  \otimes  
C^1_{gq} \otimes D^h_{g}  \\
&& + g  \otimes
C^1_{qg} \otimes 
D^h_{q} \bigg] 
(x, z_{h}, Q^{2})\! 
\Bigg\},
\end{eqnarray}

\begin{eqnarray}
\label{eq:FL}
\nonumber
F_{L}^{h}(x, z_{h}, Q^{2}) & 
\!\!=\!\! & 
\frac{\alpha_s(Q^2)}{2 \pi}
\sum_{q, \overline{q}} 
e_q^{2}  \bigg[ q
\otimes  C^L_{qq} 
\otimes D^h_{q}   \\
\nonumber
&& + q  \otimes  
C^L_{gq} \otimes D^h_{g} \\
&& + g  \otimes
C^L_{qg} \otimes 
D^h_{q} \bigg] (x, z_{h}, Q^{2})\,.
\end{eqnarray}

The  convolution symbol $\otimes$ in equations above is defined as, 

\begin{eqnarray}
\label{eq:convolution}
\nonumber
q(x)  &\otimes &  C(x, z_{h}) 
\otimes D^h(z_{h}) \nonumber\\
&=&\int _x^{1} \frac{dx^{\prime}}
{x^{\prime}} 
\int _{z_{h}}^1\frac{d{z^
	{\prime}_{h}}}
{z^{\prime}_{h}}q(\frac{x}
{x^{\prime}})
c(x^{\prime},z^{\prime}_{h})
D(\frac{z_{h}}{z^{\prime}_{h}})\,.
\end{eqnarray}

In Eqs.~\ref{eq:F1} and \ref{eq:FL}, the PDFs inside the 
nucleon are denoted by $q$, $\bar{q}$ and 
$g$, and $D_q^h$, $D_{\bar{q}}$, and $D_g$ denote the FFs. 
The hard scattering coefficient functions $C_{ij}^{1,L}$ related to 
the $F^h_1$ and $F^h_L$ structure functions 
admit the usual perturbative expansion. 
Currently, these coefficient functions are known up to $\mathcal{O}(\alpha_s)$, i.e., 
NLO and can be found for example in Refs.~\cite{Altarelli:1979kv,Graudenz:1994dq}. 
Although the coefficient functions for structure function 
in SIA process is known up to the $\mathcal{O}(\alpha_s^2)$, i.e., NNLO, 
the full set of NNLO accuracy corrections are not known for the 
SIDIS process, and the  structure functions
only known up to NLO. 
Hence, our QCD calculations in the perturbative part 
are limited to the $\mathcal{O}(\alpha_s)$, i.e. NLO, accuracy. 

It should be noted here that we determine the charged hadron FFs in 
the Zero-Mass Variable-Flavor-Number Scheme (ZM-VFNS) 
in which all the active flavors are considered to be massless. 
However, the masses of heavy quarks require to be introduced
during the subschemes to determine the number of 
active flavors based on the heavy-quark thresholds. 
In this analysis, the charm and bottom masses 
are considered to be fixed at $m_c = 1.51$ GeV and $m_b = 4.92$ GeV, respectively. 

As a final point, we should highlight here that 
we have used the proton PDF set 
\texttt{NNPDF31}~\cite{NNPDF:2017mvq} at NLO accuracy to 
calculate the cross-section in the SIDIS process.
The SIDIS data included in our QCD fit, 
has been measured  by the 
COMPASS Collaboration in which the muon beam collides with the 
Lithium ($^6$LiD) target. 
In our present study, we focus on the analysis with the proton PDFs 
without considering the nuclear corrections. 
We plan to revisit this analysis  in the near future 
to study the impact of such nuclear effect  
along with the target mass corrections (TMC) and 
hadron mass corrections.

%
\subsection{pQCD and time-like evolution}\label{sec:pQCD}
%

The integrated FFs are universal and process independent quantity 
in the sense that 
$D^h(z,Q)$ is the same in processes 
like $e^+e^-$ annihilation, SIDIS, and hadronic collisions.
FFs depend on an additional parameter called the 
renormalization scale $\mu$ because of the QCD dynamics. 
Based on the pQCD approach, the structure of the 
evolution equations for the 
unpolarized integrated FFs generally is given by,

\begin{eqnarray} 
\label{eq:evol}
&&\frac{dD^{h}_{i}(z, \mu^{2})}
{d \ln \mu^{2}}  
= 
\nonumber\\ 
&&\frac{\alpha_s(\mu^2)}{2\pi} 
\sum_j \int_z^{1} 
\, \frac{dm}{m} \, P_{ji}(m, 
\alpha_s(\mu^{2})) 
\, D_j^{h} \Big( \frac{z}{m}, 
\mu^{2} 
\Big) \,, \nonumber\\
\end{eqnarray}

where $P_{ji}$ is the matrix for the time-like splitting functions 
and have a perturbative expansion of the form, 

\begin{eqnarray} 
\label{eq:splitting}
P_{ji}(m,\alpha_s(\mu^{2})) &=& 
P_{ji}^{(0)}(m) 
+ \frac{\alpha_s(\mu^{2})}{2\pi} 
\, P_{ji}^{(1)}(m)\nonumber\\
&+& \Big( \frac{\alpha_s(\mu^{2})}
{2 \pi} \Big)^{2} P_{ji}^{(2)}(m) 
+ \ldots \,.
\end{eqnarray}

The NLO time-like splitting functions $P_{ji}$ 
have been computed in Refs.~\cite{Rijken:1996vr,Mitov:2006wy}.
Usually the evolution equation is decomposed into a singlet sector 
comprising the gluon and the sum of all quark and antiquark FFs, and the
non-singlet sector for quark-antiquark and flavor differences.

The range of applicability for the FFs is limited to the 
medium-to-large range of $z$ value. 
There are two reasons for such limitation. 
First, the strong singular behavior in the time-like 
splitting functions when $z \rightarrow 0$. 
Second, the produced hadrons in the final state are 
considered to be massless. 
In supplementary to this, the time-like splitting functions have a 
logarithmic piece $\simeq \ln^2 z / z$ in the NLO part which leads to negative 
FFs for $z\ll 1$ in the influence of 
the $Q^2$ evolution and it leads to unphysical, negative cross sections. 
In addition, at small $z$, the finite mass corrections become more important. 
Therefore, in this global QCD analysis we limit ourselves  to
the kinematic regions in which mass corrections and 
the singularity of small-$z$ behavior in the evolution
kernels are negligible, as discussed in Sec.~\ref{sec:data}.

%
\subsection{Neural Network and flavour decomposition}\label{sec:NN}
%

As we discussed in Ref.~\cite{Soleymaninia:2018uiv}, 
inclusive SIA data allow for 
the determination of only the summed quark and antiquark FFs 
by including the total inclusive, light-, charm- and bottom- quark 
tagged cross sections

\begin{eqnarray}
\label{eq:SIAcombinations}
D^{h^{\pm}}_{u^+}, ~D^{h^{\pm}}_{d^++s^+}, 
~D^{h^{\pm}}_{c^+}, ~D^{h^{\pm}}_{b^+}, 
~D^{h^{\pm}}_g.
\end{eqnarray}

It has been shown that adding the SIDIS data 
sets to the data sample could provide a direct 
constraint on the individual $q$ and $\bar{q}$ FFs for light quarks. 
We adopt the following parametrization basis for $h^+$, 

\begin{eqnarray}
\label{eq:combinations}
D^{h^+}_{u}, ~D^{h^+}_{\bar{u}}, ~D^{h^+}_{d+s}, 
~D^{h^+}_{\bar{d}+\bar{s}}, ~ D^{h^+}_{c^+}, 
~D^{h^+}_{b^+}, ~ D^{h^+}_g\,.
\end{eqnarray}

Taking into account the SIDIS data in the QCD fit, 
the above combinations of quark FFs $D^{h^{\pm}}_{u^+}$ and $D^{h^{\pm}}_{d^++s^+}$  
in Eq.~\ref{eq:SIAcombinations}
are considered to be decomposed.
The heavy distributions are assumed to be symmetric. It reads, 

\begin{eqnarray}
\label{eq:Symmetry}
 D^{h^+}_q = D^{h^+}_{\bar{q}},~~~~~~~~ q = c, b\,.
\end{eqnarray}

Hence, by 
adding SIDIS COMPASS data, the number of independent distributions increases to the seven. 
We observed that under these flavor combinations of quark FFs for hadron production, 
the best fit quality and accuracy
can be achieved. 
In order to choose the best parametrization basis, 
we study different scenarios.
However, in the most general case, 
we disentangle all light flavors. 
The data sets included can not constrain well enough all 6 light, 2 heavy quarks, and gluon FFs. 
In another case, we assume symmetry just between $D^{h^+}_d = D^{h^+}_s$ and a 
separate parametrization are implied for the 
$D^{h^+}_{\bar{d}}$ and $D^{h^+}_{\bar{s}}$. 
In particular, we find this assumption leads to a deterioration of 
the quality of FFs and the fit, and then we omitted such assumptions.

We finally note that the COMPASS measurements have been reported for 
both positive $(h^+)$ and negative $(h^-)$ charged hadron separately. 
We consider a relation based on charge conjugation between $h^+$ and $h^-$ as follow,

\begin{eqnarray}
\label{eq:Conjugate}
 D^{h^-}_{q(\bar{q})}(z,Q)&=&
 D^{h^+}_{\bar{q}(q)}(z,Q),
 \nonumber\\
 D^{h^-}_g(z,Q)&=&D^{h^+}_g(z,Q).
\end{eqnarray}

Hence, one can obtain the $h^-$ FFs in terms of the $h^+$ FFs.

The $h^+$ FFs for every parton flavor in terms 
of a Neural Network is defined at the initial scale of 
$Q_0 = 5$ GeV, which is given by,

\begin{eqnarray}
\label{NN}
zD^{h^+}_i(z,Q_0)
= 
(N_i(z, \theta) - N_i(1, \theta))^2 \, ,
\end{eqnarray}

Here $N_i(z,\theta)$ denotes the output of the Neural Network  
and $\theta$ stands for the (internal) parameters of the Neural Network. 
Note that the result of Neural Network  at $z=1$ is subtracted in order to 
satisfy the requirement that FFs should vanish at this point. 
Also, the result is squared to make sure that FFs always stay positive. 
A few remarks on the construction of Neural Networks are in order. 
First, in this analysis we use a simple yet efficient\footnote{ 
This happens because of the universal approximation theorem~\cite{Csaji:2001} states that a 
simple feed forward Neural Networks such as one used here can 
represent any function in a specific interval.} 
Neural Network  structure which has only one hidden layer.  
Second, we choose to use 20 neurons (nodes) in the hidden layer, 
admittedly this number is 
somewhat arbitrary, smaller number of nodes may result in an 
equally accurate fit. In an effort to examine the effect of choosing an alternative NN architecture on the obtained results, we performed another analysis with different configurations, i.e. \{1-9-9-7\} architecture.
Our examination shows that the results are basically unchanged, which is in agreement with other studies available in the litrature~\cite{Khalek:2021gon}. We believe that this stability despite changing the NN shows that our FFs are not driven by hyperparameters such as the number of hidden layers or the number of nodes in each layer but by input experimental data.
Third, number of replicas in this analysis is 200, regardless, 
requirements of replica method can be achieved by smaller number of replicas, 
for example 100~\cite{Khalek:2021gxf}.

\section{Fitting procedure}\label{sec:Fit}
%

In this section, we first discuss the minimization 
strategy and the uncertainty 
propagation estimation using the  Monte Carlo method. 
Then we illustrate a comprehensive set of measurements 
of the charged hadron 
production in electron-positron annihilation SIA, and 
the lepton-nucleon SIDIS processes 
as well. We also discuss the kinematic cuts on the 
experimental data in which a description 
in the framework of pQCD can be expected to work well. 
Finally, we comment on the tension between 
COMPASS data with some of other SIA data sets 
analyzed in {\tt SHK22.h} study.

%
\subsection{Minimization and uncertainty propagation method}\label{sec:UNCERTAINTY}
%

It goes without saying that any measurement in 
high-energy particle physics has 
an uncertainty associated with it, this introduces the problem of 
understanding the effect that this produces 
on other quantities referred to as uncertainty propagation. 
Two widely used methods to propagate the experimental uncertainties 
to the FFs or observables are; one the Hessian method and 
second the Monte Carlo(MC) or replica method. 
The Monte Carlo approach is nowadays widely used in various QCD 
analyses~\cite{Sato:2016tuz,Sato:2016wqj,Moutarde:2019tqa,NNPDF:2017mvq,AbdulKhalek:2019mzd}. 
This method estimates the parameters posterior probability 
distribution by performing a number of fits. 
Every fit is independent and performed on a 
pseudo data set (a replica) resulting in an optimal set. 
The results of all fits performed then \emph{learn} 
the probability distribution which defines both the central
value(mean of the probability distribution) and the 
uncertainties of FFs (standard deviation of the 
probability distribution). In order to properly into account the uncertainty of the PDFs used in SIDIS observables, we use the similar method as developed and adopted in Ref.~\cite{Khalek:2021gxf} to ensure that the PDF uncertainty is propagated into FFs. In SIDIS calculations each time a different proton replica of \texttt{NNPDF3.1} is chosen at random from \texttt{NNPDF31\_nlo\_pch\_as\_0118} set.
	
In order to perform the QCD analysis, one mainly 
applies the maximum log-likelihood method which in 
turn reduces to minimum $\chi^2$ under usual assumptions. 
In this case, the problem of finding the optimal parameters of a 
parametric form or optimal Neural Network parameters is 
equivalent to minimizing the $\chi^2$ function at hand. 
There are a few ways to minimize a $\chi^2$ 
function in a QCD fit
which is based on Neural Network; one that readily comes to mind is 
explicit differentiation and calculation of global 
minimum directly, this approach is inefficient and 
sometimes straight impossible. 
It is therefore natural to look for numerical methods 
such as genetic algorithm used by \texttt{NNPDF} for PDFs~\cite{NNPDF:2014otw}, 
stochastic gradient descent methods used by 
\texttt{nNNPDF} for nuclear PDFs~\cite{AbdulKhalek:2020yuc}, and 
trust-region methods as provided by \texttt{Ceres Solver}~\cite{ceres-solver} and 
utilized by \texttt{MAPFF}~\cite{Khalek:2021gxf} and {\tt SHKS22}~\cite{Soleymaninia:2022qjf}.
In this analysis, we adopt the later method which is 
implemented in the \texttt{MontBlanc} package~\cite{MontBlanc}. 
The $\chi^2$ function that is subject to minimization is defined as follows,
	
\begin{equation}
\chi^{2(k)} \equiv  
\left(\boldsymbol {\rm T}(\boldsymbol 
{\theta}^{(k)})- 
\boldsymbol {x}^{(k)}\right)^{\rm T} 
\cdot \boldsymbol {\rm C}^{-1}\cdot 
\left(\boldsymbol{\rm T}
(\boldsymbol {\theta}^{(k)})- 
\boldsymbol {x}^{(k)}\right)\,.
\end{equation}

In the equation above, the $\boldsymbol{x}^{(k)}$ is the $k$-th replica, 
$\boldsymbol{\rm T}(\boldsymbol{\theta}^{(k)})$ is 
the theoretical prediction for the $k$-th replica 
based on the parameters of Neural Network ($\boldsymbol{\theta}^{(k)}$), and 
$\boldsymbol{\rm C}$ is the covariance matrix of 
the data which contains all information on 
the uncertainties and correlations. 
In view of the fact that Neural Networks by construction are 
redundant i.e. number of parameters is 
typically much bigger than that of a functional 
form parametrization. 
For this reason the $\chi^2$ is by convention normalized to 
$N_\mathrm{dat}$, number of data points.

%
\subsection{Data sets selection}\label{sec:data}
%

In the present analysis, we make use of all  
available experimental data on the 
charged hadron production in SIA and 
SIDIS processes to determine the 
unidentified light-charge hadron FFs. 
In our previous analyses~\cite{Soleymaninia:2020bsq,Soleymaninia:2022xnx,Soleymaninia:2018uiv}, 
we have included 
all available SIA experimental data sets to determine the 
FFs of charged hadron production.
The SIA measurements are reported 
as a sum of the observables for the 
positive and negative charged 
hadron production. 
However, the 
observables in SIDIS process are separated into positive and negative 
charged hadron production.

%
\begin{figure}[htb]
\vspace{0.50cm}
\resizebox{0.50\textwidth}{!}{\includegraphics{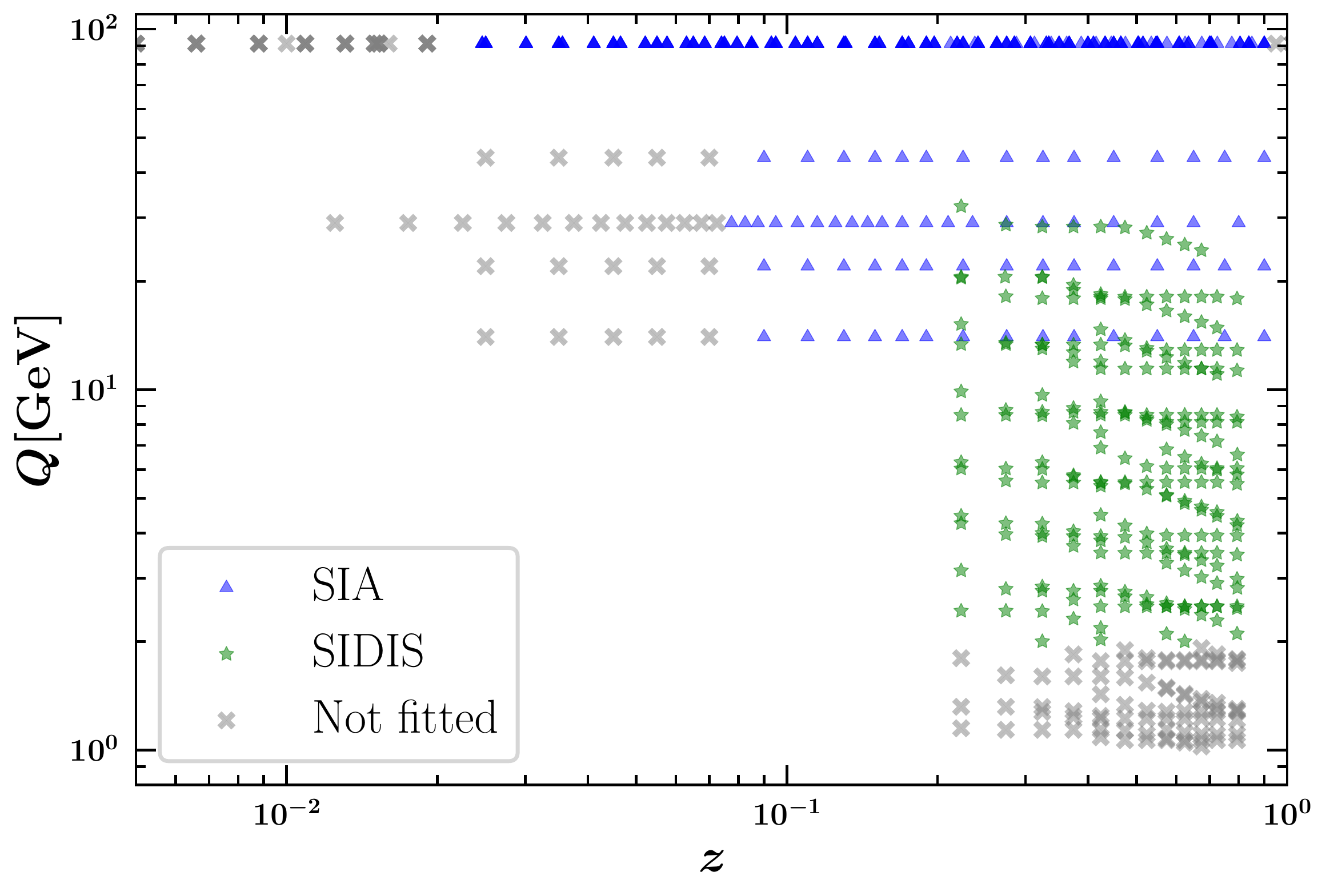}}
\begin{center}
\caption{ 
\small 
Kinematic coverage in the ($z$, $Q$) plane of 
the SIA and SIDIS data sets analyzed in 
{\tt SHK22.h} study. 
The data points for SIA are shown as blue, 
the SIDIS data points are shown as green; and the 
gray points are excluded by kinematic cuts as 
discussed in the text.
}
\label{fig:zQ}
\end{center}
\end{figure}
 
The kinematic coverage in the ($z$, $Q$) plane of 
the SIA and SIDIS data sets analyzed in {\tt SHK22.h} 
analysis are shown in Fig.~\ref{fig:zQ}.
The data points for SIA are shown as blue, 
the SIDIS data points are shown as green; and the 
gray points are excluded by kinematic cuts as 
discussed in the text.
They data sets contain all analyzed flavor-untagged and tagged 
measurements which are reported by different experiments. 
These data sets include the {\tt TASSO}~\cite{Braunschweig:1990yd} 
experiment at DESY; the {\tt TPC}~\cite{Aihara:1988su} 
experiment at SLAC, {\tt ALEPH}~\cite{Buskulic:1995aw}, 
{\tt DELPHI}~\cite{Abreu:1998vq} and 
{\tt OPAL}~\cite{Ackerstaff:1998hz} experiments at CERN 
and {\tt SLD}~\cite{Abe:2003iy} experiments at SLAC.

As one can see from Table~\ref{tab:chi2-SHK22.h}, 
there are several different measured observables available 
for these data sets. 
The experimental collaborations have reported 
total inclusive and light-, charm-, and bottom-tagged cross sections. 
Determination of the separate FFs for light and 
heavy quark flavors is provided by the light 
and heavy flavor tagged measurements.
For a detailed discussion of the SIA data sets, we refer the reader to our 
previous study on the light-charged hadrons FFs.~\cite{Soleymaninia:2018uiv}.

One of our main aim and motivations in this analysis is to revisit our
 previous QCD analysis~\cite{Soleymaninia:2018uiv} by 
including the available SIDIS data to the SIA data sample.
The COMPASS Collaboration has measured 
the multiplicities of the charged hadrons 
produced in semi-inclusive scattering. 
They have used a $160$ GeV muon beam and a 
target ($^6$LiD). 
COMPASS has measured the differential 
multiplicity for positive and 
negative charged hadrons separately, which is given by.

\begin{eqnarray}
\label{eq:mult}
\nonumber
\frac{dM^h(x,z,Q^2)}{dz}=
\frac{d^3\sigma ^h(x,z,Q^2)/
dxdQ^2dz}{d^2\sigma 
^{DIS}(x,Q^2)/dxdQ^2}\,,
\end{eqnarray}

where the numerator is given by the differential SIDIS cross 
section for charged hadron production and the denominator is given 
by the differential inclusive DIS cross section. 
The cross sections at leading-order (LO) can be expressed in 
terms of PDFs $q(x,Q^2)$ and FFs $D^h_q(z,Q^2)$, 

\begin{eqnarray}
\label{eq:mult_PDF_FF}
\nonumber
\frac{dM^h(x,z,Q^2)}{dz}=
\frac{\sum _qe_q^2q
(x,Q^2)D^h_q(z,Q^2)}
{\sum _q e_q^2q(x,Q^2)}.
\end{eqnarray}

The kinematic cuts have been imposed on 
the photon virtuality $Q^2>1$ {\rm (GeV/c)}$^2$, 
on the Bjorken scaling variable $0.004 < x < 0.4$, 
on the scale variable in the final state $0.2 < z < 0.85$ 
and on the inelasticity $0.1 < y < 0.7$.

Although the cleanest way to access the FFs for 
hadron production in the final state is 
electron-positron annihilation process 
and also the FFs are the only non-perturbative 
objects in the observables, 
SIA has several limitations which can be addressed by SIDIS process. 
On one hand, while the extraction of flavor-separated FFs 
is difficult in a QCD analysis based on the SIA only, 
the data from SIDIS experiments are 
crucial to getting direct constrain on the 
separation of quark and anti-quark FFs.
On the other hand, the range of the center-of-mass energy at 
which FFs are probed in SIA covers
from $Q=10$ GeV to the $Q=M_Z$. 
However, SIDIS data cover lower scales of energy, 
from $Q\sim 1$ GeV to the $Q\sim 6$ GeV.

Considering the discussions presented in Sec.~\ref{sec:pQCD}, 
we apply kinematic cuts on the 
experimental data for 
which a description in the framework of pQCD  and time-like evolution can be 
expected to work well. 
Hence,  we exclude the range of very small values of $z$ from SIA data sets. 
For the SIA data points, we use the kinematic cuts on 
$z$ as $0.02\leq z\leq 0.9$ for data at a 
center-of-mass energy of $M_Z$, and $0.075\leq z\leq 0.9$ 
for other data points.  
As a matter of fact, at low- energy scale $Q$, higher order purterbative corrections are 
necessary to have acceptable theory predictions. So the perturbative QCD corrections up to 
NLO accuracy are unreliable for low $Q$. Hence,  we exclude the range of very small values of $Q$ 
from SIDIS data sets. 
For the COMPASS SIDIS, we implemented 
cuts of $Q>2$ GeV.

Finally, we include in total $N_{\rm dat}=684$ data points in our analysis after 
kinematic cuts in which include the  $N_{\rm dat}=314$  data points for  
SIDIS, and $N_{\rm dat}=370$ for the SIA.

%
\subsection{Compatibility of TASSO 35 GeV}\label{sec:TASSO}
%

In this section, we comment on the possible tensions
between the SIA data sets. 
As one can see from data sets reported in Table.~\ref{tab:chi2-SHK22.h},
one of the sources of SIA data sets is the 
TASSO 35 GeV which we do not include in the list of the data sets.
Our detailed study on the individual $\chi^2$ 
shows that there is a tension between this data set 
and the COMPASS data sets.
We first perform the calculation of the 
light-charged hadron FFs using the SIA experimental data only, and 
we achieve an acceptable description 
for the TASSO dataset at 35 GeV 
with the individual $\chi^2$ per data point of 1.56. 
However, when the COMPASS SIDIS data are added to the QCD fit, 
one can not obtain an optimal description of the TASSO 35 GeV data 
set in the fit, and a large $\chi^2$ per data point is achieved
for it, $\chi^2/N_{\rm dat} = 8.37$. 
The origin of this behavior, can be related to a 
tension between the TASSO 35 GeV and the COMPASS data.

We should note here that the drop of matching 
between the theoretical predictions and all other TASSO 
experimental data are also examined,  and have been seen for all 
other scales of energy 14, 22, 35, and 44 GeV.  As a matter of 
fact, the extracted values of $\chi^2$ per data points for other 
data points are also worsened, which in order of 2.3 for the  
TASSO 14 GeV, 1.9 for the TASSO 22 GeV, 4.5 the for TASSO 44 GeV, 
and 5.3 for the TASSO 35 GeV after the inclusion of the COMPASS data.  
Since the matching drop for the TASSO measurements at 14, 22 and 
44 GeV are milder than those of the 35 GeV one,  and the extracted 
ranges of $\chi^2$ per data point are seem to be acceptable for them, 
we decided to exclude only the TASSO 35 GeV measurements in the fit. Notwithstanding, we have performed a separate analysis which included the \texttt{TASSO35} dataset and noticed that the central value of the distributions does not affect, whereas the uncertainty estimates are now larger at small $z$ values. We presume that this large uncertainty band is an overestimate and not truthful. This finding also indicates to the tension between \texttt{TASSO35} and \texttt{COMPASS} datasets.

%
\section{{\tt SHK22.h} numerical Results}\label{sec:results}
%

In this section, we present the main results and findings for the 
determination of the FFs of light charged hadron,
called {\tt SHK22.h}, in which most of available and updated 
SIA and SIDIS measurements are
added to the data sample and analyzed up to the NLO accuracy in perturbative QCD.

We first present the fit quality and discuss in details the 
term of both the individual and the total datasets included in our analysis.
Then we present the data and theory comparison, both for the SIA and SIDIS 
data sets analyzed  in 
{\tt SHK22.h}. 

We also illustrate the resulting light charged hadron FFs 
and their uncertainties,  for  
all parton species, focusing on the comparison of 
the extracted NLO FFs with the 
publicly available {\tt JAM20} and {\tt NNFF1.1h} analyses.
We discuss the interplay between the SIA and SIDIS experimental data, and the stability of 
the light charged hadron FFs upon inclusion of SIDIS data sets.

%
\subsection{Fit quality}\label{sec:Fit-quality}
%

In Table.~\ref{tab:chi2-SHK22.h} we report the 
value of the $\chi^2$ per data point,
$\chi^2/N_{\mathrm dat}$, for the individual data sets 
for both SIA and SIDIS included in the {\tt SHK22.h} analysis.
This table also includes the number of data points that pass the
kinematic cuts. 
The values of the $\chi^2$ per data point for the 
total datasets also shown as well.

Considering the numbers presented in this table, a 
few remarks for the individual and total datasets are in order.
As can be seen, the global $\chi^2$ per data point in our fit, equal to 1.080, indicates, in general, a very good description of the entire data sets.
Remarkably, a comparable fit quality is observed for both the SIA and SIDIS
data sets separately.

Concerning the fit quality  of the individual SIDIS experiments, we see that 
for the $h^-$ production at COMPASS, 
we obtain a better $\chi^2$ per data point with
respect to the COMPASS $h^+$. 

A closer look to the $\chi^2$ per data point presented in this table 
reveals that  
acceptable descriptions are achieved almost for all 
of the individual data sets 
analyzed in the {\tt SHK22.h} fit, with two main exceptions.

First, for some data sets reported in the table, 
the $\chi^2/N_{\mathrm dat}$ value is still large: 
this specifically happens for the $h^\pm$ light charged 
hadron production in  TASSO 14 GeV, TASSO 44 GeV and OPAL total inclusive. 

From this table we also observe that the
$\chi^2/N_{\mathrm dat}$ value for the DELPHI $uds$ and 
OPAL bottom$h^\pm$ is anomalously
small.
This finding was already observed and reported in some previous FFs 
analyses which is 
likely to be due to the overestimate of 
the uncorrelated systematic uncertainty. We refer the reader 
to the Refs.~\cite{deFlorian:2014xna,deFlorian:2017lwf,Hirai:2016loo,Sato:2016wqj} for more details. 

%
%
\begin{table}[!t]
	\renewcommand{\arraystretch}{1.25}
	\centering
	\scriptsize
	\begin{tabular}{lccccccccc}
		\toprule
		Experiment ~&~ $\chi^2/N_{\rm dat}$ ~&~ $N_{\rm dat}$  \\ \hline
		TASSO 14 GeV $h^\pm$ & 1.791 &  14 \\
		TASSO 22 GeV $h^\pm$ & 1.254 &  14 \\
		TASSO 44 GeV $h^\pm$ & 2.912 &  14 \\
		TPC $h^\pm$ & 0.659 &  21 \\
		ALEPH $h^\pm$ & 0.825 &  32 \\
		DELPHI total $h^\pm$ & 0.610 &  21 \\
		DELPHI $uds$ $h^\pm$ & 0.380 &  21 \\
		DELPHI bottom $h^\pm$ & 1.028 &  21 \\
		OPAL total $h^\pm$ & 1.821 &  19 \\
		OPAL $uds$ $h^\pm$ & 0.794 &  19 \\
		OPAL charm $h^\pm$ & 0.599 &  19 \\
		OPAL bottom$h^\pm$ & 0.299 &  19 \\
		SLD total $h^\pm$ & 1.047 &  34 \\
		SLD $uds$ $h^\pm$ & 0.946 &  34 \\
		SLD charm $h^\pm$ & 1.034 &  34 \\
		SLD bottom $h^\pm$ & 1.102 &  34 \\ \hline
		COMPASS $h^-$ & 0.907 &  157 \\
		COMPASS $h^+$ & 1.338 &  157 \\ \hline 
		\textbf{Global data set} & \textbf{1.079} & \textbf{684}  \\ \hline \hline
\end{tabular}
\caption{\small The $\chi^2$ values per data point for the individual 
data sets, total SIA and total SIDIS included in the {\tt SHK22.h} analysis. The
number of data points $N_{\mathrm dat}$ after the kinematic cuts and the
global $\chi^2$ values are also displayed. }
\label{tab:chi2-SHK22.h}
\end{table}
%

%
%
\subsection{Theory and data comparison}\label{sec:datatheo}

In order to assess, it would be instructive to 
look at the comparison between the data
and the NLO theory predictions obtained with the 
{\tt SHK22.h} light charged hadron FFs for 
all SIA and some selected SIDIS data sets.

\begin{figure*}[htb]
\vspace{0.50cm}
\resizebox{0.480\textwidth}{!}{\includegraphics{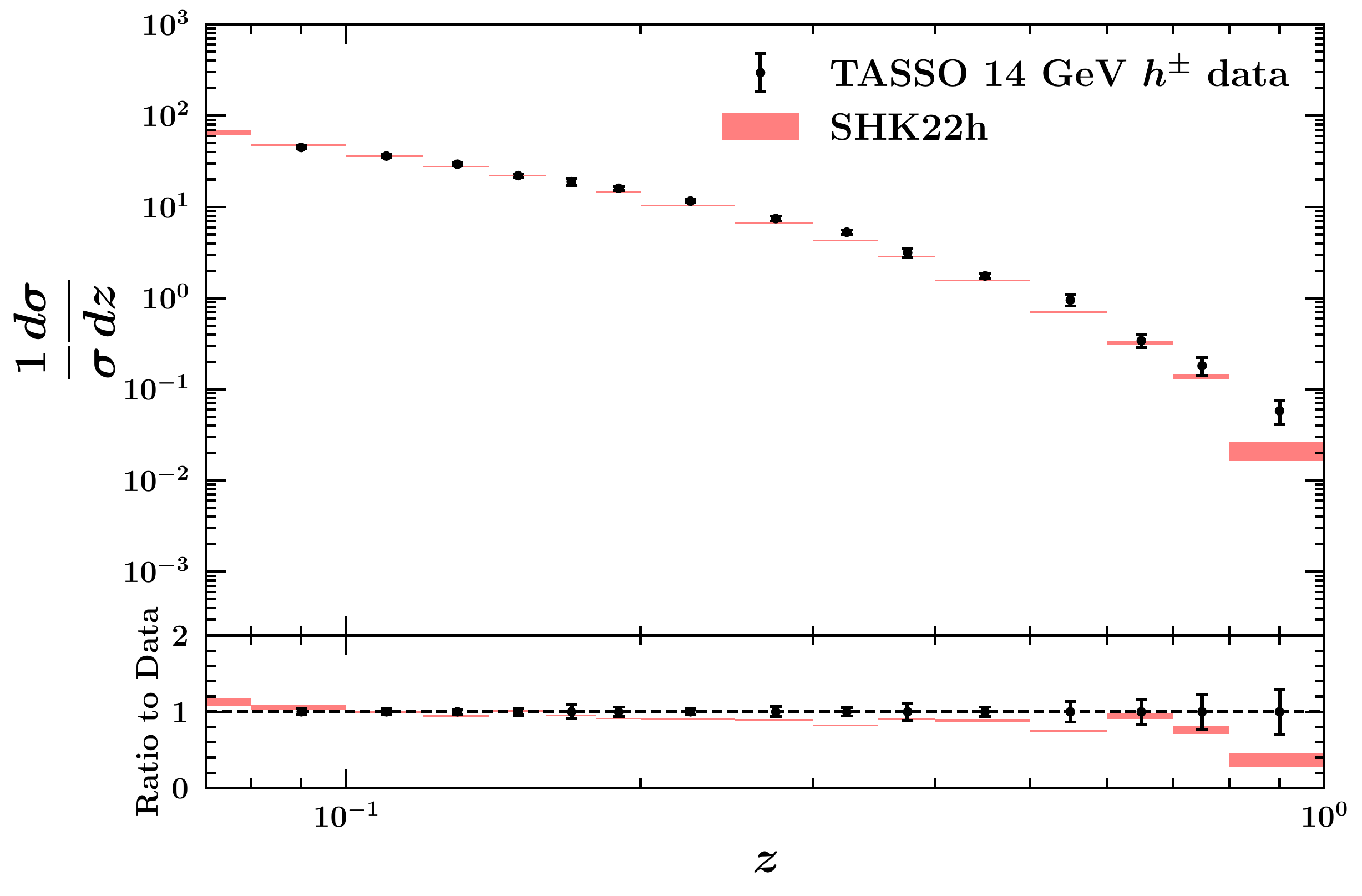}} 	
\resizebox{0.480\textwidth}{!}{\includegraphics{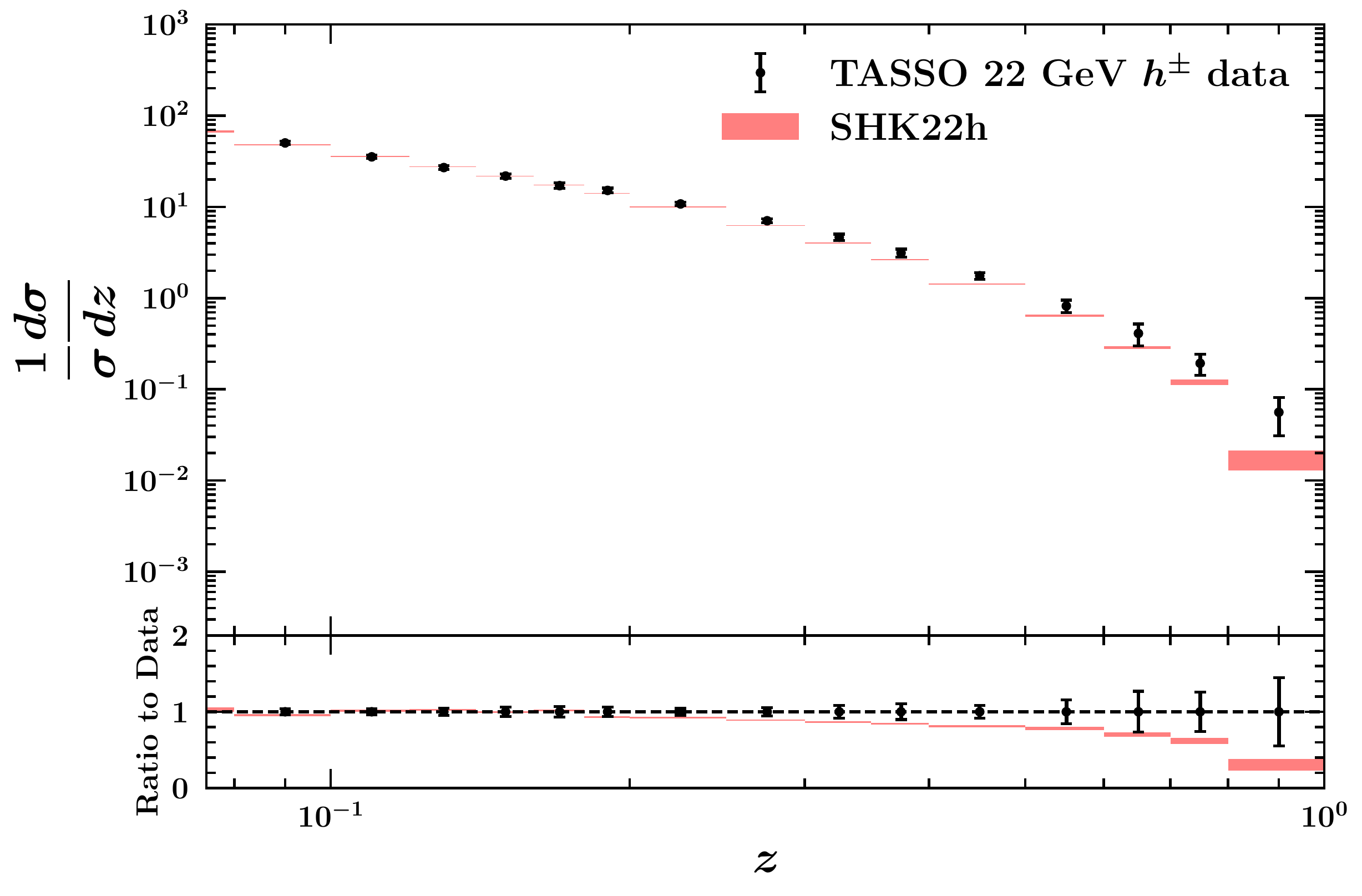}}  	
\resizebox{0.480\textwidth}{!}{\includegraphics{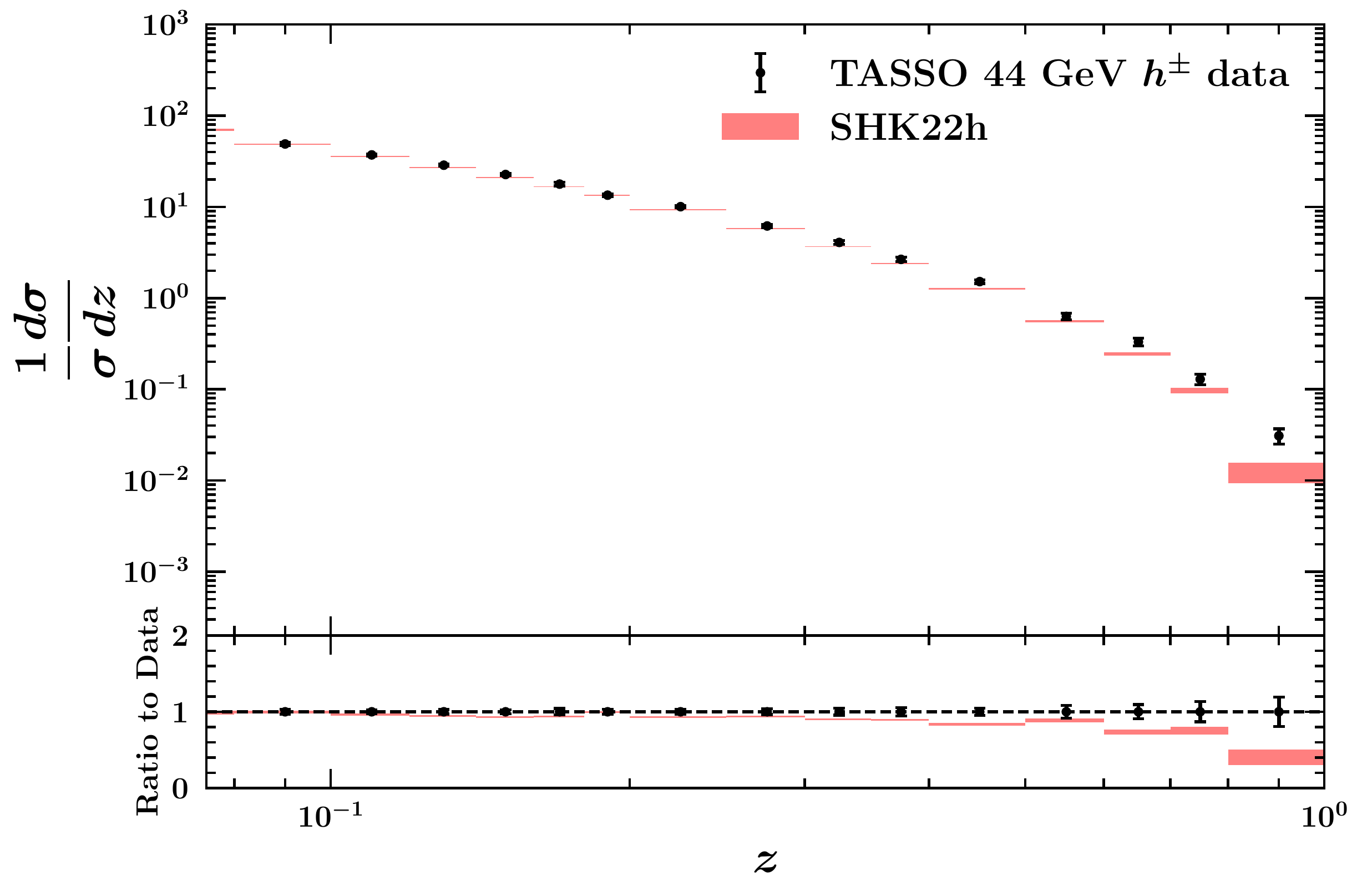}}
\resizebox{0.480\textwidth}{!}{\includegraphics{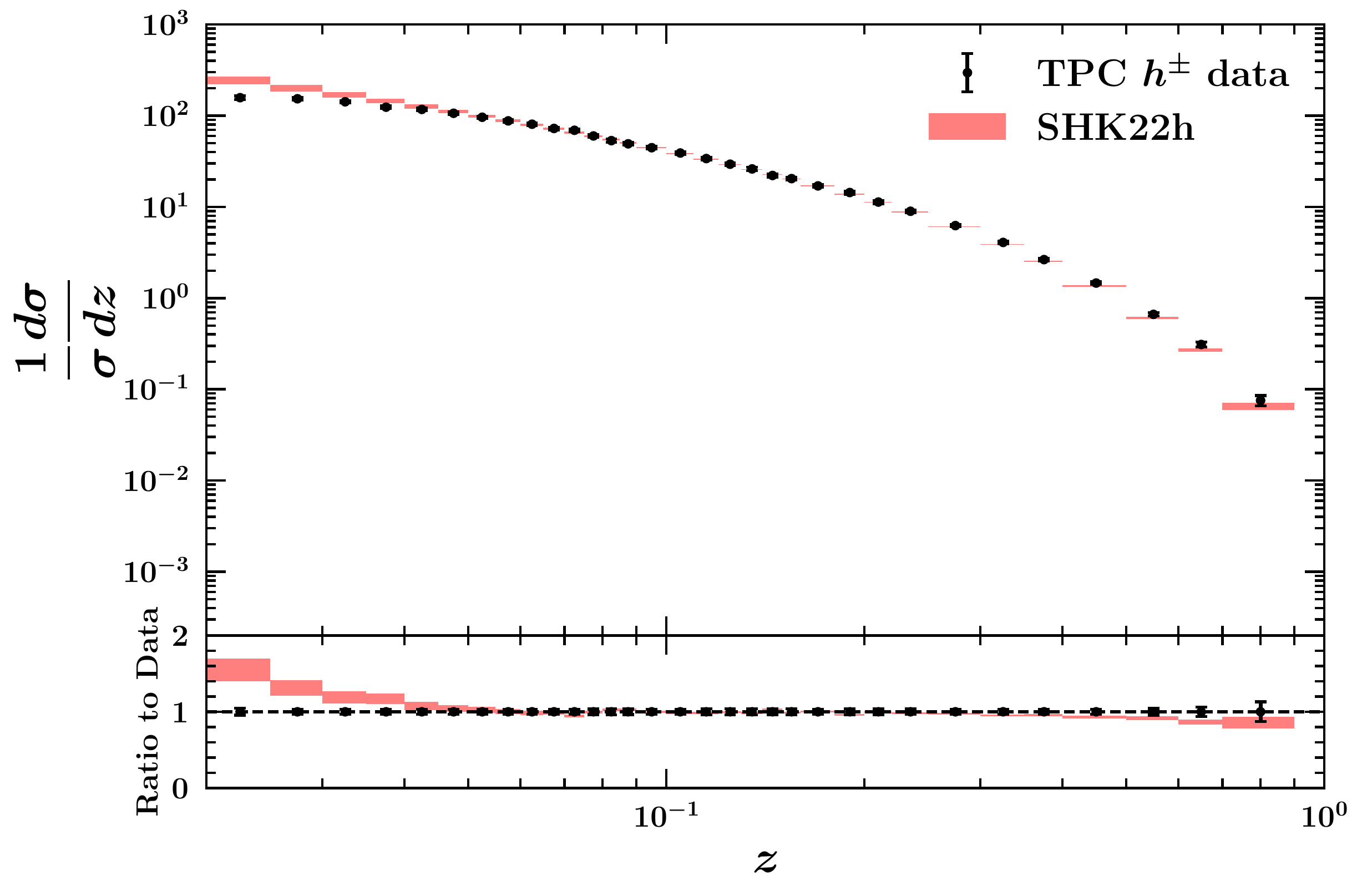}}  
\begin{center}
\caption{ 
\small 
The data/theory comparison for the TASSO 14, TASSO 22 
and TASSO 44 GeV Collaborations at
 $\sqrt{s} < M_Z$ for inclusive data sets at NLO.  
The same comparison also are shown for the TPC data.  
 The lower panels display the ratio to the experimental central values. 
}
\label{fig:TASSO-TPC}
\end{center}
\end{figure*}

We start with detailed comparisons with the SIA 
data analyzed in this work.
For all results presented in {\tt SHK22.h}, 
the upper panels represent the absolute
distributions while the lower ones display 
the ratio to the experimental central values analyzed in {\tt SHK22.h}.
In Fig.~\ref{fig:TASSO-TPC}, we compare the 
NLO theory predictions with the  inclusive data sets 
form TASSO 14, TASSO 22 and TASSO 44 GeV Collaborations. 
The same comparison also are shown for the TPC data.
As one can see,  overall,  satisfying agreements are achieved, 
however, with exception at high $z$.
Our theory predictions do not satisfy the high-$z$ TASSO data, 
 which is the origin of 
the slightly high-$\chi^2$ value reported in Table.~\ref{tab:chi2-SHK22.h} 
for these data sets.
This specific feature is particularly pronounced for 
TASSO 44 data, and more moderate, but still significant, for TASSO 14 and 22. 
For the TPC data, some deviation can bee seen for small value of $z$, but 
with the $\chi^2$ value reported in Table.~\ref{tab:chi2-SHK22.h}, 
the description of TPC data seems to be still convincing.

%

\begin{figure*}[htb]
\vspace{0.50cm}
\resizebox{0.480\textwidth}{!}{\includegraphics{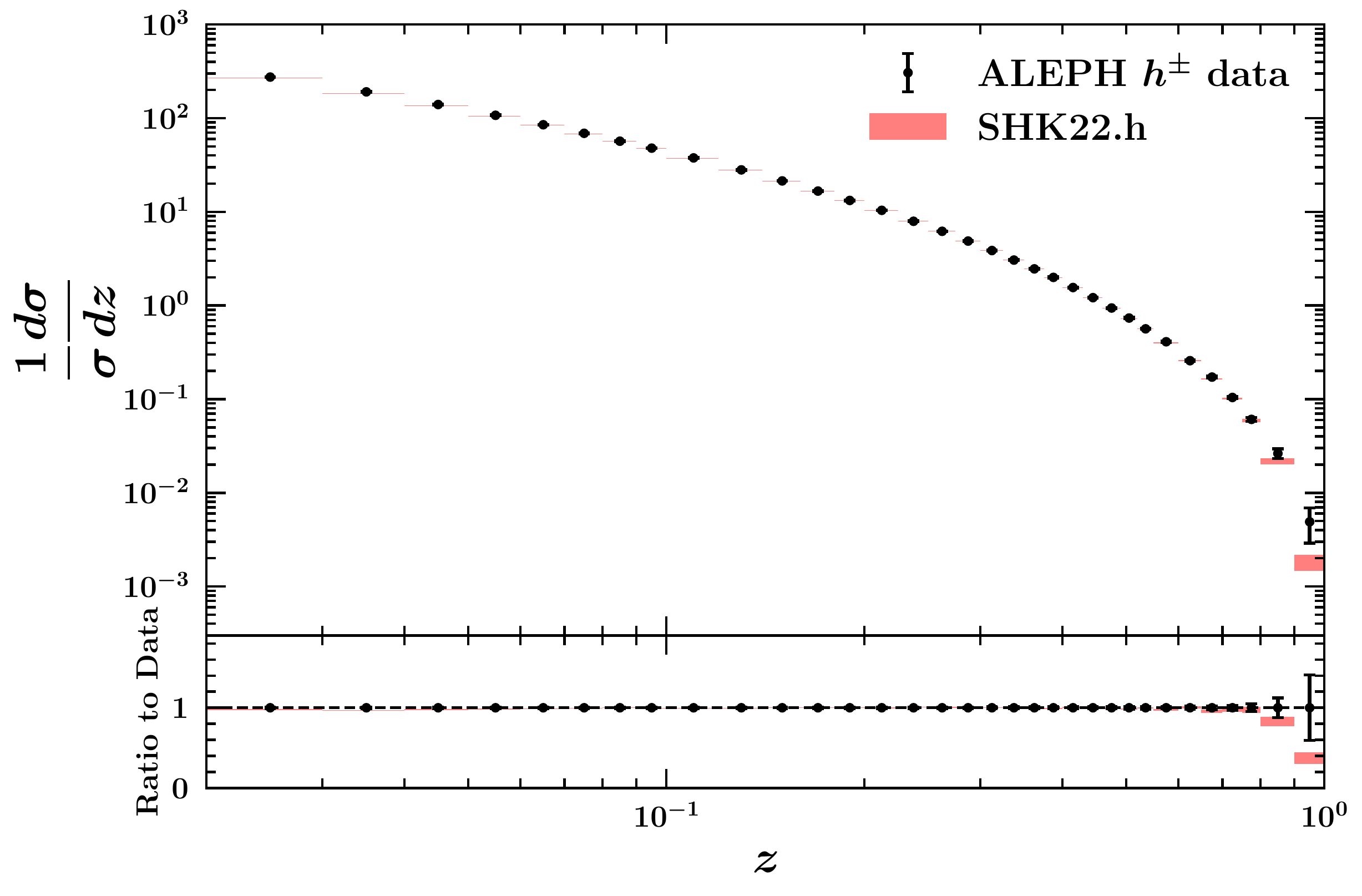}} 	
\resizebox{0.480\textwidth}{!}{\includegraphics{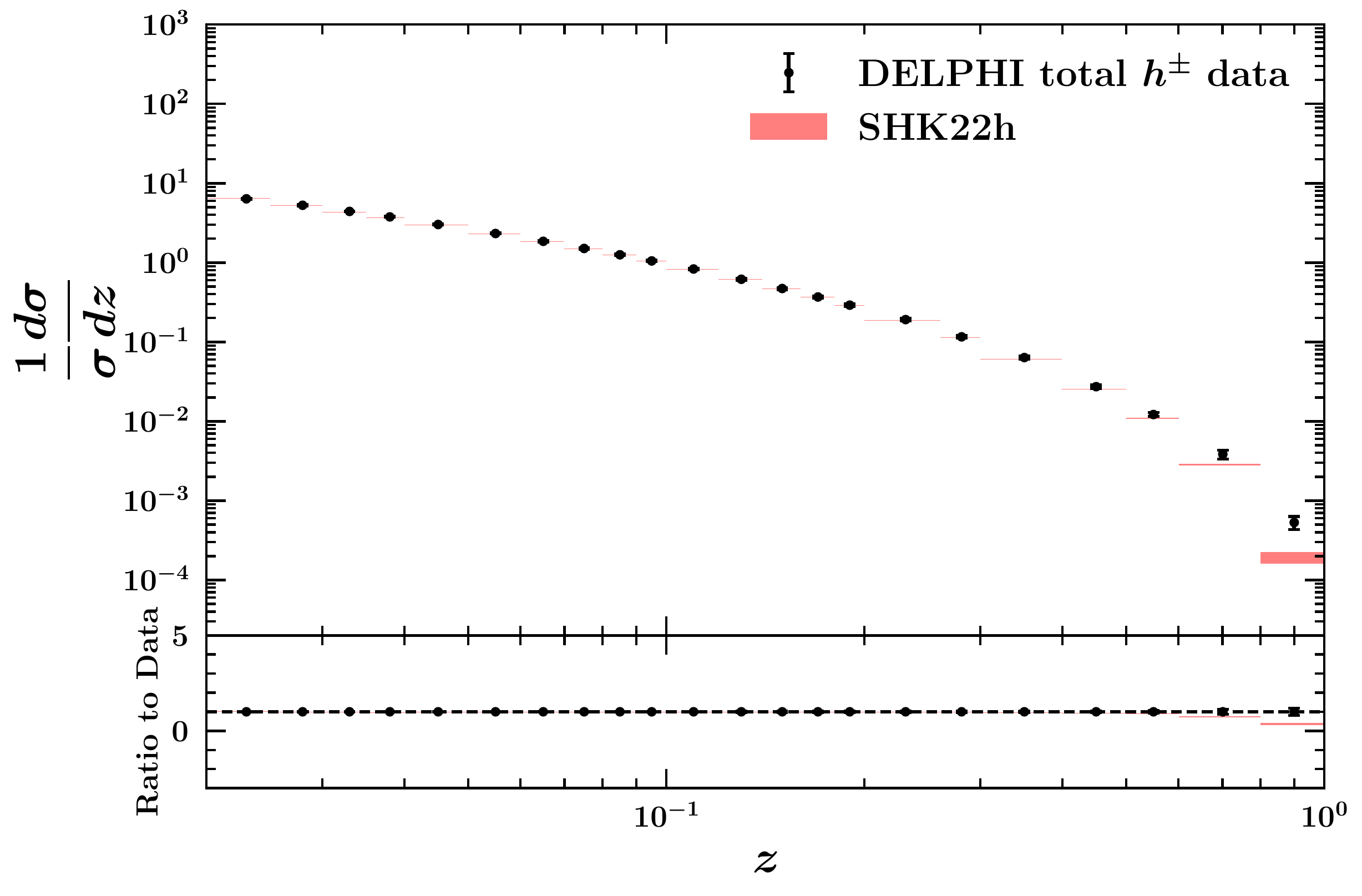}}	
\resizebox{0.480\textwidth}{!}{\includegraphics{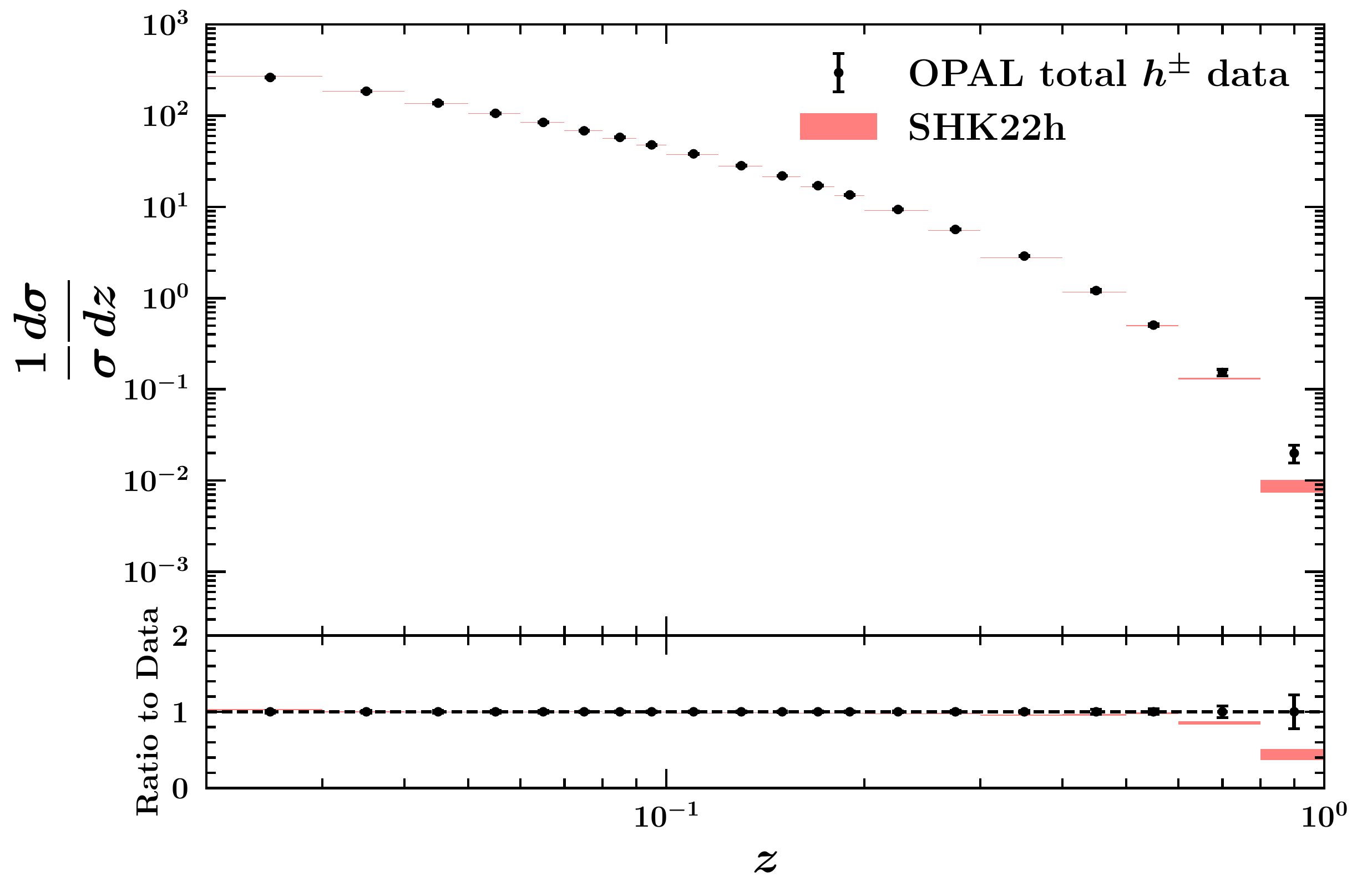}}
\resizebox{0.480\textwidth}{!}{\includegraphics{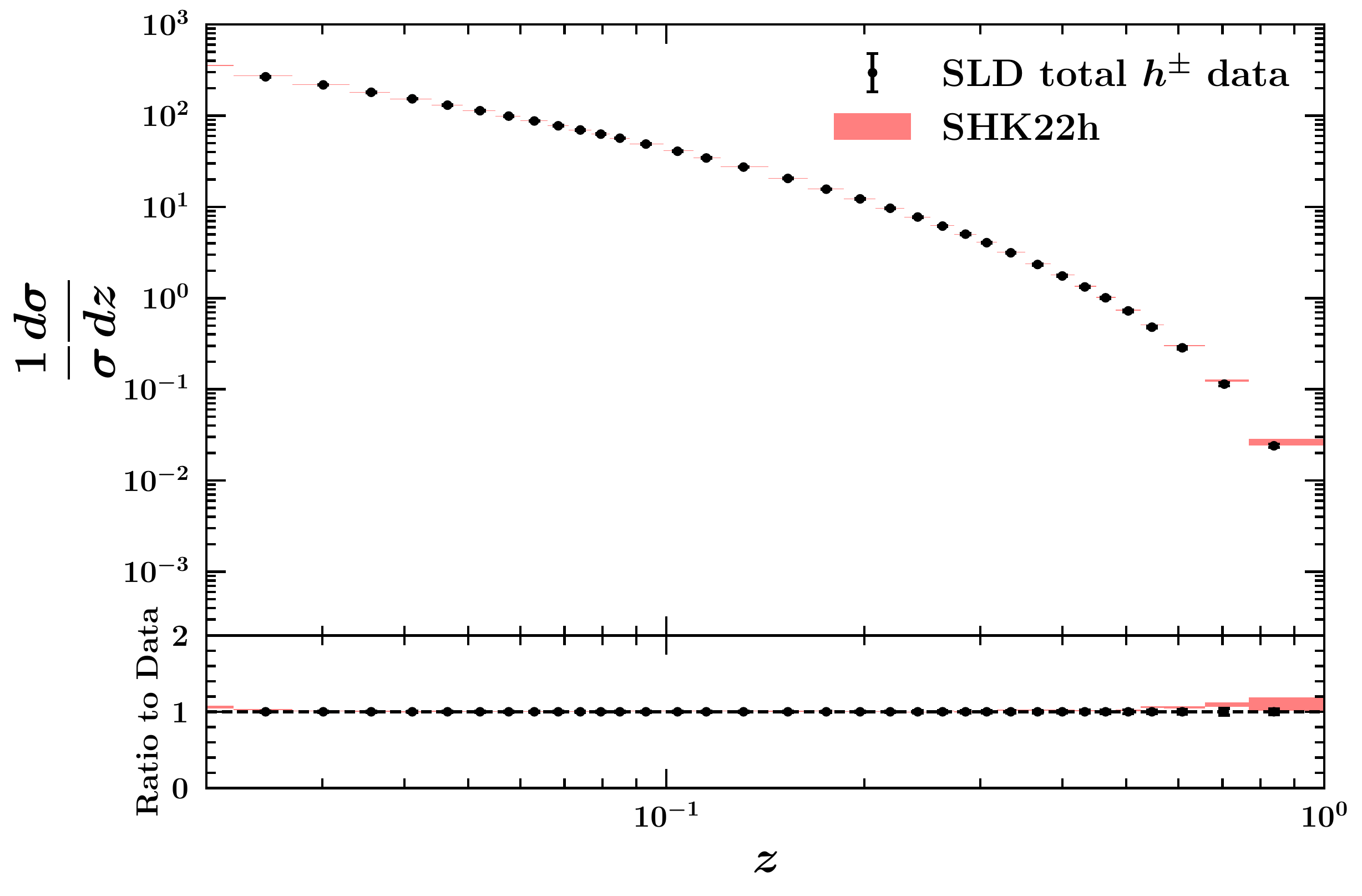}}  
\resizebox{0.480\textwidth}{!}{\includegraphics{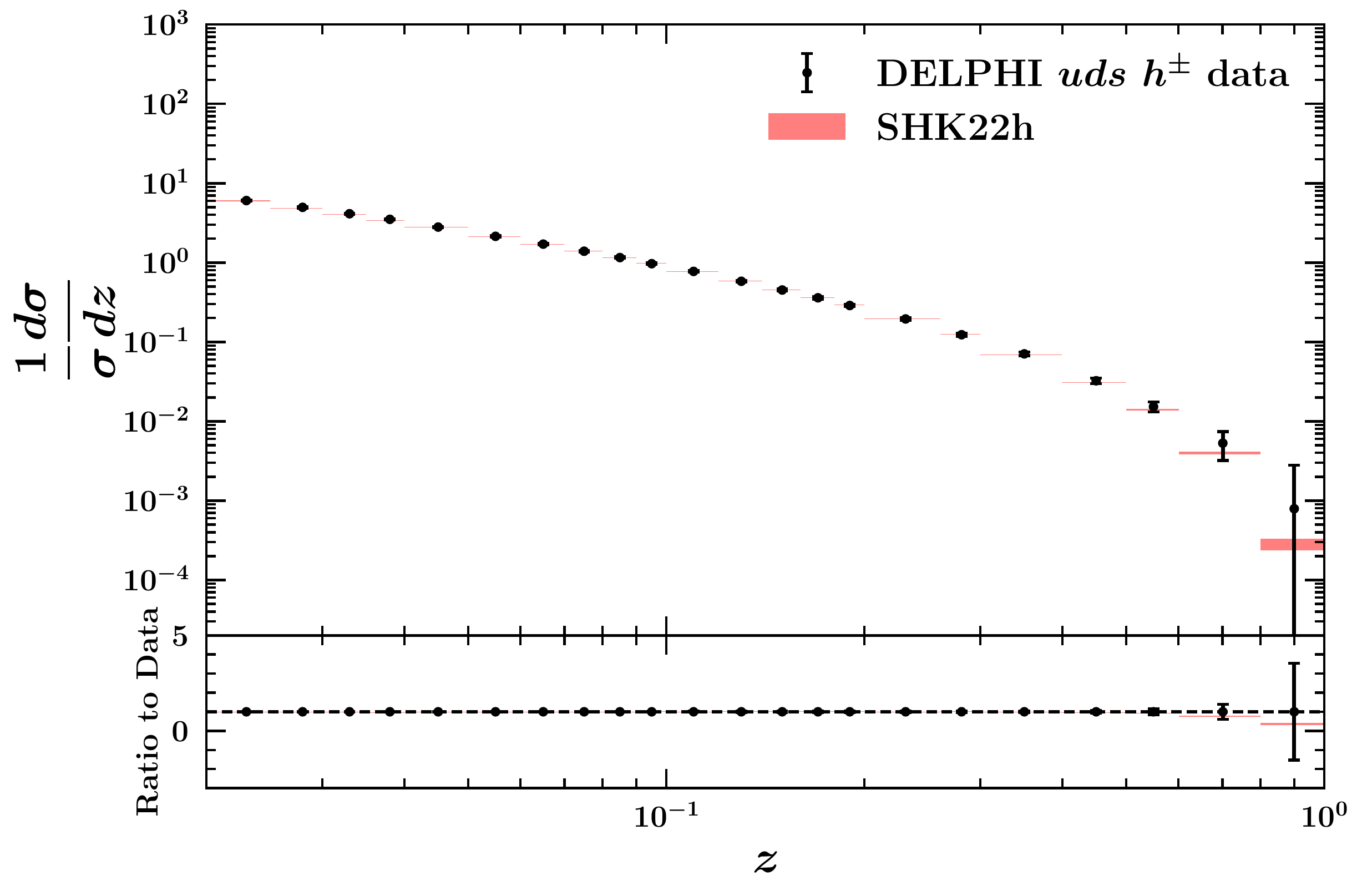}}  
\resizebox{0.480\textwidth}{!}{\includegraphics{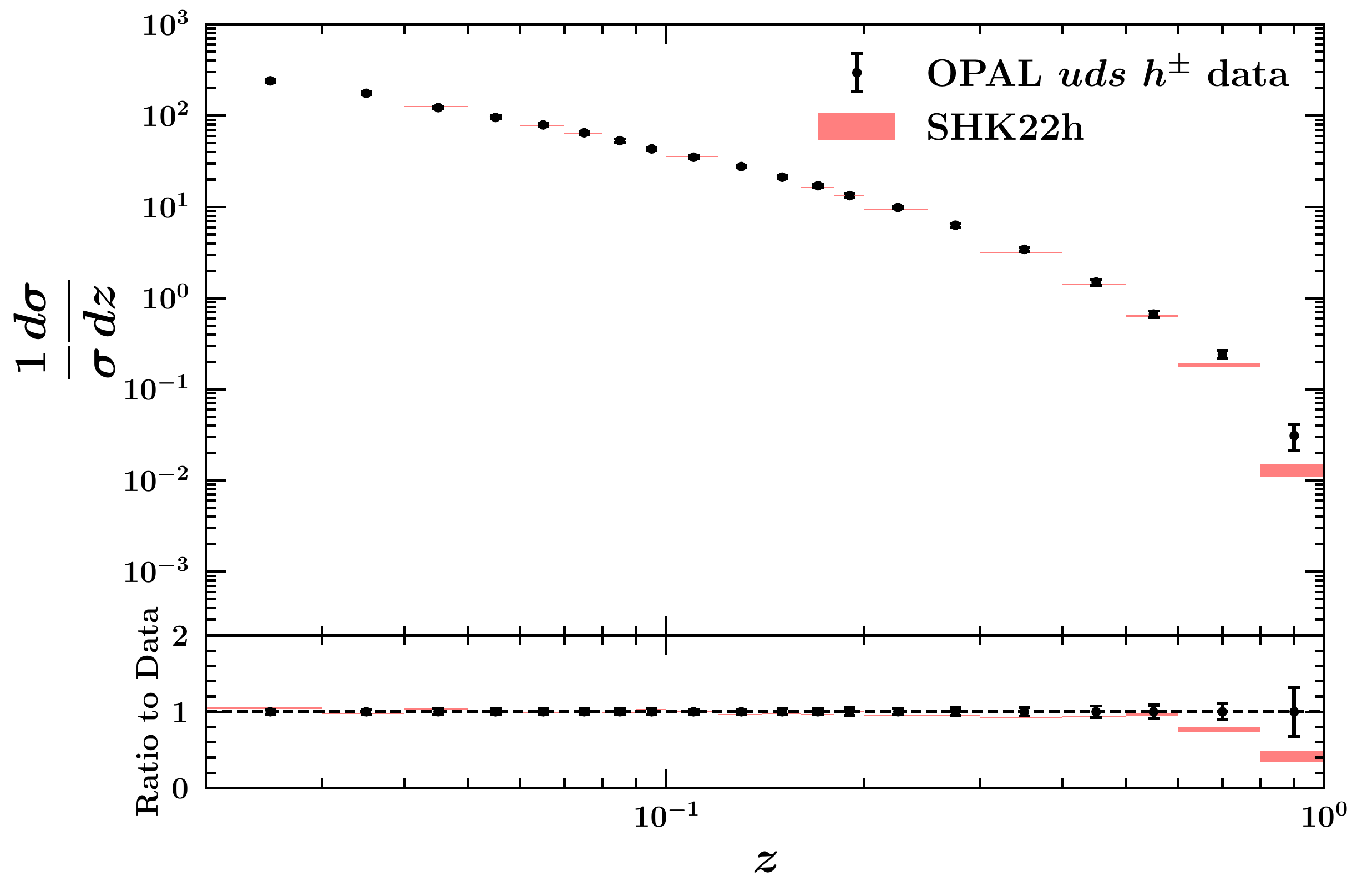}}	
\resizebox{0.480\textwidth}{!}{\includegraphics{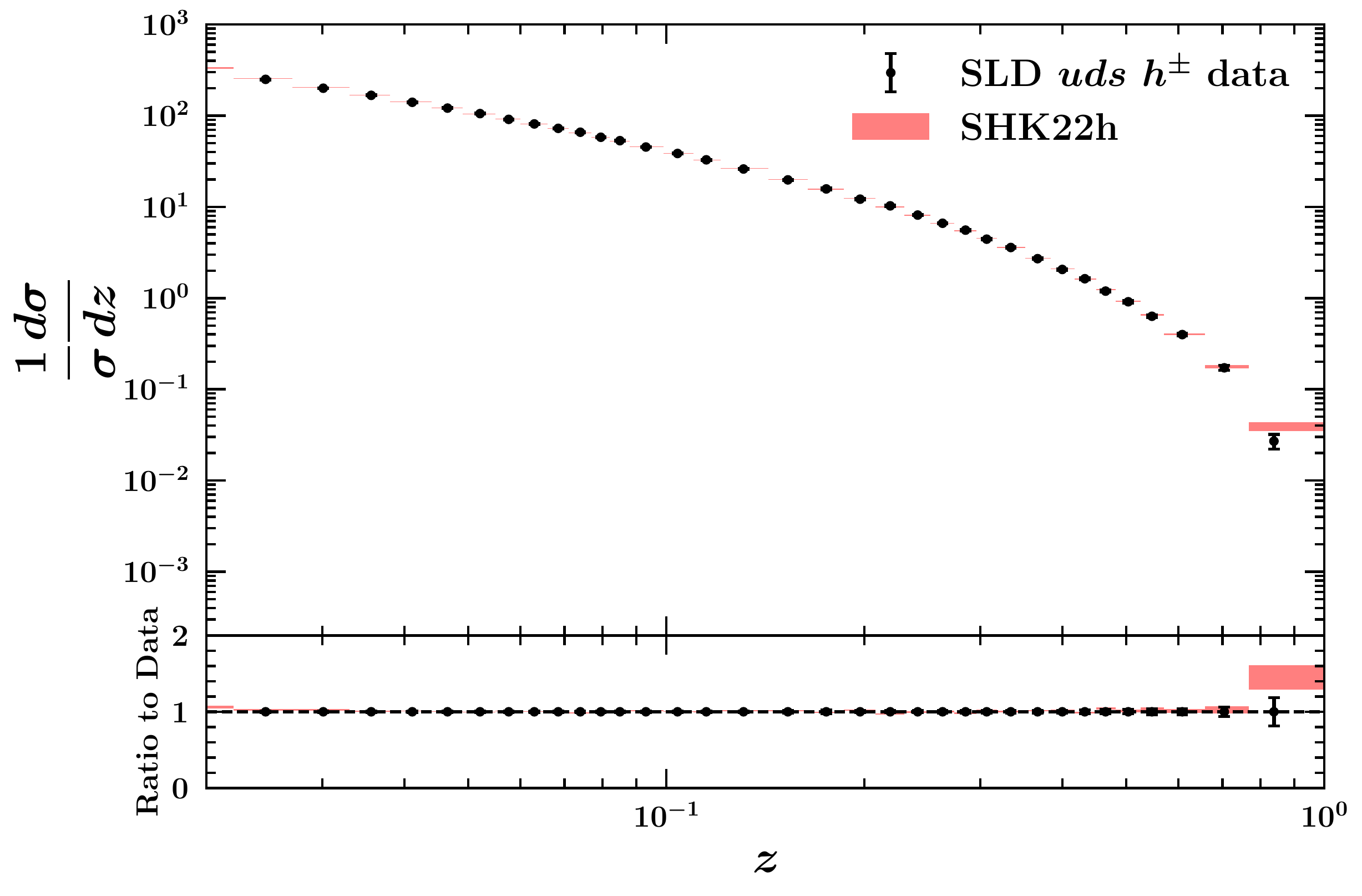}} 	 	
\begin{center}
\caption{ 
\small 
The data/theory comparison for the ALEPH, DELPHI, 
OPAL and SLD Collaborations at
$\sqrt{s}=M_Z$ for inclusive and light flavor-tagged data sets. 
 The lower panels display the ratio to the 
 experimental central values. 
}
\label{fig:LEP-SLAC}
\end{center}
\end{figure*}

In Fig.~\ref{fig:LEP-SLAC}, we present detailed comparisons of the NLO
theory predictions with the total inclusive ALEPH, DELPHI, OPAL and SLD data.
Comparisons with the uds-tagged data from DELPHI, OPAL and SLD data 
also presented as well. 
Consistently with the $\chi^2$ values reported
in Table.~\ref{tab:chi2-SHK22.h}, the description of these data sets is desirable, 
with one exception for the OPAL inclusive data. 
Some small deviation for large-$z$ data can be seen from the plots presented in  Fig.~\ref{fig:LEP-SLAC}.
Another important finding emerges for the comparison is that, 
the experimental data points 
for the inclusive measurements by OPAL Collaboration 
fluctuate at high-$z$ around the 
theoretical predictions by an amount that is 
seems to be typically larger than the calculated 
uncertainties. 
This should explain the poor $\chi^2$ values reported in 
Table.~\ref{tab:chi2-SHK22.h} 
for this specific data sets. 

\begin{figure*}[htb]
\vspace{0.50cm}
\resizebox{0.480\textwidth}{!}{\includegraphics{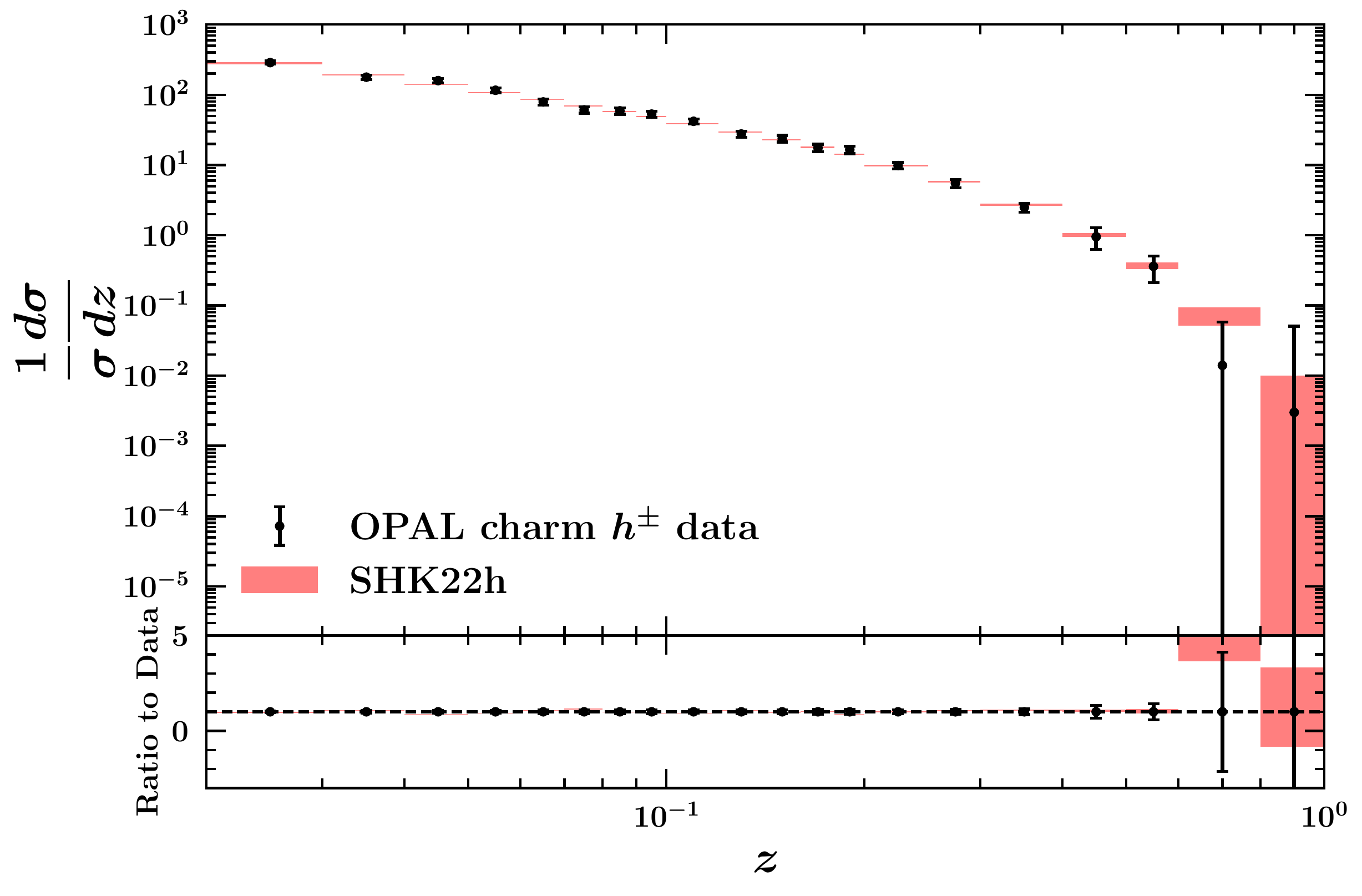}} 	
\resizebox{0.480\textwidth}{!}{\includegraphics{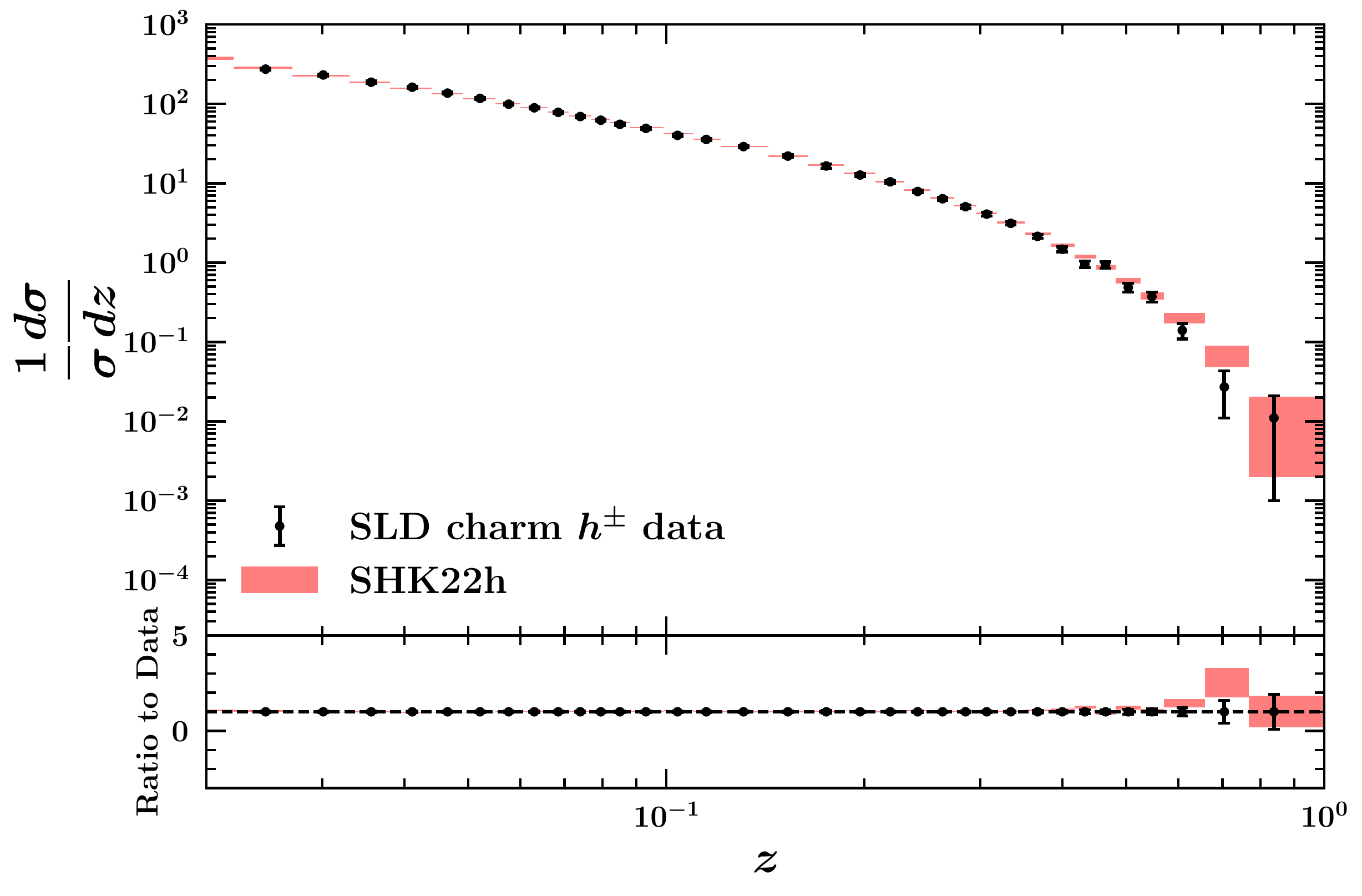}}  	
\resizebox{0.480\textwidth}{!}{\includegraphics{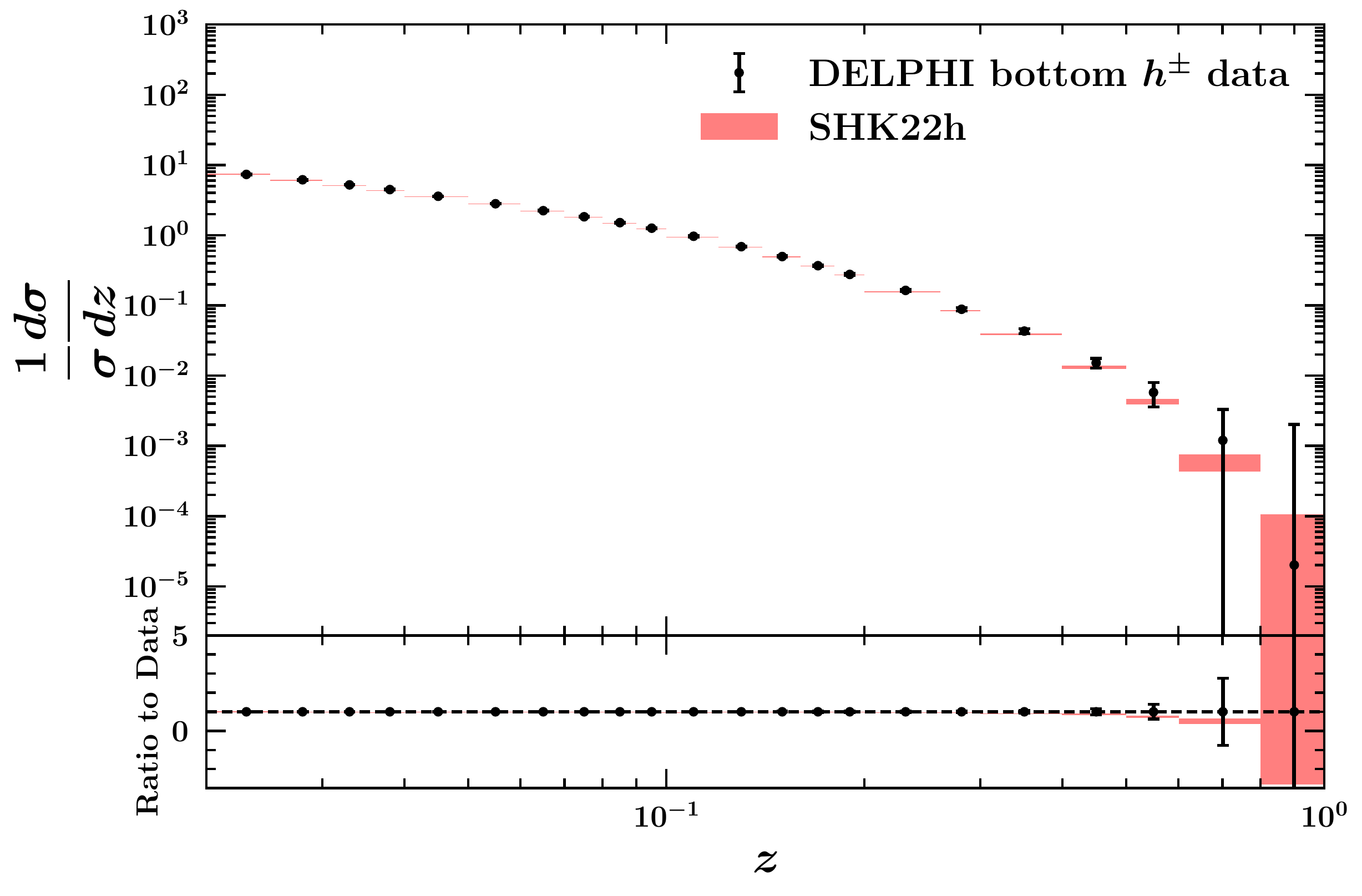}}
\resizebox{0.480\textwidth}{!}{\includegraphics{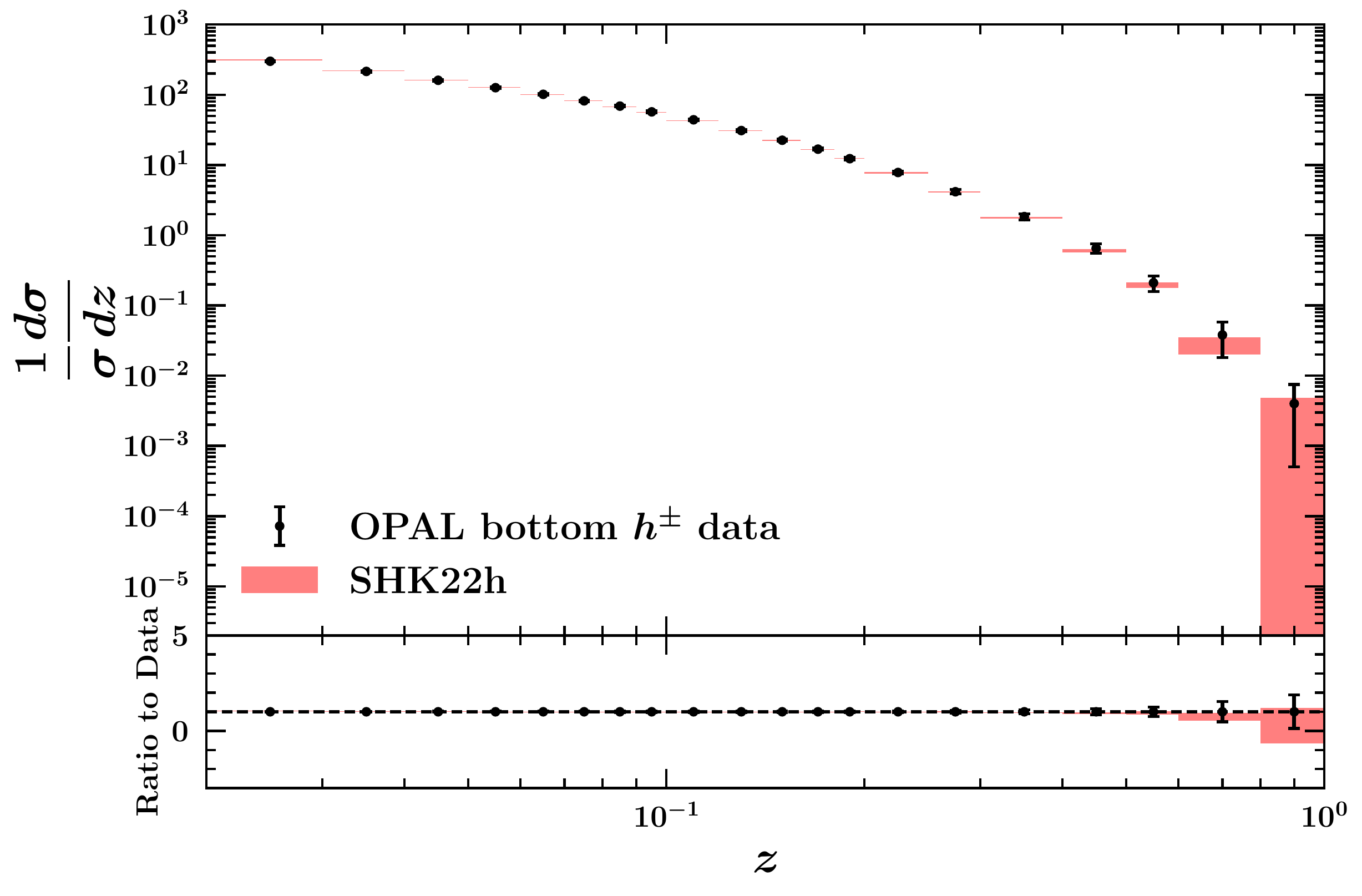}}  
\resizebox{0.480\textwidth}{!}{\includegraphics{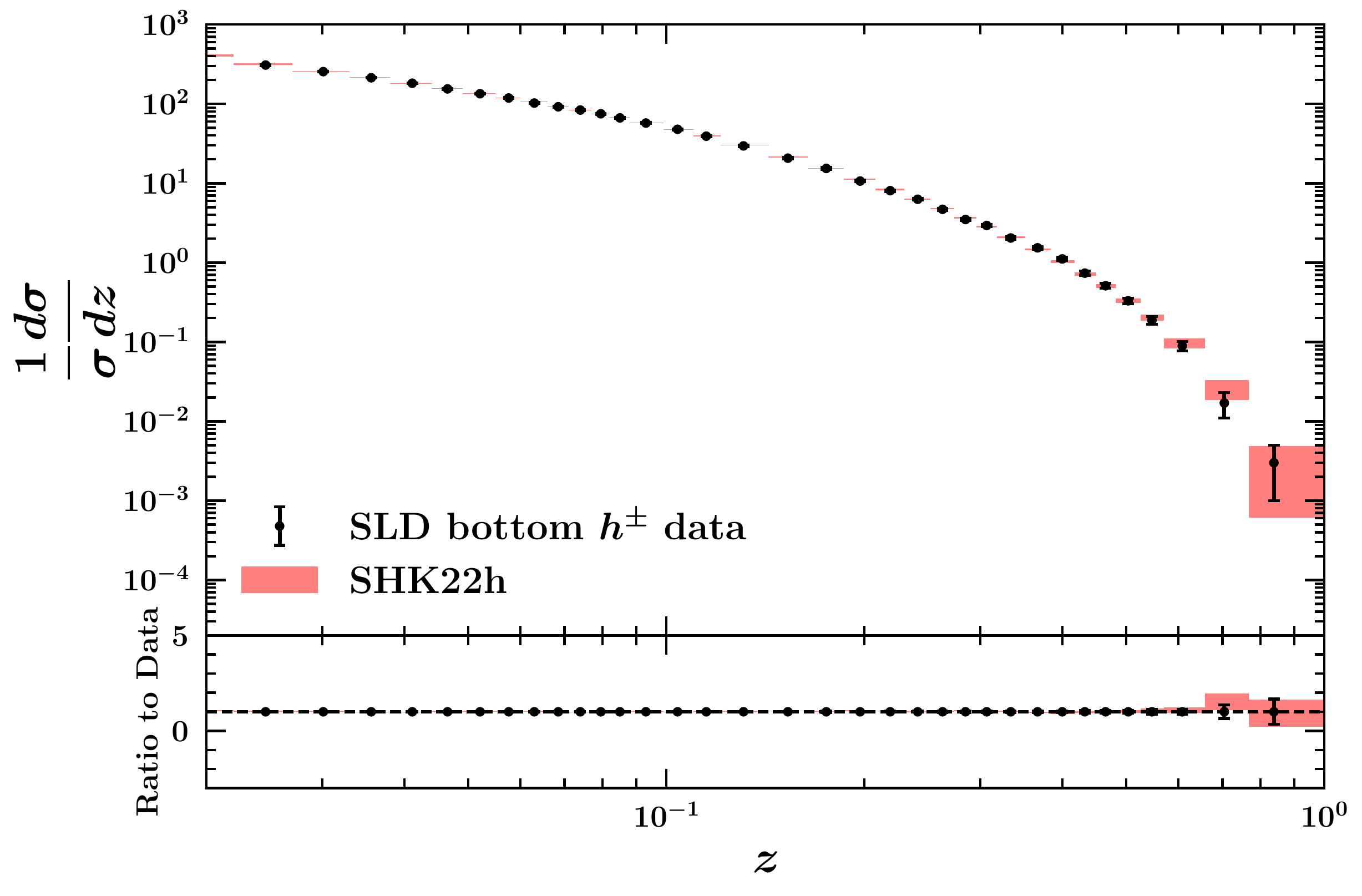}} 	
\begin{center}
\caption{ 
\small 
The data/theory comparison for the charm and 
bottom-tagged data from OPAL, SLD, DELPHI and SLD. 
The lower panels display the ratio to the 
experimental central values.
}
\label{fig:flavor-tag}
\end{center}
\end{figure*}

We now turn to the comparisons of our NLO theory predictions
with the charm and bottom-tagged data from OPAL, SLD, DELPHI and SLD. 
The 
corresponding plots are shown in Fig.~\ref{fig:flavor-tag}.
Once again, the goodness of the $\chi^2$ values reported in Table.~\ref{tab:chi2-SHK22.h} 
is reflected in a general good
description of the charm and bottom-tagged data.

%

\begin{figure*}[htb]
\vspace{0.50cm}
\resizebox{0.480\textwidth}{!}{\includegraphics{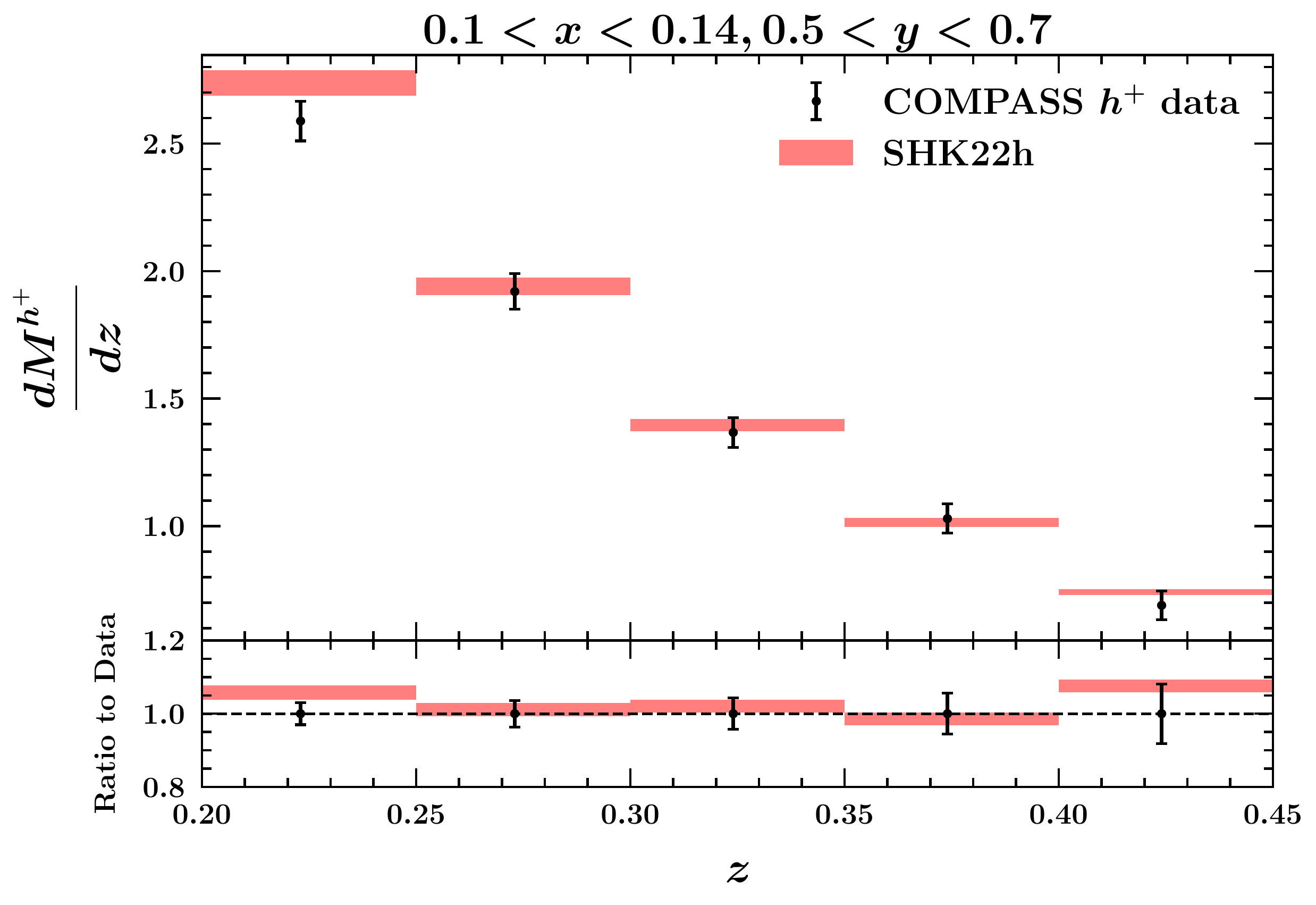}} 	
\resizebox{0.480\textwidth}{!}{\includegraphics{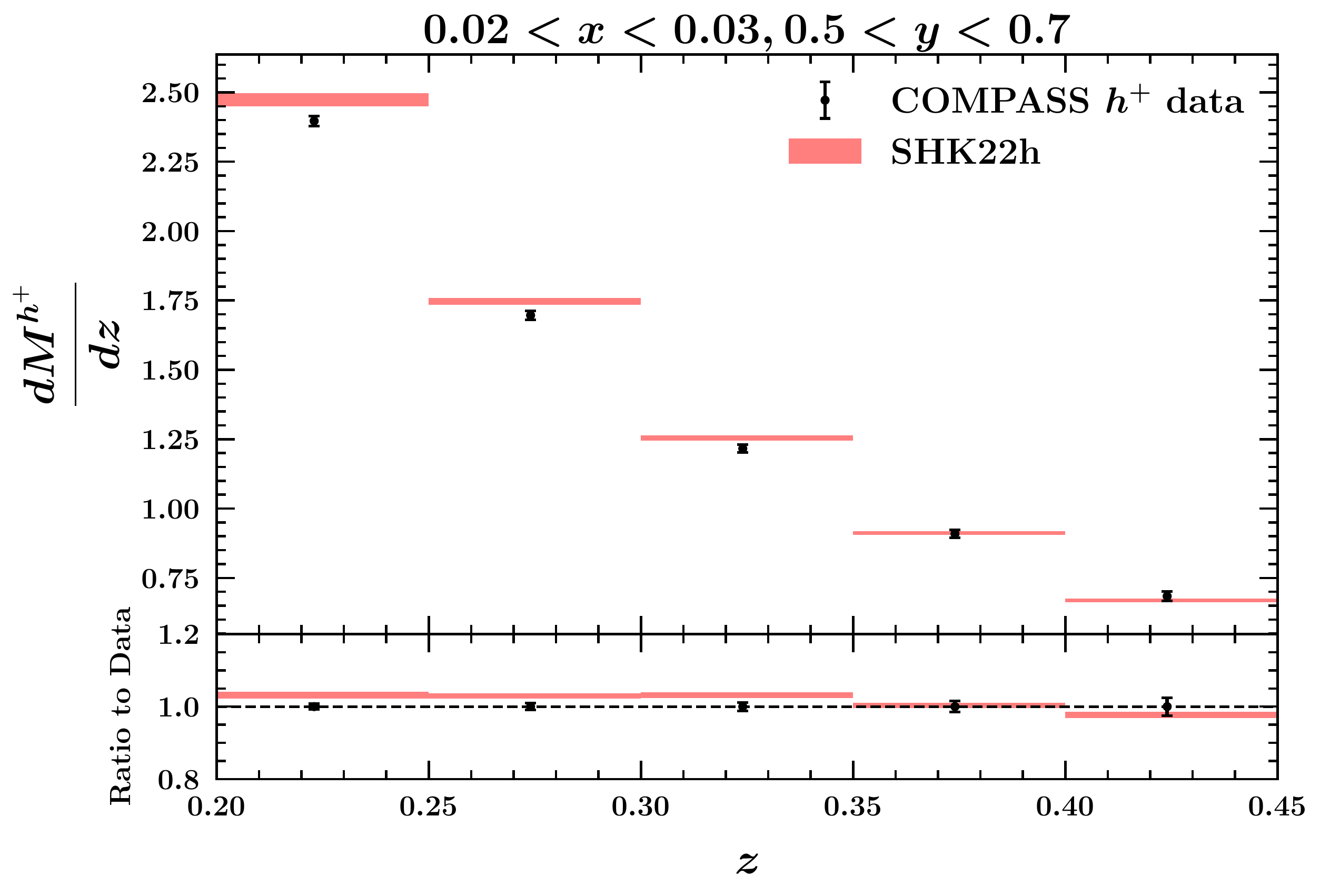}}  	
\resizebox{0.480\textwidth}{!}{\includegraphics{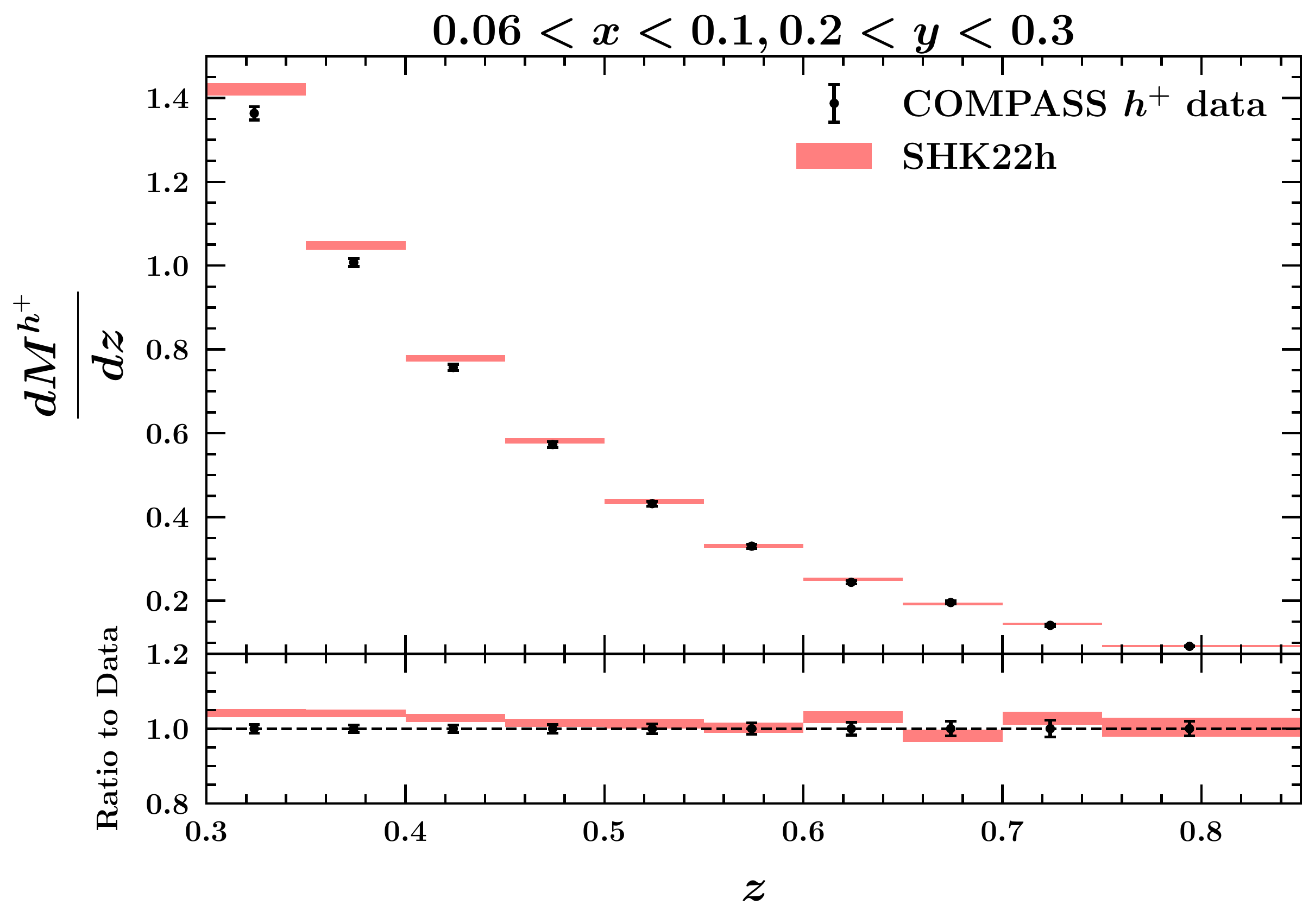}}
\resizebox{0.480\textwidth}{!}{\includegraphics{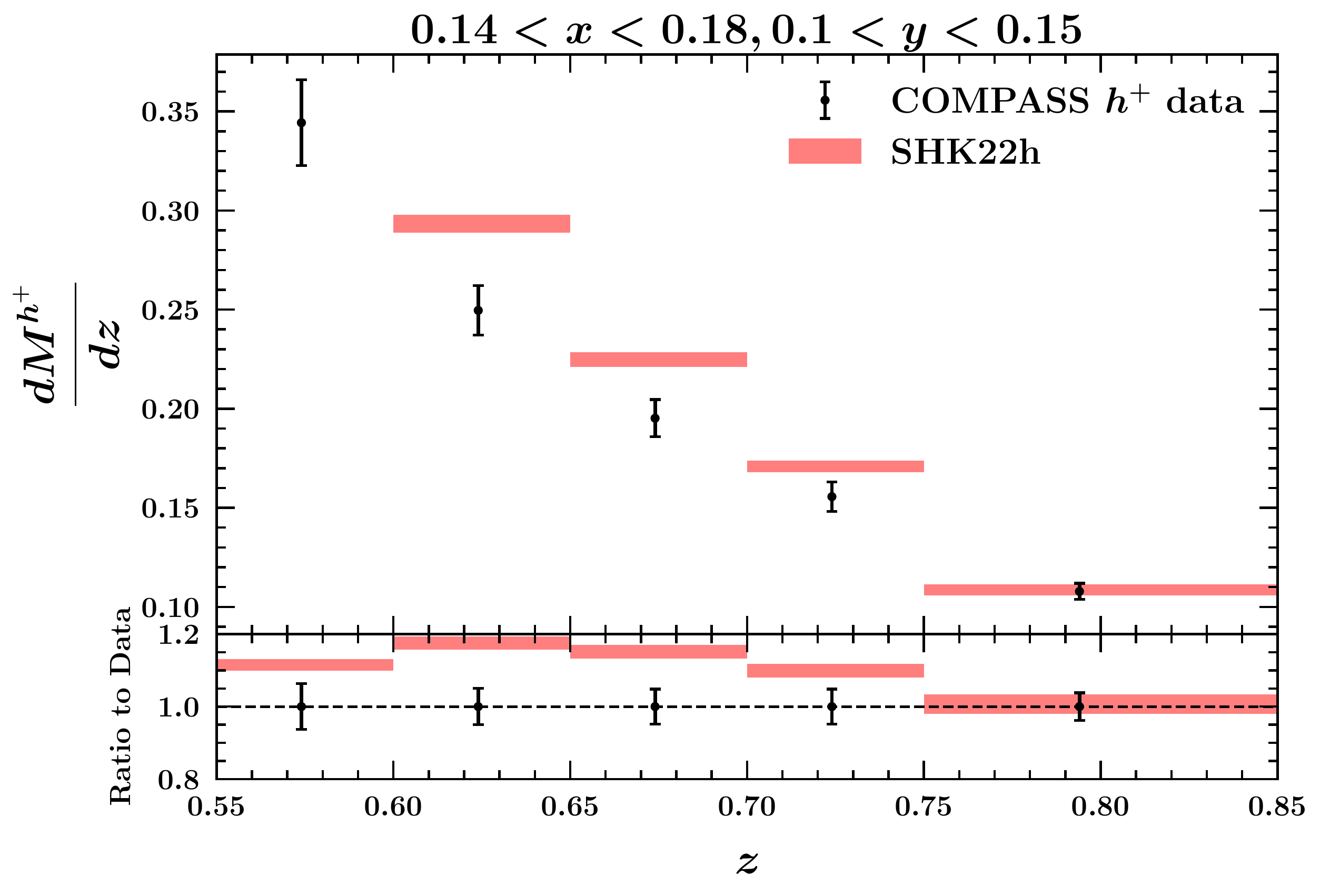}}  
\resizebox{0.480\textwidth}{!}{\includegraphics{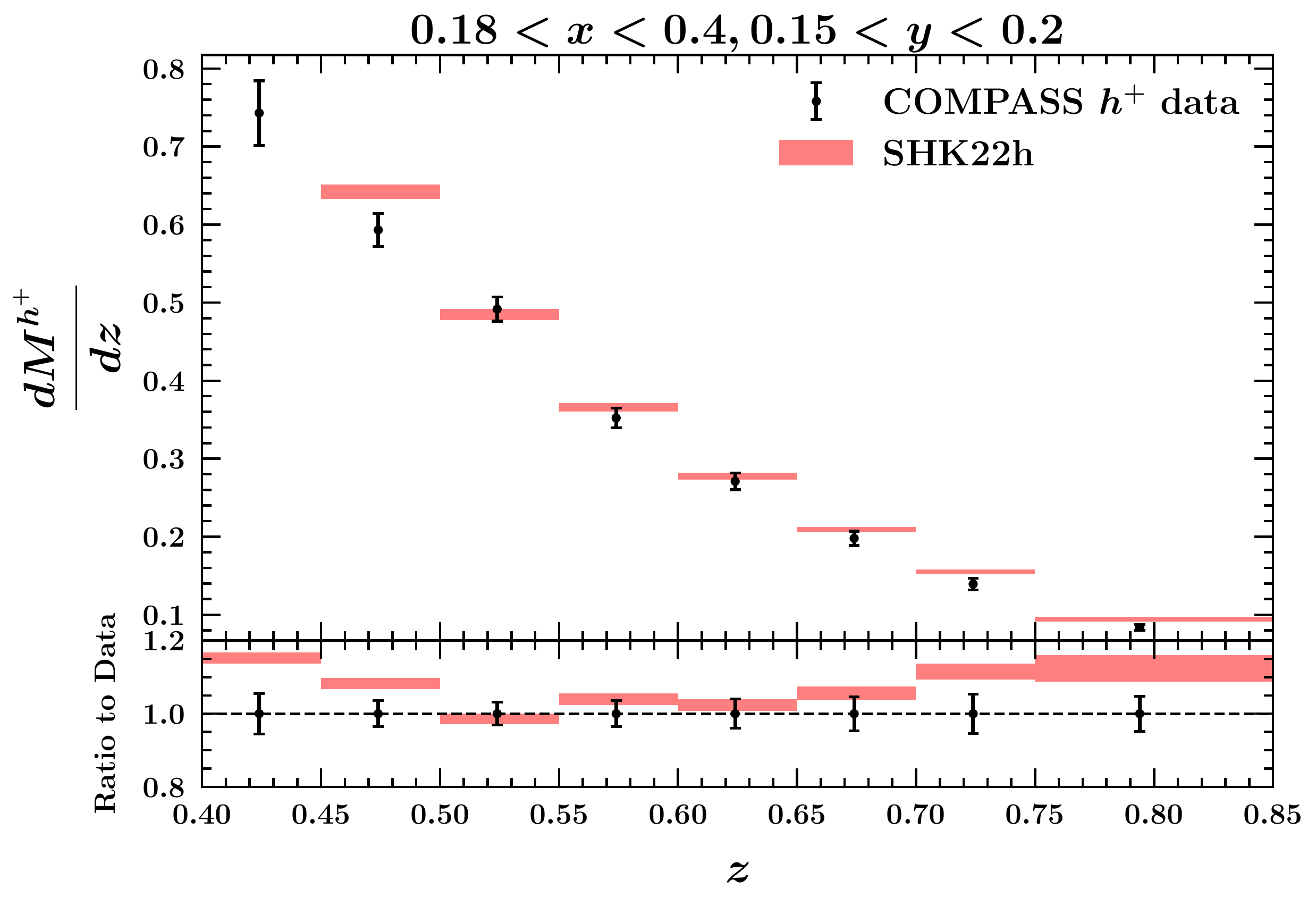}} 
\resizebox{0.480\textwidth}{!}{\includegraphics{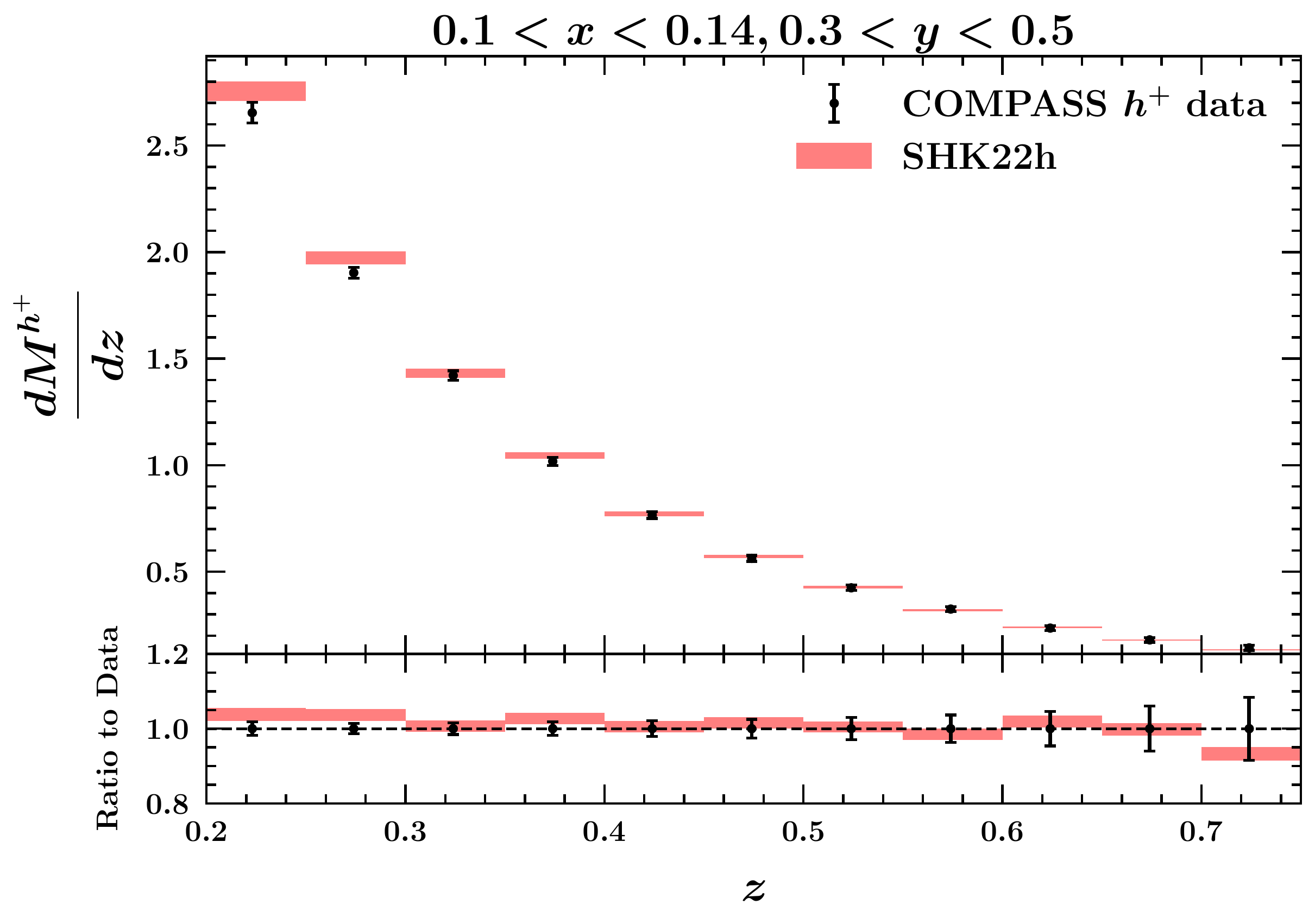}}	
\begin{center}
\caption{ 
\small 
The data/theory comparison for COMPASS 
multiplicities data sets for $h^+$. Each panel shows a distribution as a function of $z$ corresponding to a bin in $x$ and $y$.  
The lower panels display the ratio to the experimental central values.
}
\label{fig:SIDISHp}
\end{center}
\end{figure*}

\begin{figure*}[htb]
\vspace{0.50cm}
\resizebox{0.480\textwidth}{!}{\includegraphics{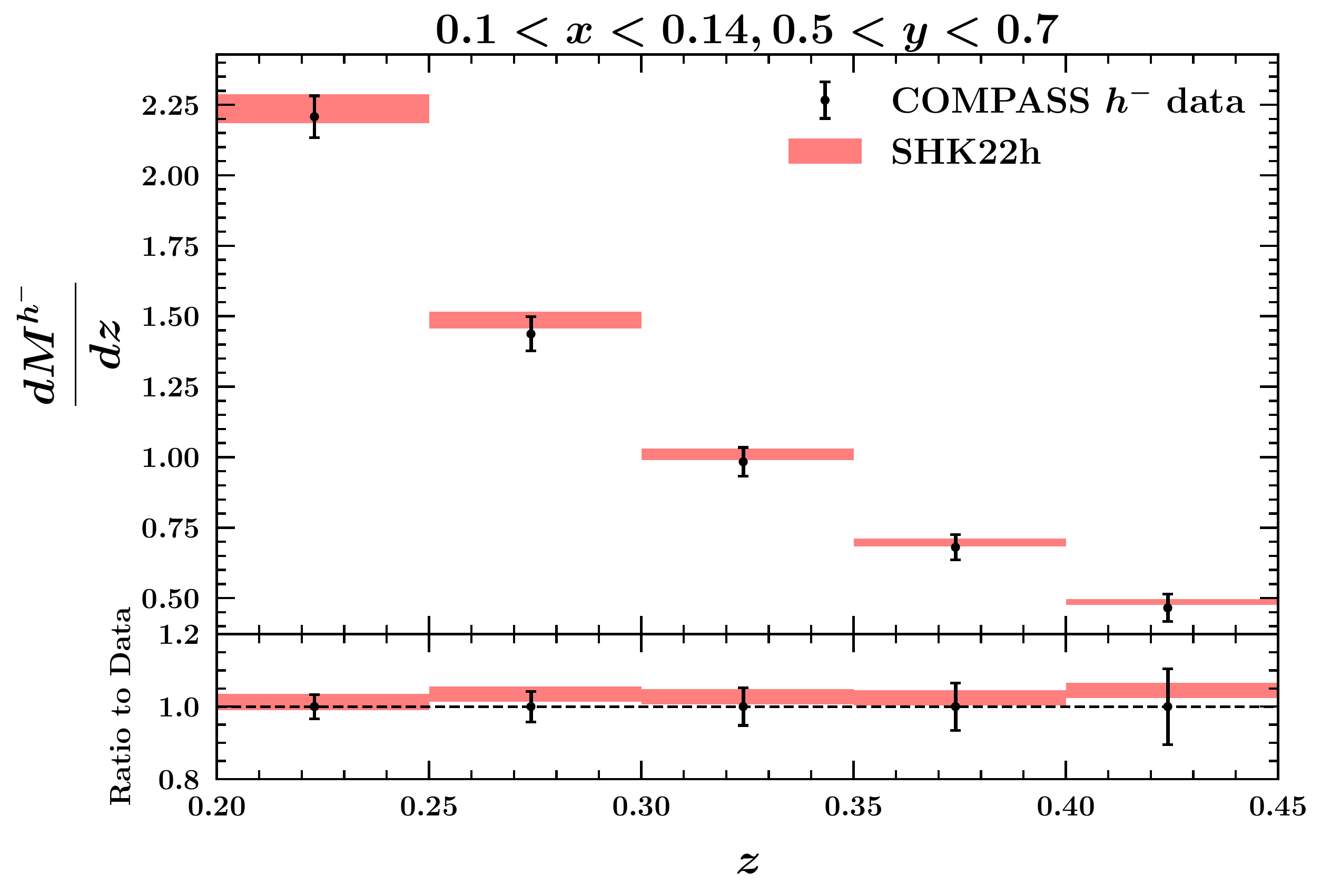}} 	
\resizebox{0.480\textwidth}{!}{\includegraphics{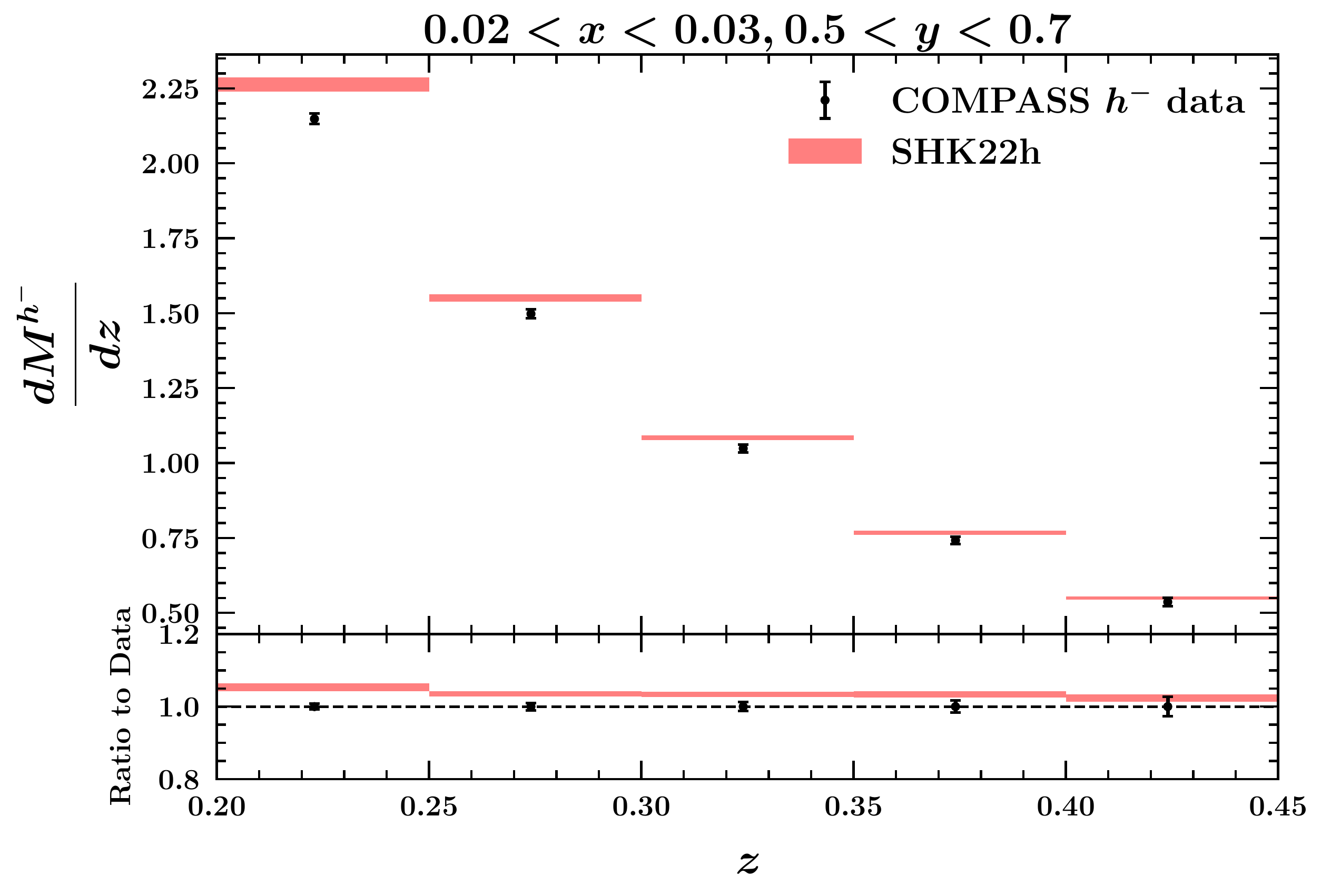}}  	
\resizebox{0.480\textwidth}{!}{\includegraphics{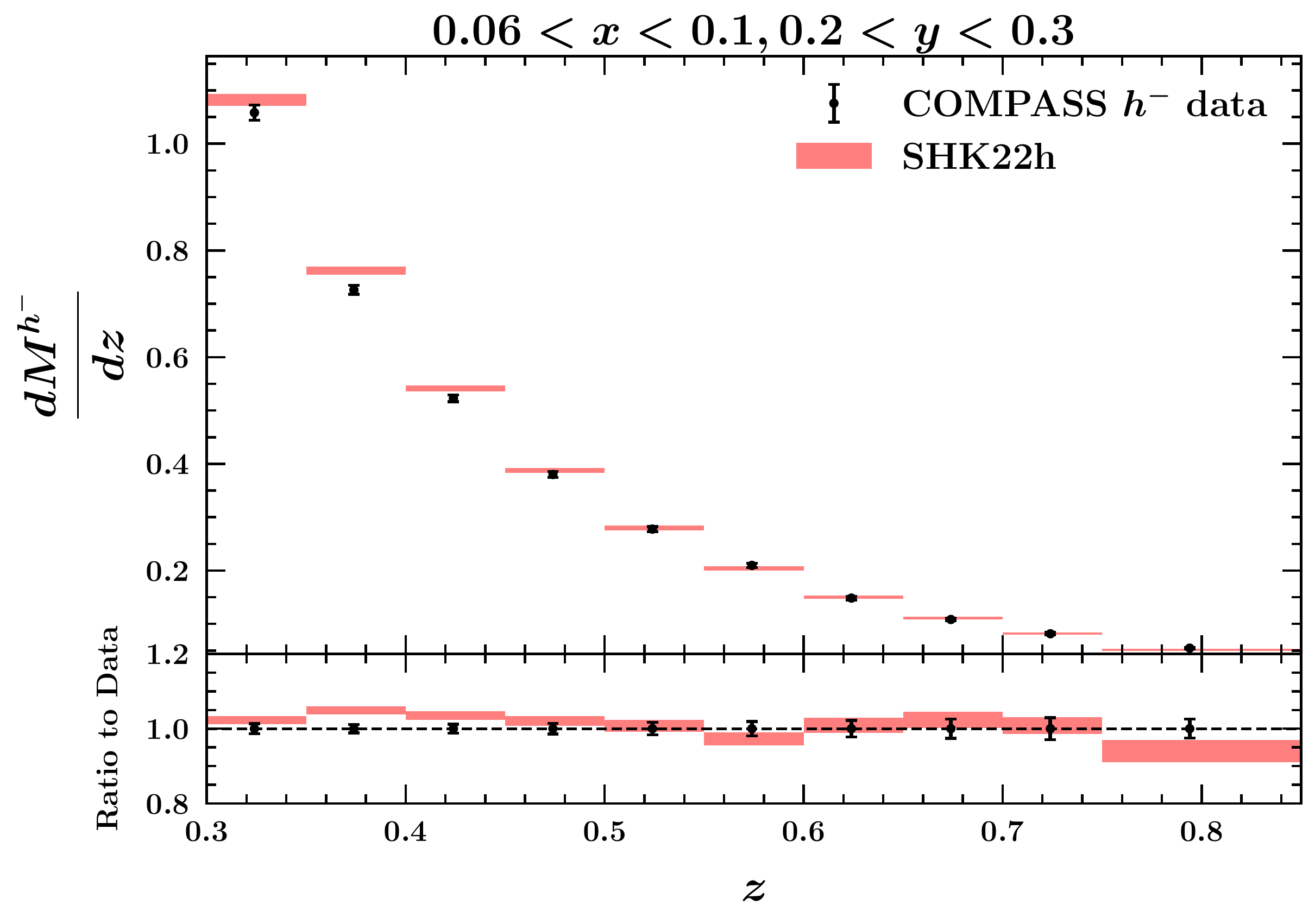}}
\resizebox{0.480\textwidth}{!}{\includegraphics{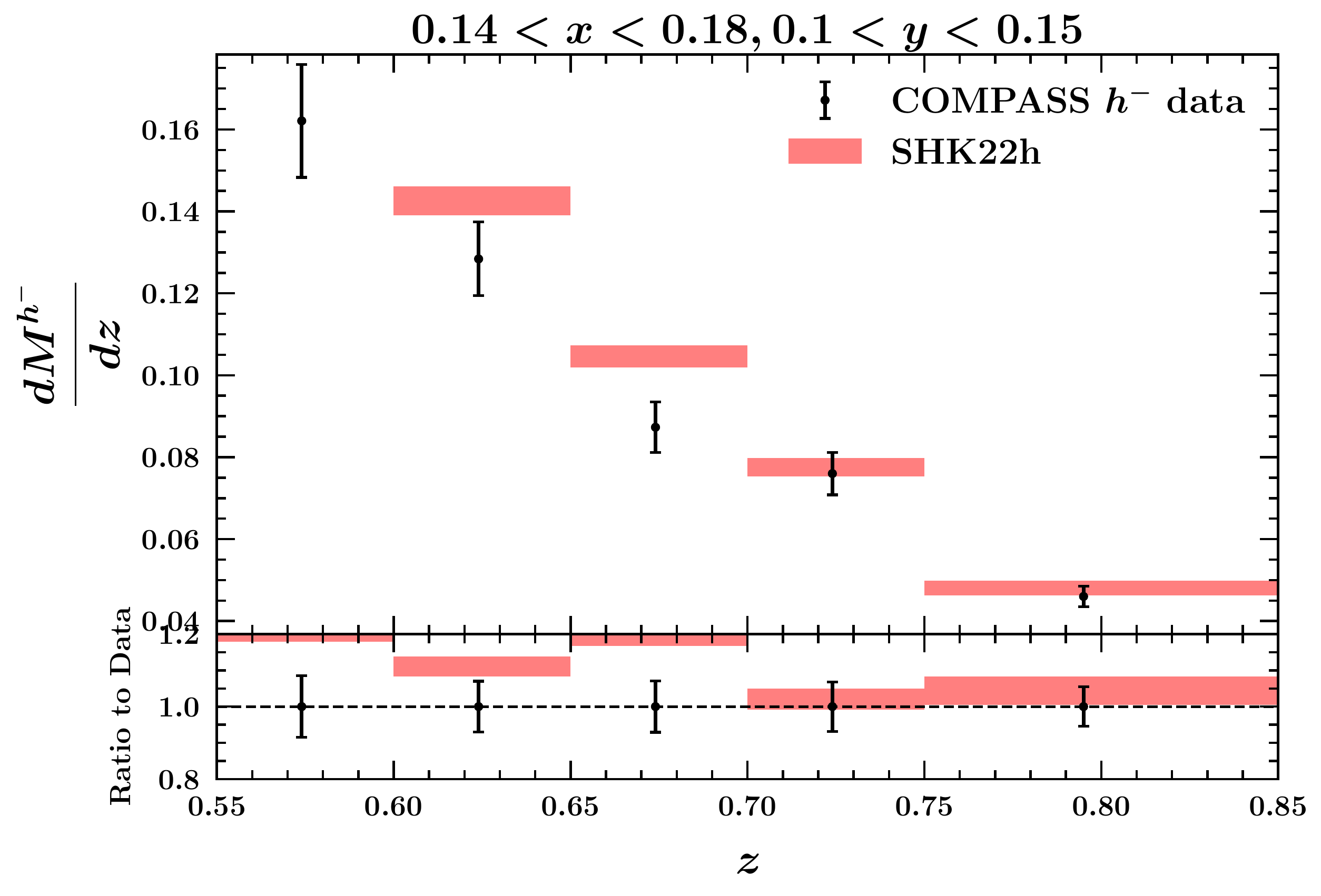}}  
\resizebox{0.480\textwidth}{!}{\includegraphics{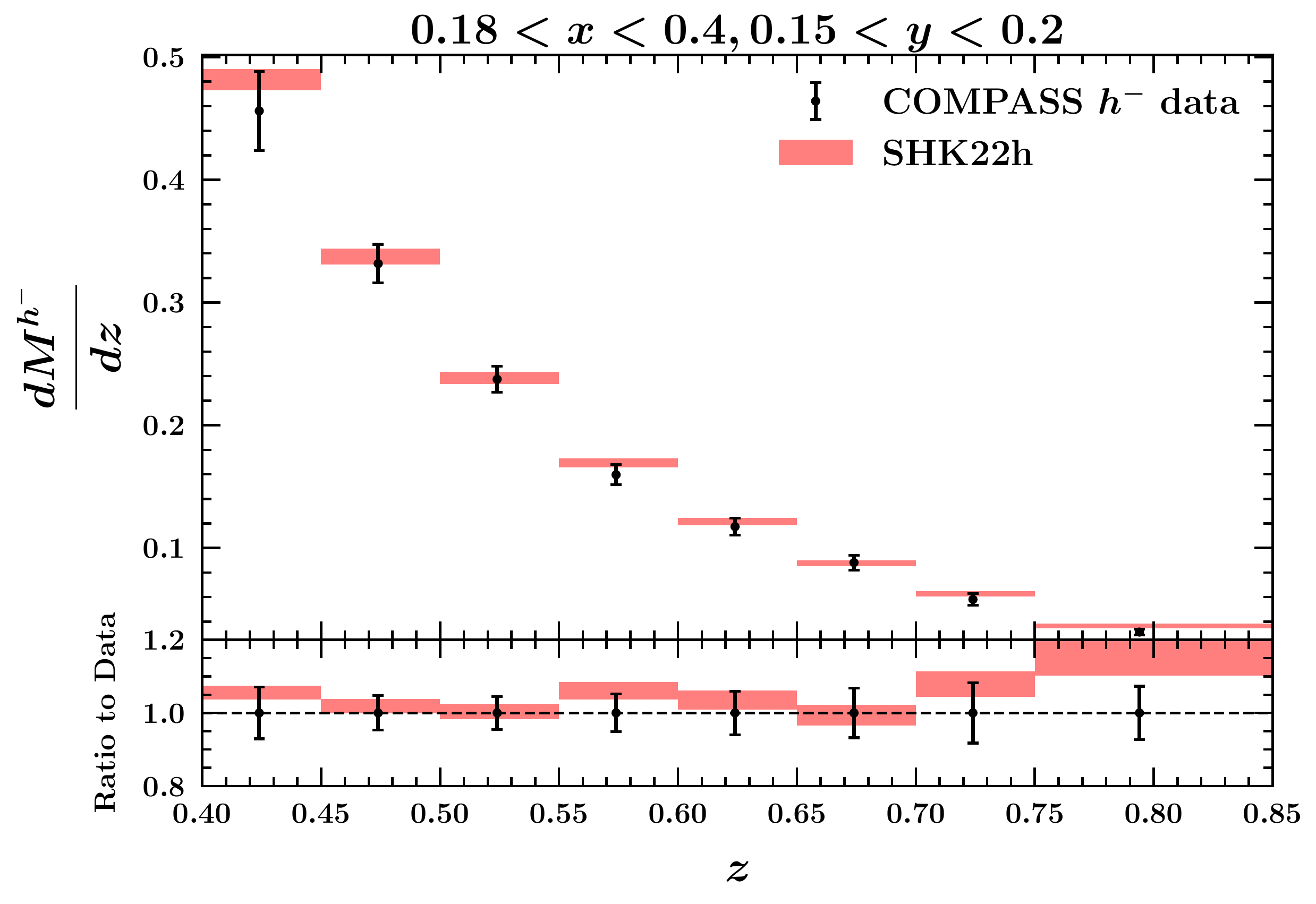}} 
\resizebox{0.480\textwidth}{!}{\includegraphics{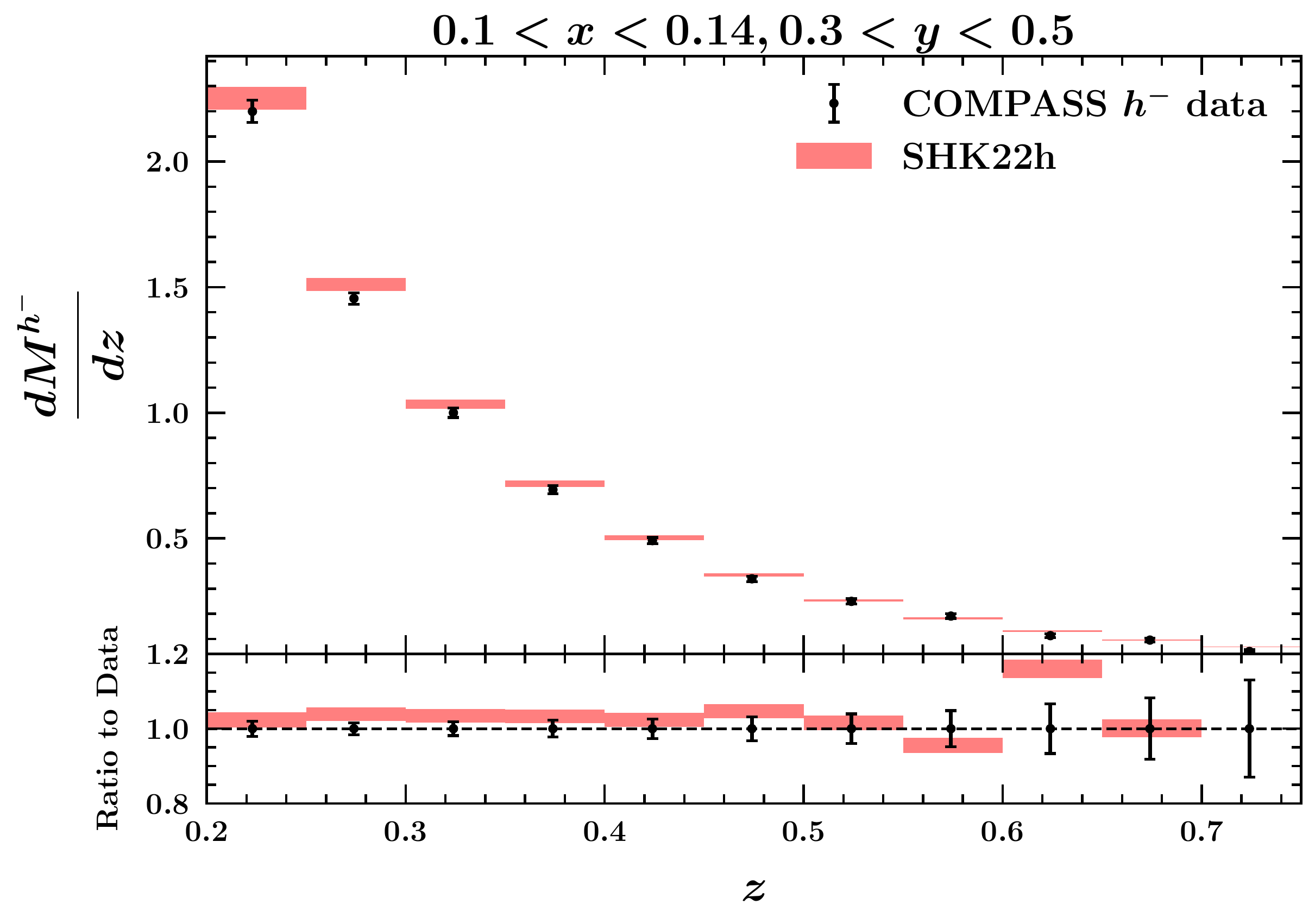}}	
\begin{center}
\caption{ 
\small 
Same as Fig.~\ref{fig:SIDISHp} but for $h^-$}
\label{fig:SIDISHm}
\end{center}
\end{figure*}

Figs.~\ref{fig:SIDISHp} and \ref{fig:SIDISHm} presents 
the data/theory comparison for some of 
the COMPASS multiplicities data sets for the $h^+$ and $h^-$, 
respectively. Each panel shows a 
distribution as a function of $z$ corresponding to a bin 
in $x$ and $y$. As above, the lower 
panels display the ratio to the experimental central values.
A remarkable feature of the distributions shown in these figures
is the very nice agreements between the light charged hardon COMPASS data 
and the {\tt SHK22.h} NLO theory prediction. 
 
While the agreements for the COMPASS $h^-$ data is noticeable 
for all range of $z$ and for
all bin in $x$ and $y$, some  
differences between the COMPASS $h^+$ data and theory predictions 
can be seen, more specifically for the small value of $z$ in which the  
theoretical predictions overshoot the data. 
This is also consistent with the poor $\chi^2$ for 
COMPASS $h^+$ data 
reported in Table.~\ref{tab:chi2-SHK22.h}.

%

%

As a short summary, in general, an overall good agreement 
between the analyzed data sets 
and the NLO theoretical predictions is achieved for all experiments, 
consistent
with the total values of $\chi^2$ reported in this section. 
Remarkably, the {\tt SHK22.h} NLO theoretical predictions and both SIA and SIDIS data are 
in reasonable agreement also in the small 
 and large-$z$ values with exception of few data 
sets that we discussed above.

%
\subsection{The {\tt SHK22.h} light-charged hadron FFs Set}\label{sec:FFs}
%

We are now in a position to discuss the {\tt SHK22.h} 
light-charged hadron FFs sets. 
In order to study in details the extracted FFs sets, 
we compare our best-fit results to other recent counterparts available in the literature, the {\tt JAM20} \cite{Moffat:2021dji} and 
{\tt NNFF1.1h} \cite{Bertone:2018ecm} analyses.

\begin{figure*}[htb]
	\vspace{0.50cm}
	\resizebox{0.470\textwidth}{!}{\includegraphics{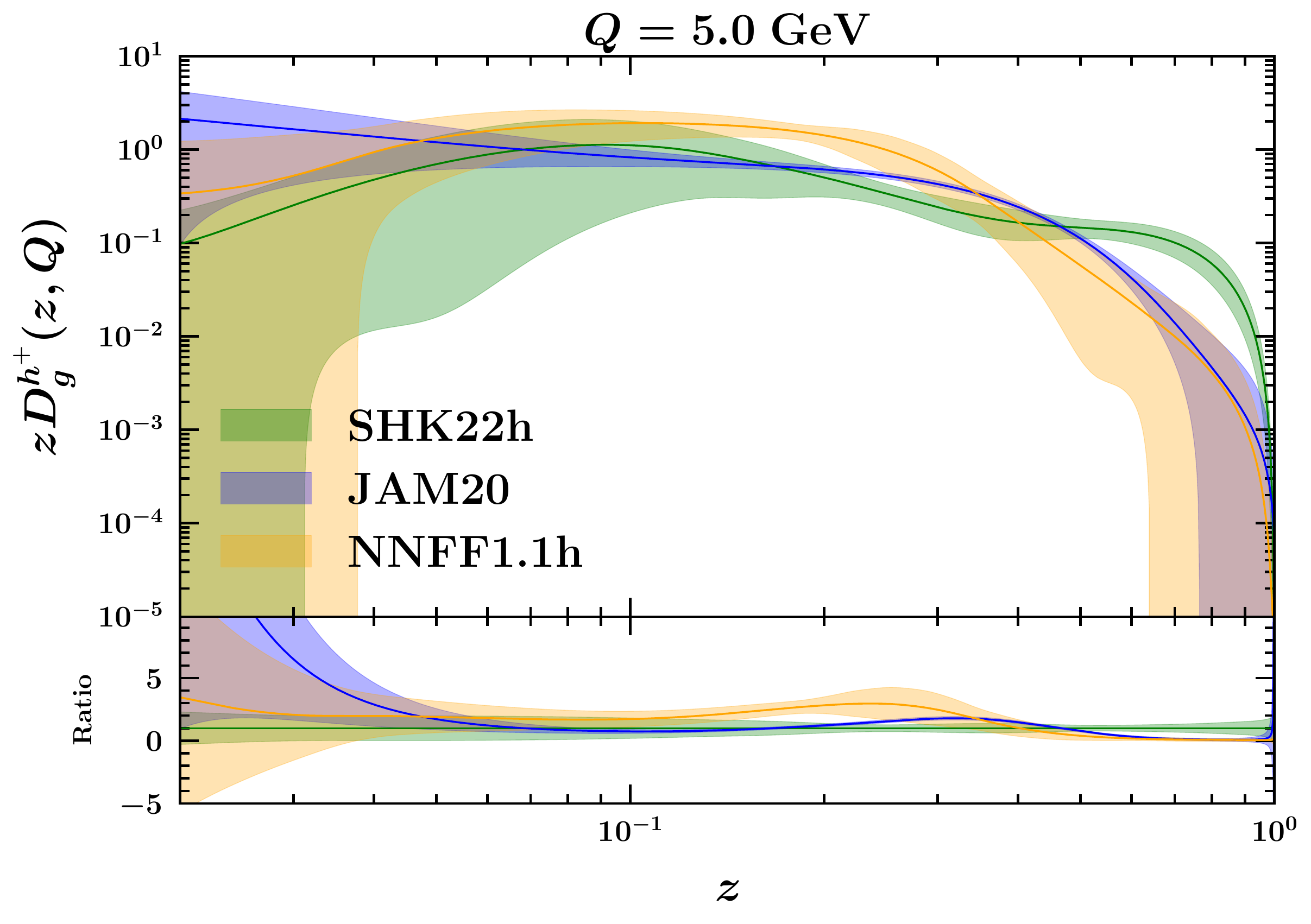}} 	
	\resizebox{0.470\textwidth}{!}{\includegraphics{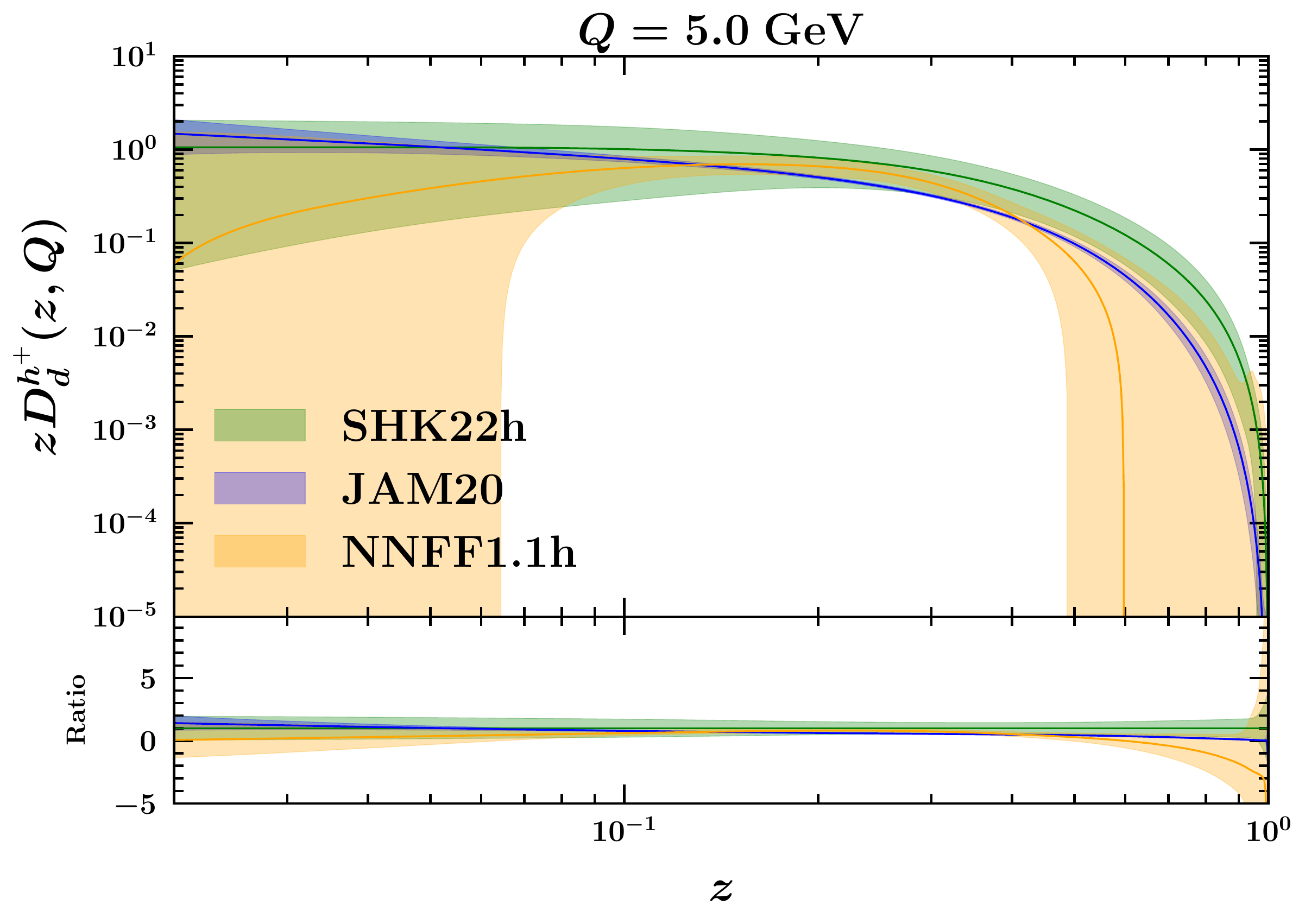}}  	
	\resizebox{0.470\textwidth}{!}{\includegraphics{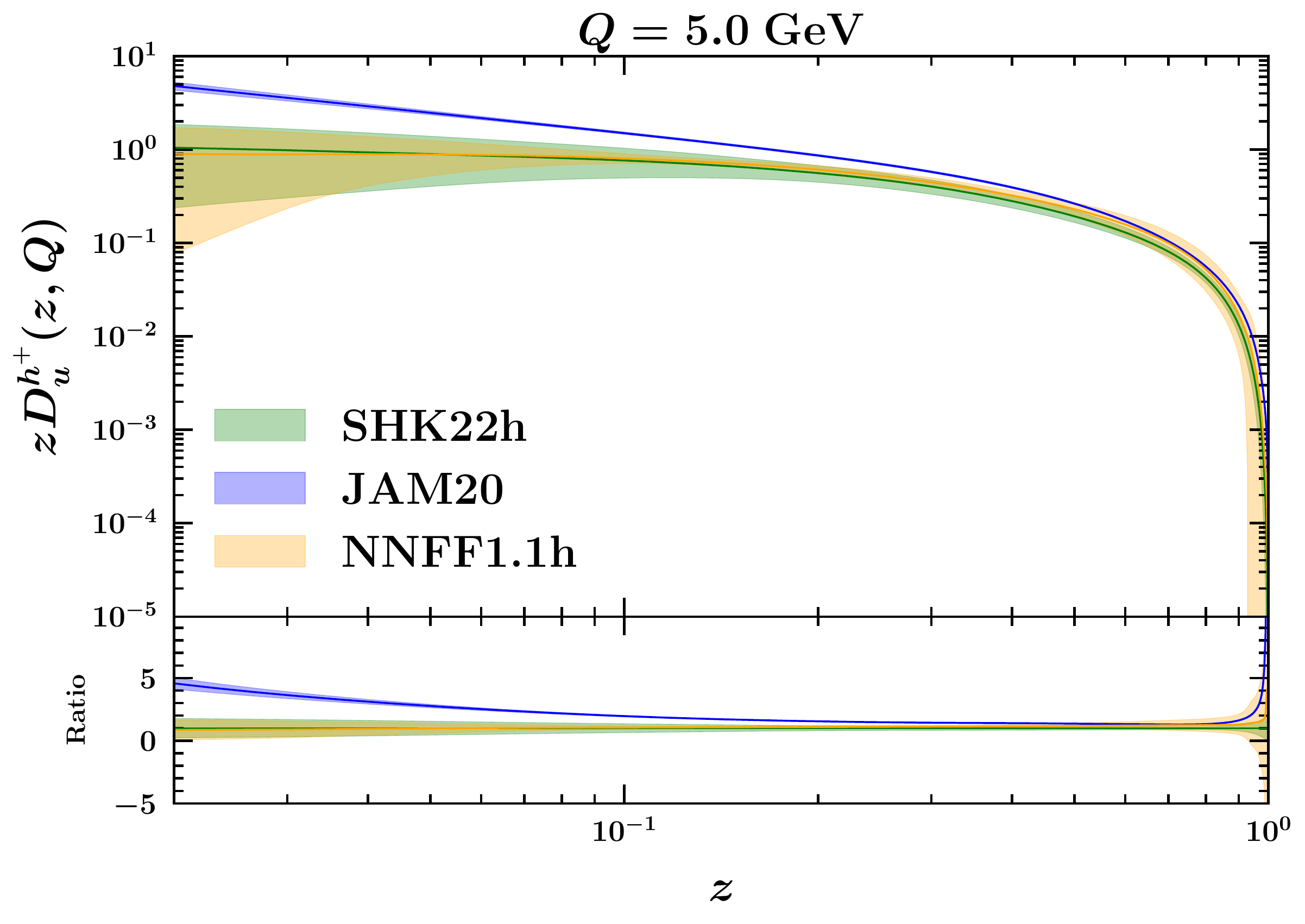}} 	
	\resizebox{0.470\textwidth}{!}{\includegraphics{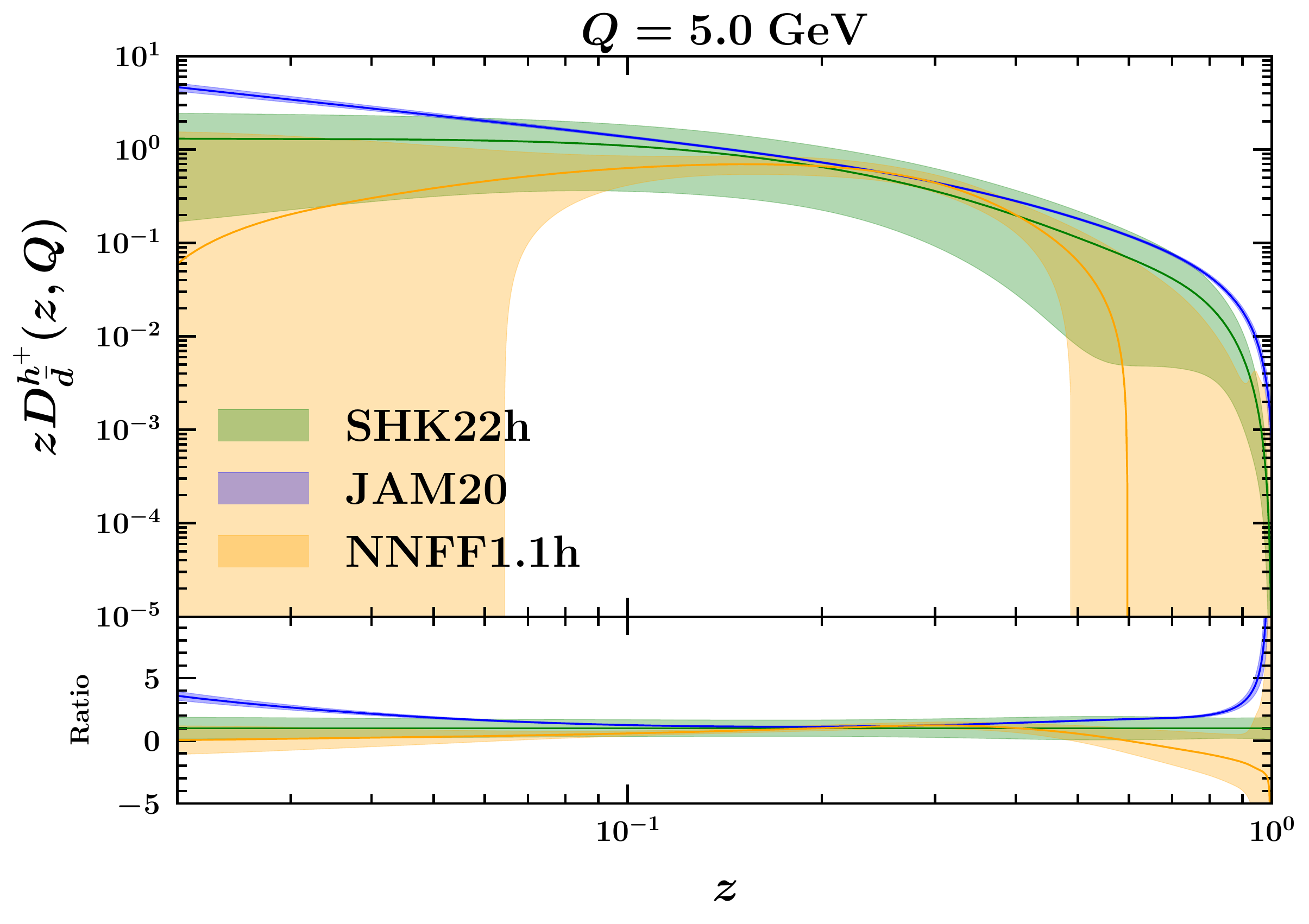}}
	\resizebox{0.470\textwidth}{!}{\includegraphics{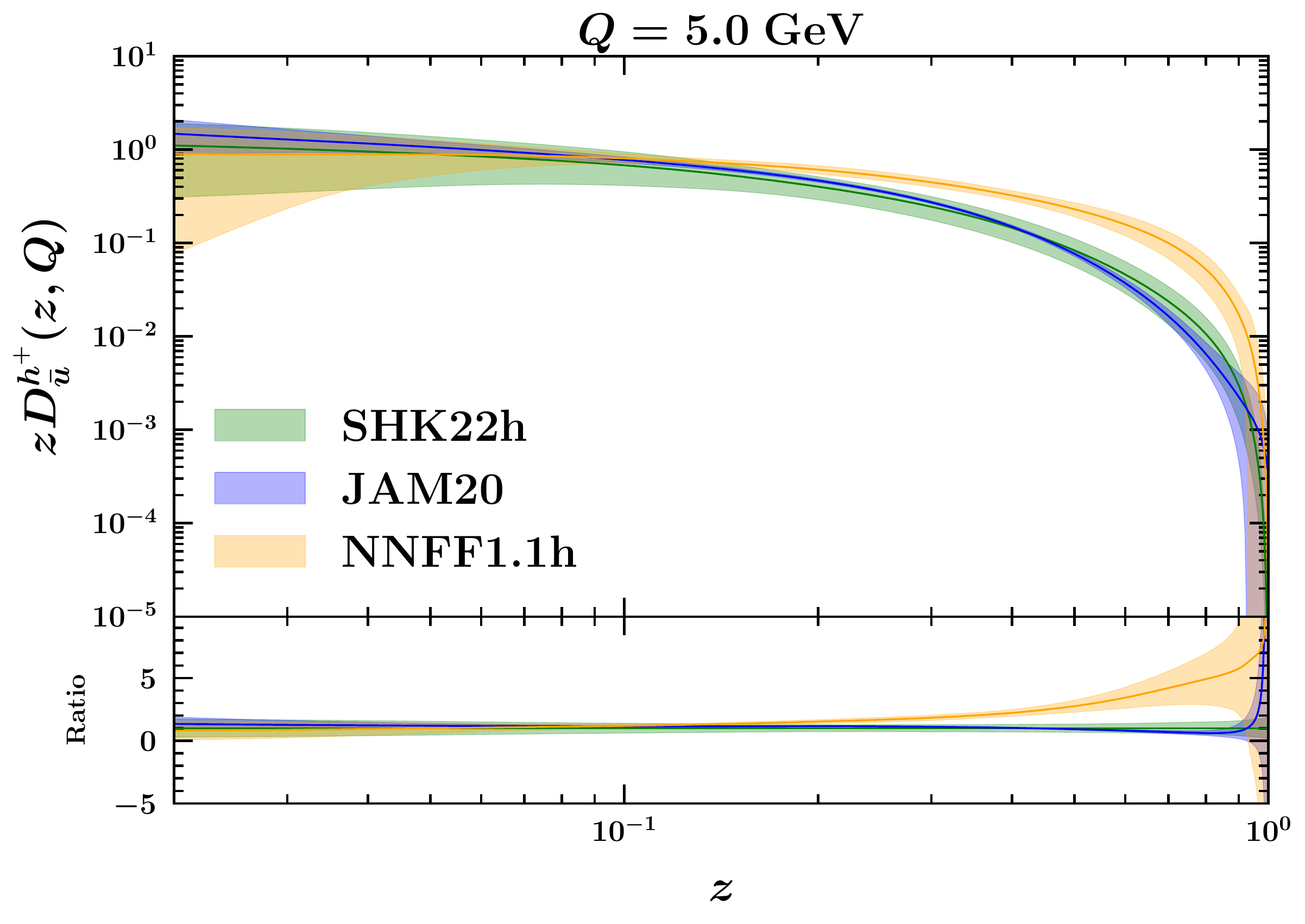}} 
	\resizebox{0.470\textwidth}{!}{\includegraphics{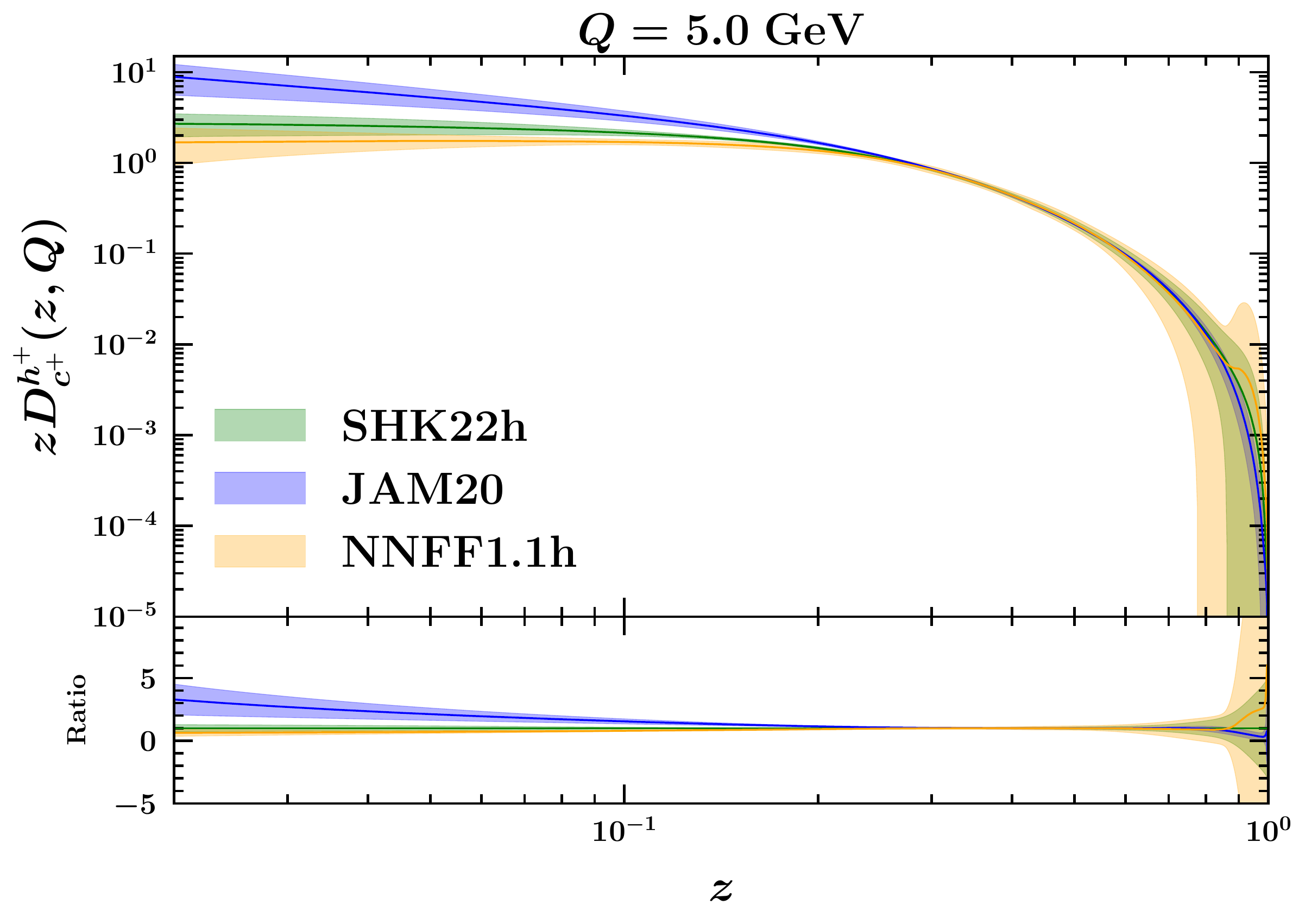}} 
	\resizebox{0.470\textwidth}{!}{\includegraphics{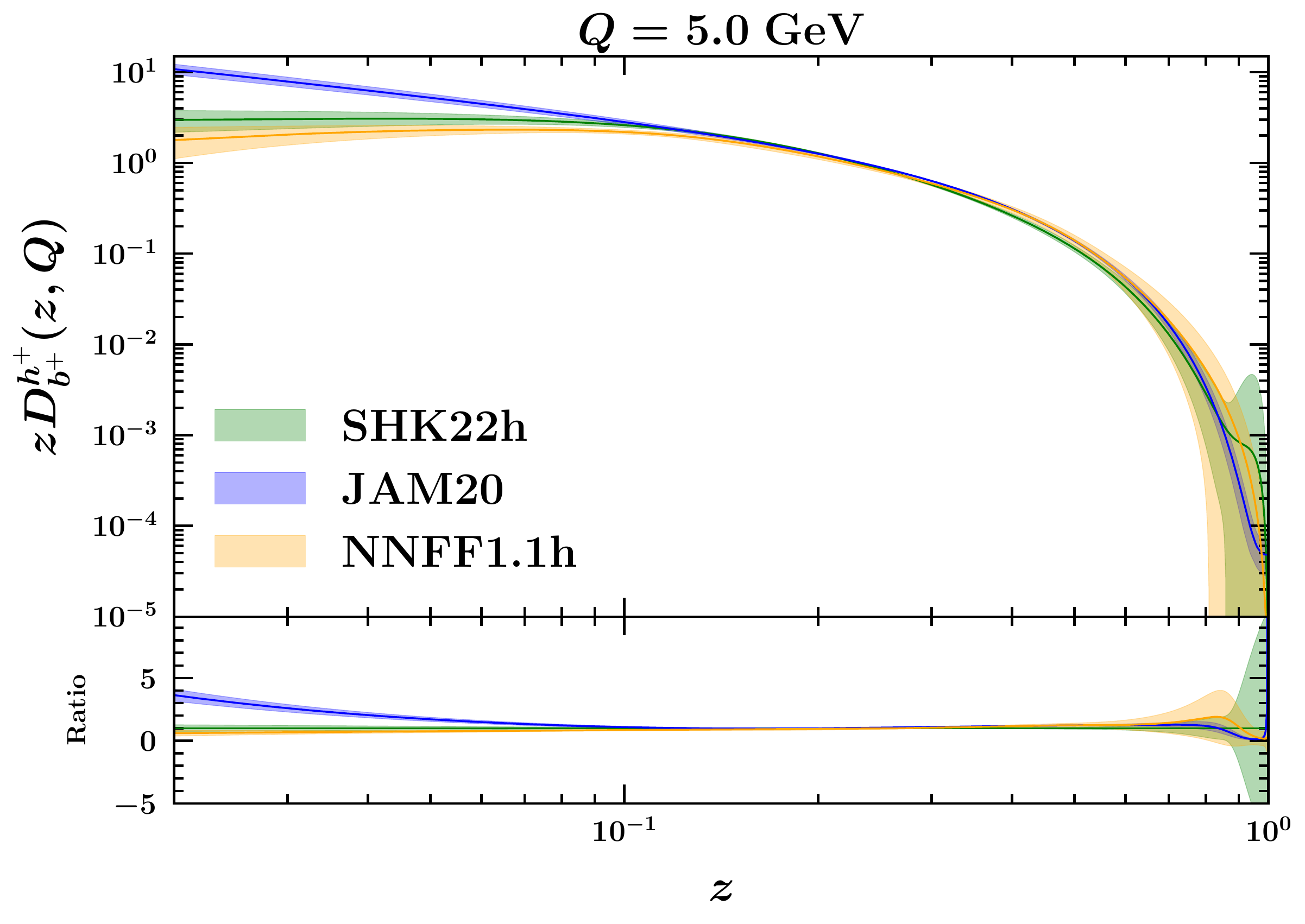}}   	 	
	\begin{center}
		\caption{ 
			\small 
Upper panel indicates the comparison of our FF 
sets for charged hadrons with {\tt JAM20}~\cite{Moffat:2021dji} and {\tt NNFF1.1h}~\cite{Bertone:2018ecm} FF sets at $Q=5$ GeV at NLO accuracy. 
Lower panel represents the ratios of the all sets to the 
central value of {\tt SHK22.h}.}
		\label{fig:light-charged-hadron-FFs}
	\end{center}
\end{figure*}

We display the light-charged hadron FFs parameterized in {\tt SHK22.h} fits,
and their uncertainties in Fig.~\ref{fig:light-charged-hadron-FFs}. 
We present the 7 hadronic species at Q=5 GeV which are $zD_g^{h^+} (z, Q)$,  $zD_d^{h^+} (z, Q)$,  $zD_{\bar d}^{h^+} (z, Q)$,  $zD_u^{h^+} (z, Q)$,  $zD_{\bar u}^{h^+} (z, Q)$,  $zD_{c^+}^{h^+} (z, Q)$ and $zD_{b^+}^{h^+} (z, Q)$. 
The upper panel of each plot presents the absolute distributions,
while the the lower panels display the ratio to the {\tt SHK22.h}. It is to be noted that \texttt{NNFF1.1h} only extracted the gluon, $c$, $b$ quark, and flavor singlet combination in their analysis. However, the authors have given instructions to disentangle the up and down contributions in Appendix~A of \cite{Bertone:2017tyb} and also produced the related \texttt{LHAPDF} format grids, and the plots that are presented here use such prescriptions.

Concerning the shapes of the light-charged hadron FFs, there are  
a number of interesting similarities and 
differences between these three different sets as 
can be seen from the comparisons presented in Fig.~\ref{fig:light-charged-hadron-FFs}. 

We start with the bottom quark FF  $zD_{b^+}^{h^+} (z, Q)$. 
As one can see from Fig.~\ref{fig:light-charged-hadron-FFs}, 
in term of the central distribution, 
these three sets are in very good agreement, except for the high-$z$ region.
The uncertainty of the $zD_{b^+}^{h^+} (z, Q)$ FFs 
deserves a separate comment.
The uncertainty for the {\tt NNFF1.1h} is relatively large over all 
region of $z$, while for the 
{\tt JAM20} is very narrow.
For  {\tt SHK22.h}, the calculated uncertainty is slightly 
large for the large value 
of $z$ due to the lack of data in this region. 
Although, it still remained smaller than {\tt NNFF1.1h} over all range of $z$. 
The same findings are hold for the charm-quark FF 
$zD_{c^+}^{h^+} (z, Q)$ with the exception that 
the central value of {\tt SHK22.h} and {\tt NNFF1.1h} is smaller than those of the 
{\tt JAM20} for low value of $z$; $z < 2 \times 10^{-1}$.
An interesting difference can be seen for the  $zD_{\bar u}^{h^+} (z, Q)$ FF between {\tt SHK22.h} 
and  {\tt NNFF1.1h}  for medium to large value of $z$ in which  the {\tt NNFF1.1h} result is larger than 
{\tt SHK22.h}, 
while for the case of $zD_u^{h^+} (z, Q)$ they are in good agreement. 
Moderate differences on up-quark FF are observed for the 
{\tt SHK22.h} and {\tt JAM20} 
for almost all region of $z$ while {\tt SHK22.h} and {\tt NNFF1.1h} FFs remain consistent.

The most pronounced differences both in shape and uncertainty 
bands are observed for the gluon $zD_g^{h^+} (z, Q)$, 
down-quark $zD_d^{h^+} (z, Q)$ and $zD_{\bar d}^{h^+} (z, Q)$ FFs.
For the $zD_d^{h^+} (z, Q)$ and $zD_{\bar d}^{h^+} (z, Q)$ FFs, 
the differences in shape and 
uncertainty bands among the three FF sets are 
more marked for large $z$ rather than the small region of $z$.
As one can see, a fair agreement for the central value is 
observed only in the region of $z < 2 \times 10^{-1}$.
For medium to large $z$ region, the {\tt SHK22.h} $zD_{\bar d}^{h^+} (z, Q)$ FF is larger than {\tt NNFF1.1h} and smaller than the {\tt JAM20}.
For the case of $zD_d^{h^+} (z, Q)$, 
our results are larger than those of the two others in most of the $z$ range.
In term of uncertainty bands, we obtained a larger error bands 
in respect to the {\tt JAM20} analysis, which is expected, considering their functional parametrization.
For the {\tt NNFF1.1h}, the central value of  
$zD_d^{h^+} (z, Q)$ and $zD_{\bar d}^{h^+} (z, Q)$ tend to 
zero for large value of $z$, $z>0.6$, with much 
wider error bands for all region of $z$.
 
The central value and uncertainty bands of the 
gluon FFs $zD_g^{h^+} (z, Q)$, deserve separate comments. 
As one can see from Fig.~\ref{fig:light-charged-hadron-FFs}, 
there are noticeable differences both in 
term of central values and uncertainty bands between these 
three different sets.

The {\tt JAM20}  analysis includes all available SIDIS 
and SIA, and the inclusive
DIS and Drell-Yan lepton-pair production as well to 
calculate the PDFs simultaneously with the FFs. 
The analysis by  {\tt NNPDF} Collaboration included the proton-proton data for 
unidentified charge hadron production by 
means of Bayesian reweighting to the 
analysis based only on the SIA data sets.
Typically, the uncertainties of the {\tt NNFF1.1h} FFs are much 
larger than our results and {\tt JAM20} at
small and large values of $z$.
The  uncertainty band for our result is wider than {\tt JAM20} over the 
small value of $z$, and smaller for the high-$z$ region. 
The smallness of the uncertainties for the {\tt JAM20}  analysis are discussed in details in Ref.~\cite{Abdolmaleki:2021yjf}.
We should stress here that the smaller uncertainty for all FFs presented in  Fig.~\ref{fig:light-charged-hadron-FFs} 
reflect the more restrictive functional form used in the {\tt JAM20}  
analysis to parameterize their FFs at the input scale.


As we previously mentioned, the SIA
cross-sections are less sensitive to the gluon FF. 
As a consequence, one would expect that, in the presence of SIA data only, the gluon FF will be 
determined with larger uncertainties than other quark FFs. 
Hence, in {\tt SHK22.h} study, the SIDIS data 
added to the data sample to provide stronger constraint for the gluon density. In the next subsection, we present our study on the effect of SIDIS data on the extracted FFs.

%
\subsection{Impact of SIDIS data on the {\tt SHK22.h} FFs}\label{sec:SIDID}
%

In this section, we discuss the impact of SIDIS data sets on the extracted FFs.
To this end, we compare the main results of the {\tt SHK22.h} 
global QCD analysis which include the SIA and SIDIS 
experimental data with the 
analysis in which based on the SIA measurements only.

\begin{figure*}[htb]
\vspace{0.50cm}
\resizebox{0.450\textwidth}{!}{\includegraphics{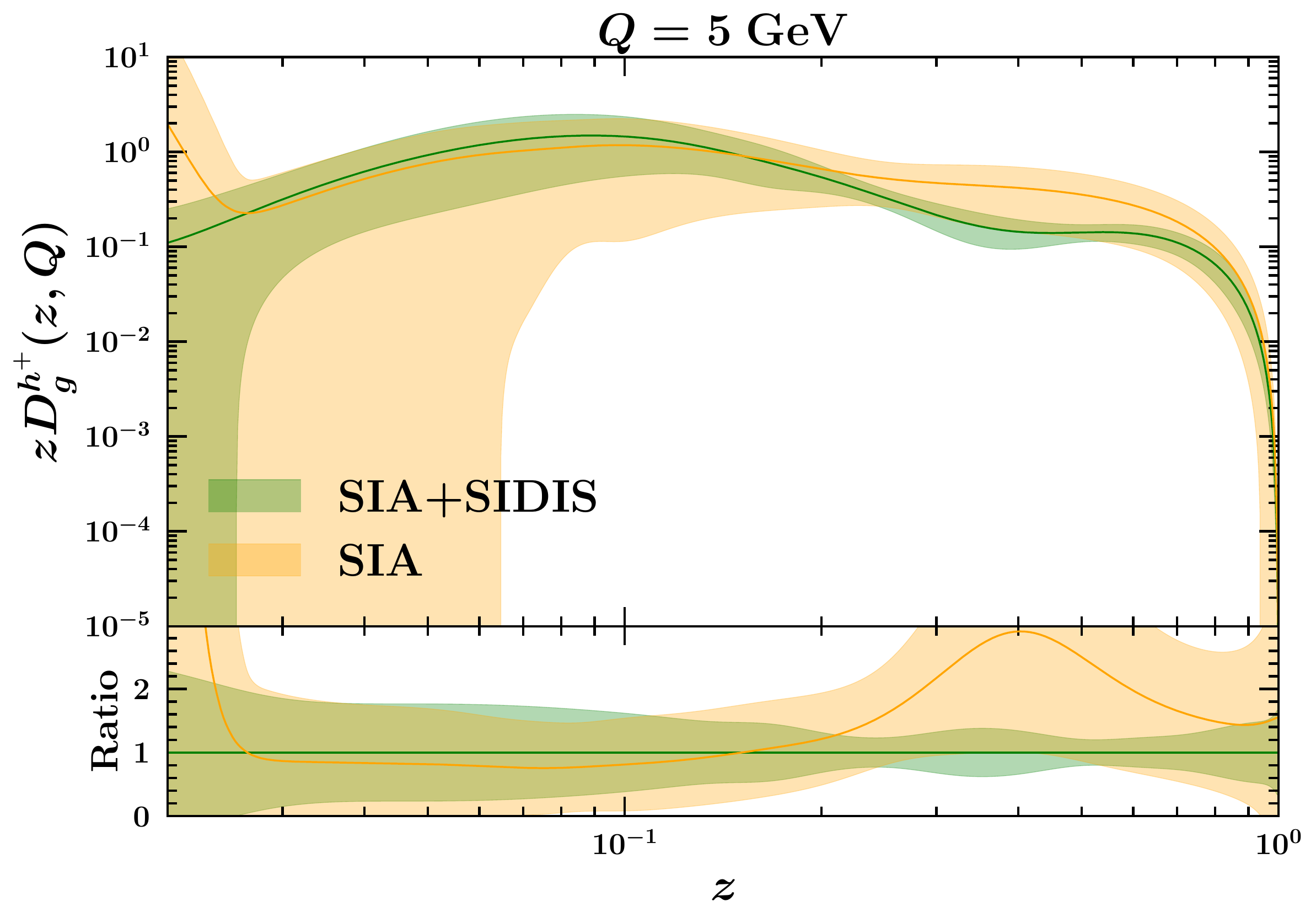}} 	
\resizebox{0.450\textwidth}{!}{\includegraphics{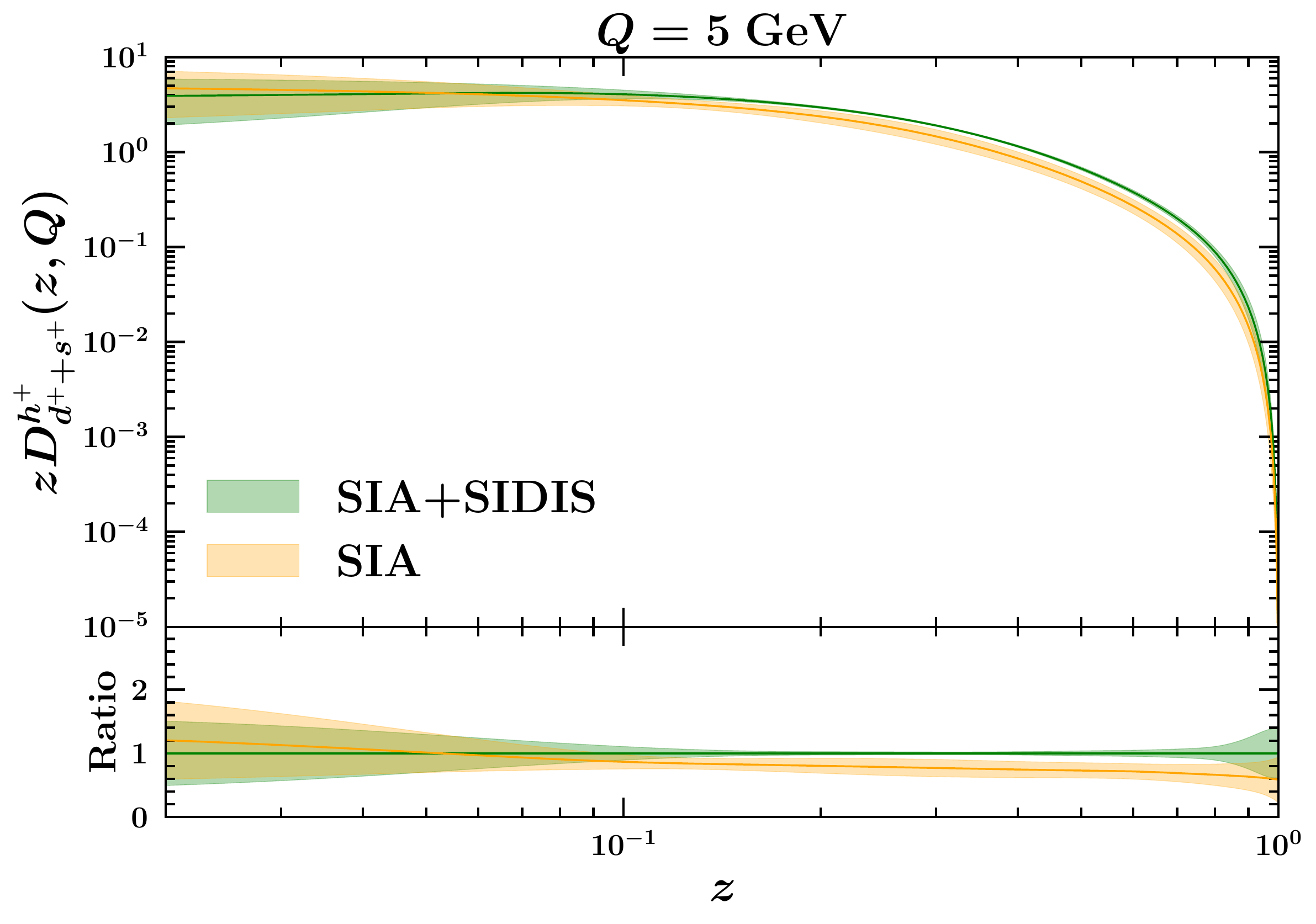}}  	
\resizebox{0.450\textwidth}{!}{\includegraphics{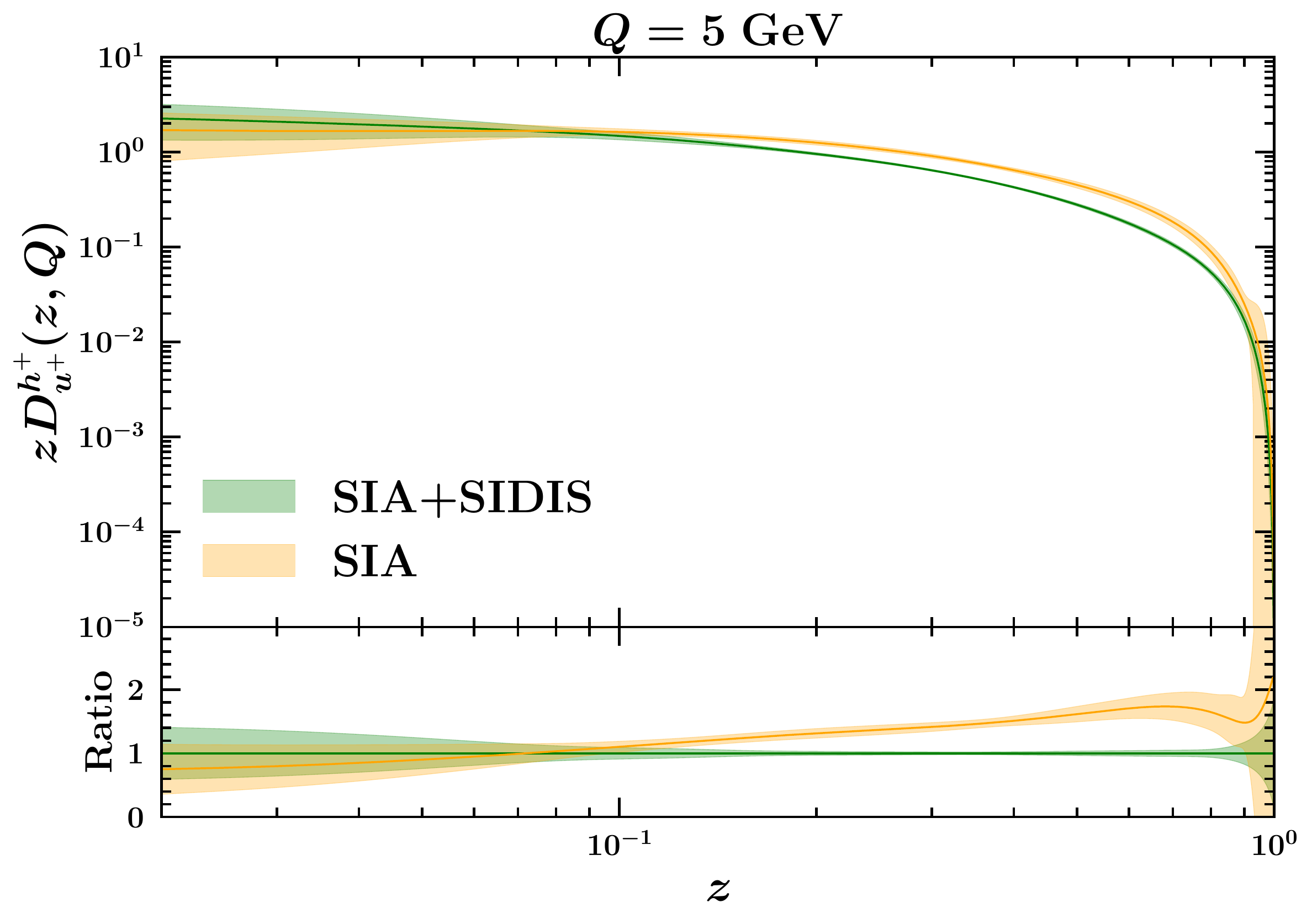}} 
\resizebox{0.450\textwidth}{!}{\includegraphics{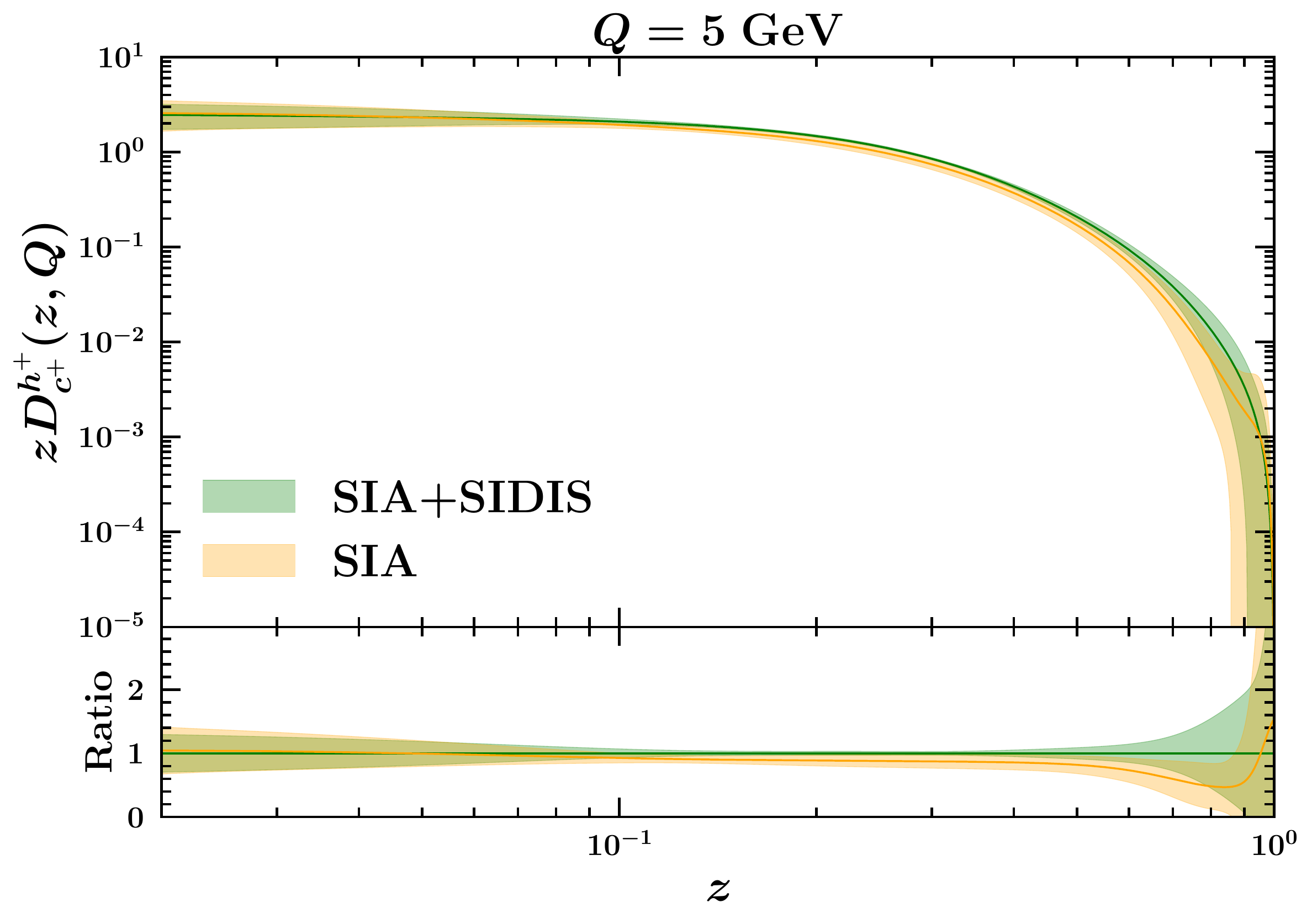}} 
\resizebox{0.450\textwidth}{!}{\includegraphics{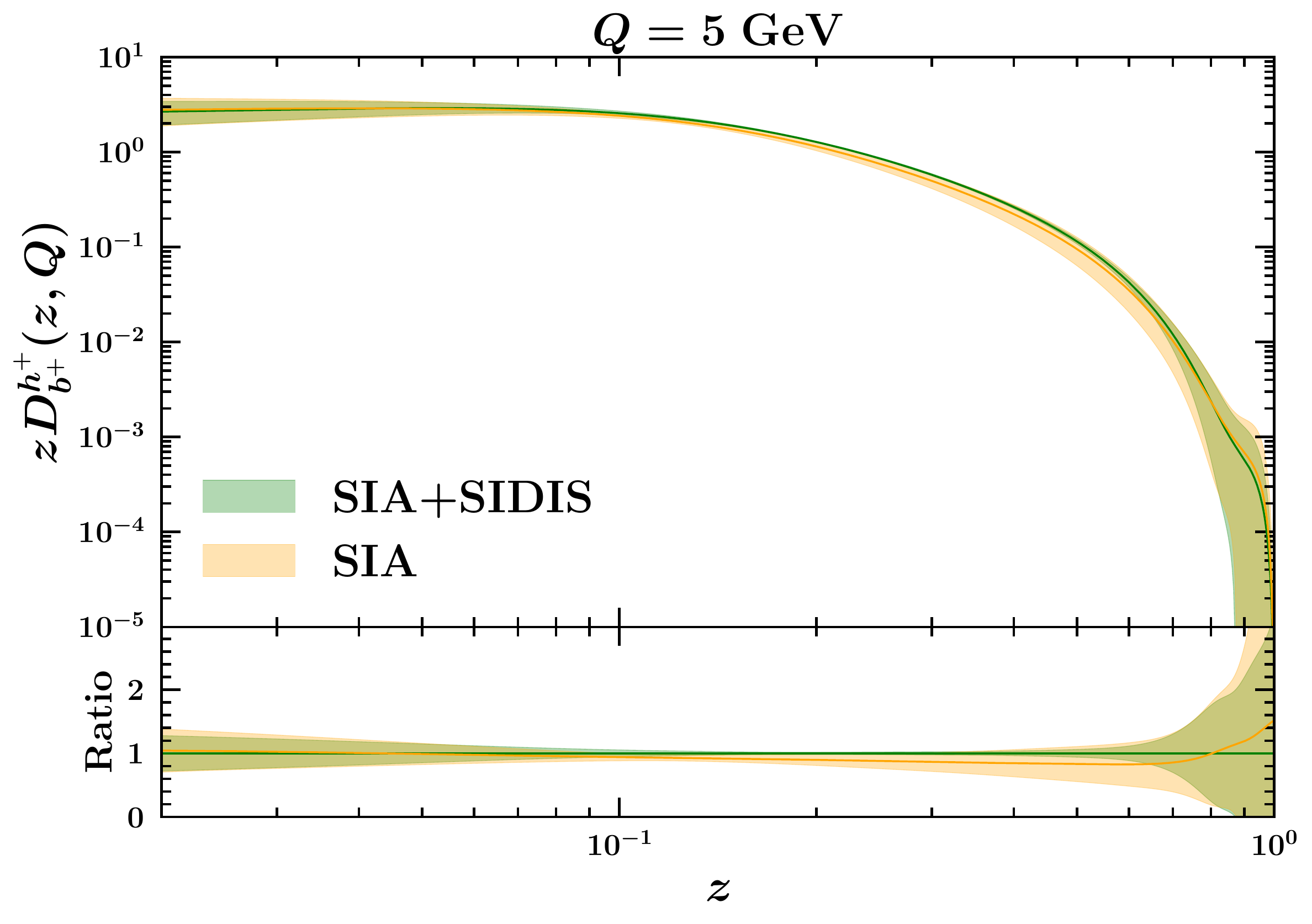}}   	 	
\begin{center}
\caption{ 
\small 
Comparison of light-charge hadron FF 
sets obtained from only SIA data with the  
{\tt SHK22.h} FFs in which obtained by a combination of SIA 
and SIDIS experimental data at NLO accuracy. 
}
\label{fig:SIA-SIDIS}
\end{center}
\end{figure*}

According to Eq.~\ref{eq:SIAcombinations} and 
Eq.~\ref{eq:combinations}, the flavor 
decompositions in the parametrization of FFs are not 
the same for the analysis with SIA data only, 
and the global analysis with both SIA and SIDIS data sets. 
Then, in order to investigate the impact of 
including SIDIS data in the {\tt SHK22.h} FFs analysis, 
the comparison 
of FFs for different flavors have been shown 
in Fig.~\ref{fig:SIA-SIDIS} at Q=5 GeV.

%

From the comparisons presented in Fg.~\ref{fig:SIA-SIDIS}, 
one observes that the inclusion of 
SIDIS COMPASS data sets affects both the central values and the 
uncertainty bands of the extracted FFs. \\

Such differences are more pronounced for the gluon FF $zD^{h^+}_{g}$ in terms of both the central 
value and for the error bands. As one can see from the comparison
between the SIA and SIDIS for the gluon FF, the inclusion of the SIDIS data 
leads to an enhancement of 
the $zD^{h^+}_g$ distribution for small value of $z$; $z < 0.2$ in comparison 
with a fit to the SIA data. We also see that at the large value of $z$, the SIA+SIDIS  
and SIA fits are in good agreement. 
For the medium  value of $z$, the  gluon distribution of 
the global SIA+SIDIS
fit get suppressed with respect to the SIA fit. 
One can also see from Fig.~\ref{fig:SIA-SIDIS} that the gluon 
FF uncertainty for all range of $z$ is reduced. However, gluon FF enter the description of SIA and SIDIS at the same order of perturbation theory, so they are not different regarding the sensitivity to the FF of gluon. In order to constrain the gluon FF one needs to include proton-proton data in analysis which is sensitive to gluon FF already at LO~\cite{Bertone:2018ecm}. Therefore, we believe that the reduction in the uncertainty of gluon FF is because of significant increase in the statistics of the included experimental information from SIDIS observables.

For other FFs, 
we find in general a reasonable agreement between 
SIA and SIDIS fits,
but also with important differences.
As one can see, SIA and SIDIS fits are in good agreement for the 
central value of $zD^{h^+}_{c^+}$ and $zD^{h^+}_{b^+}$ FFs 
for all range of $z$. Reductions on the uncertainty of 
SIA+SIDIS also can be seen in respect to the SIA for both FFs. 
For the case of $zD^{h^+}_{u^+}$ FF, the SIA fit are 
larger than SIA+SIDIS for the range of $z$ down to $z\sim 0.1$ and smaller elsewhere, however similar 
size of error bands are obtained for both SIA and SIA+SIDIS fits.
For the case of $zD^{h^+}_{d^+ + s^+}$,  a smaller uncertainty band
is obtained and the central value for the  SIA+SIDIS fit is 
larger than SIA over the
whole range of $z$. 
Generally speaking, we find that the inclusion of the SIDIS data could affect 
the central value of the extracted FFs and leads to 
significant reductions of the uncertainty, and more specifically for 
the gluon FF.

We finally note that the value of total $\chi ^2$ per data point 
increases from 0.8 in SIA data only fit to the 1.079 in the 
global analysis of SIA+SIDIS data.
As we mentioned before in section~\ref{sec:TASSO}, the 
increasing of the value of the total $\chi^2$ is related to the 
tension between TASSO and the COMPASS experimental data sets. 
As we reported, the values of $\chi^2$ per data points for the 
TASSO data, more specifically TASSO 35 GeV, 
sets significantly increases after the inclusion the 
COMPASS data to the data sample.

\section{Summary and Conclusions}\label{sec:Summary}

In summary, we have presented a new global QCD analysis of 
light-charged hadron FFs, {\tt SHK22.h}, by 
introducing several new features and some methodological improvements.
On the methodological front, we have used the Machine Learning 
framework to extract the {\tt SHK22.h} FFs sets, along with the 
Monte Carlo uncertainty analysis.
This well-established fitting methodology is specifically 
designed to provide a faithful representation of the experimental
uncertainties. 
This methodology is also useful to minimize any bias 
related to the parametrization of the light-charged 
hadron FFs and to the minimization procedure as well.

In terms of the input data sets, in addition to the comprehensive set 
of high-energy lepton-lepton annihilation (SIA), we have added the 
lepton-hadron scattering (SIDIS) data sets to our data sample.
We have shown that SIDIS data sets have 
significant effect on the FFs, and more specifically on the gluon FFs 
and the reduction of its uncertainty.
The tension among some of the data sets included in our analysis 
also studied and discussed in details. 

The detailed comparisons to the existing light-charged hadron FFs sets 
({\tt NNFF1.1h} and {\tt JAM20}) fully demonstrate a 
reasonable agreement within the FFs error bands.
Although, some discrepancies in flavor dependence were 
observed, more specifically for the gluon and down-quark FFs. 
The resulting NLO theory predictions for the SIA and SIDIS 
cross-sections show very good agreement with the corresponding 
analyzed experimental data sets, as confirmed by 
the reported total $\chi^2$ per data point. 

Based on our findings in this study, one can conclude that 
adding the SIDIS data in the light-charged hadron study 
could lead to a much better 
level of precision of the extracted FFs.

In terms of future work, it would be interesting to revisit this analysis and 
study in detail the light-charged hadron FFs analysis 
described here considering the nuclear corrections in which we expect that 
it could affects the resulting FFs and their uncertainty, and 
could improve the description of the SIDIS data as well.  
Exploring the implications of such correction is left for the future work.

The parametrizations of the {\tt SHK22.h} light-charged hadron 
FFs presented in this paper are available in the standard {\tt LHAPDF} format~\cite{Buckley:2014ana},
and can be obtained from the authors upon request.

%
\begin{acknowledgments}
%

The authors gratefully acknowledge many helpful discussions and comments by Hubert Spiesberger and Valerio Bertone  which elevated and enriched the paper significantly.
Authors thank the School of Particles and Accelerators, 
Institute for Research in Fundamental Sciences (IPM) 
for financial support provided for this project. M. Soleymaninia is thankful
to the Iran Science Elites Federation for the financial support.
H.Khanpour also is thankful to the  
Theoretical Physics Department of Maynooth 
University and, University of Science and 
Technology of Mazandaran for financial support provided for this 
research. 

\end{acknowledgments}
%



\end{document}